"Alexandru Ioan Cuza" University of Iaşi, Romania

Faculty of Computer Science

# Habilitation Thesis

Exploiting Social Networks. Technological Trends

Adrian Iftene

- 2019 -

# Acknowledgments

**To my collaborators and to my family**

# Content

# I. Introduction

This thesis presents the author's research activity after March 2009 when he defended his Ph.D. Thesis "Textual Entailment" from the artificial intelligence domain, related to *natural language processing* (NLP).

NLP is a field of research that covers computer understanding and manipulation of human language, trying to make the machine derive meaning from human language in a smart and useful way, and performing difficult tasks such as *information retrieval and extraction*, *question answering*, *exam marking*, *document classification*, *report generation*, *automatic summarization and translation*, *speech recognition*, *dialogs between human and machine*, or other tasks currently performed by humans such as help-desk jobs. Continuing the work from this domain another great challenge of the NLP was approached, the one related to the creation of large textual resources, where the notion of *credibility* was introduced. When this activity is done manually by human experts, it is costly in terms of the time required to create them and in terms of the human resource to be involved. At the same time, these resources are the basic elements of NLP software applications, their quality depending on the size of the resources and their quality. For this reason, in recent years, automated methods involving social networks have proved to be a worthy method to consider, because here we have access on the one hand to many data, and on the other hand to the thoughts and feelings of users, their comments on events or products etc..

A new direction that emerged after supporting the Ph.D. thesis is related to the use of new technologies in applications, which will come to the aid of the users who use them. Applications have been made for smartphones or tablets most of the time, and have used more natural modes of user interaction with applications. We mention here, the *interpretation of user gestures*, *speech recognition*, the *exploitation of images made with the help of video cameras* and *the use of information taken by sensors*. *Augmented reality* and *virtual reality* are two current research directions that allow the creation of applications for increasingly diverse fields, such as elearning, games, interior design, museums, botanical gardens, medicine, etc..

There are two main directions: *exploiting data from social networks* and *using new technologies to improve the quality of life*, which corresponds to the next two sections. The last section contains conclusions and proposals for future work.

## I.1 Social Networks

Due to the increasing popularity of social networks, over the past few years, users have begun to expose themselves more and more as ever. Humans started to use a lot of pictures, different news to get informed about what their friends do or what is happening in the world. Many questions arise (1) *can we use all this information to create our resources*? (2) *can we have trust in all news or in all users that spread them*? (3) *can we improve the quality of our tools exploiting all this information*? We will see how we build our resources based on images from the Flickr network and what is our proposal to deal with the credibility of these resources. Currently, on Twitter, we do not have an automatic method of figuring out in real-time how to monitor the user's credibility and credibility of tweet issues that thoroughly clarify what fake news means.

Big companies like Facebook and Twitter have repeatedly stated that they have real problems because of fake accounts and messages posted by them. Mark Zuckerberg has claimed that there are more than 1 billion fake accounts on Facebook[1]. With these accounts, opinions can be created, voters can be influenced, questions can be asked about events presented in the press or on television, and the honest users can be misinformed, etc. We will see our different approaches to identify fake accounts and fake information and what are our plans for the future.

## I.2 New Technologies

The technological evolution and the evolution of the way we think about the applications and the way they interact with them they have changed into alert rhythm in recent years. Already smart phones take the place of key phones, we use a lot of voice and gesture interaction, and visualization is done with the help of special glasses that show us a virtual or augmented world. Again, many questions arise: (1) *Where are we going in the next period*? (2) *How will the interaction mode evolve with the applications that will be created in the future*? (3) *How will the*

---

[1] https://www.ccn.com/facebook-billion-fake-account-zuckerberg-con-man/

*lessons be and how will students interact with teachers and how virtual or augmented teaching materials look like*? We will see how we can create more attractive lessons that use augmented and virtual reality, that combine visual information (images, videos and animations), with textual and sound information. Voice interaction is a more natural way to interact with applications and it is preferred by children and the elderly, in classes or in smart houses.

After Google invested 542 million dollars in 2014[2] in a start-up on augmented reality, after heavy investments of Facebook over the years[3], in 2018 the biggest investment in augmented reality was done by Epic Games which raised 1.25 billion dollars[4] for that. The great interest of the important companies makes us think that in the coming years the applications in this area will be more and more and in different fields. We will see details about some applications created by us and what our next proposals are.

## I.3 Thesis Structure

After this introductory part, the thesis is divided into three main sections "Exploiting Social Networking Data", "Technological Trends" and "Final Conclusions and Future Work". Every section begins with context and state of the art, followed by our contributions and results. The sections are closed with conclusions and proposals for future work and with bibliography.

**Exploiting Social Networking Data** section presents:

- How we create **textual** and **visual resources** in MUCKE (Multimedia and User Credibility Knowledge Extraction) project and in international competitions and evaluation campaigns where we deal with social networks;
- Applications created to identify **social media credibility** and **sentiments** or to perform **image retrieval** or to show in a multi-dimensional manner information from Twitter;
- The software components created for image or text processing were tested in **evaluation campaigns**, where we obtained very good results during the time.

---

[2] https://www.theverge.com/2014/10/21/7026889/magic-leap-google-leads-542-million-investment-in-augmented-reality-startup
[3] https://newsfeed.org/facebook-continues-to-invest-more-in-augmented-reality/
[4] https://next.reality.news/news/10-biggest-ar-investments-2018-0191870/

**Technological Trends** section contains:

- A first big subsection presents the notion of **augmented reality**, where we contribute with applications for eLearning, gastronomy and museum domains;
- Next section is allocated to **virtual reality** and to an application from astronomy domain;
- Another big section is allocated to the voice applications created for Amazon Alexa, from domains like education, tourism and Internet of things.

**Final Conclusions and Future Work** section draws the main future directions of research:

- Identification of fake news in real time using semantic resources like DBpedia, Wikidata or YAGO;
- Collaboration with Botanical Garden from Iasi to create an augmented reality application for identification of flora and fauna elements;
- Collaboration with colleagues from "Grigore T. Popa" University of Medicine and Pharmacy of Iasi to create applications with new technology and to create an application that uses artificial intelligence to provide a better quality of life for patients or to identify faster problems in evolution of diseases.

# II. Exploiting Social Networking Data

## II.1 Context

**Introduction**

In recent years we have had a significant increase in interest for social networks. On the one hand, users use these networks to create and distribute **textual information** (thoughts, opinions, experiences, etc.), **multimedia content** (images, movies, audio files) taken during holidays, concerts, trips, events with various occasions (weddings, baptisms, anniversaries, etc.), at the restaurant or in parks, etc. Their purpose is to communicate with friends and family, to share experiences, feelings and to engage remotely with those who cannot be near those who post. On the other hand, the researchers found an area full of information and resources that they can exploit and use for various purposes, most often free of charge. For example, it's very easy to use the Twitter API[5] to access the latest posts on this network. Posts can be filtered according to the *keywords* that appear in them, depending on the *language* of the posts and can be a resource of applications in the medical field or for identifying feelings.

**The Most Popular Social Networks[6]**

- **Blogs**[7] is a platform that allows users to discuss on a specific topic and to express their opinions and sentiments.
- **Facebook**[8] is currently the largest social network in the world, with over 2.375 billion active monthly users, 1.49 billion active daily users, in 2019. Interesting are the facts that 45% of the users take daily news from Facebook, which generates 4 new petabytes of data per day[9]. A user can create a personal profile, add other existing users as friends, and exchange messages in a chat component. Large companies can create their pages, and usual Facebook users can follow the latest news through these pages.

---

[5] https://developer.twitter.com/en/docs.html
[6] https://communications.tufts.edu/marketing-and-branding/social-media-overview/
[7] https://www.blogs.com/
[8] https://www.facebook.com/
[9] https://www.brandwatch.com/blog/facebook-statistics/

- **Twitter**[10] is a social network platform that allows users and groups to post short messages (140 character limit). There are now 1.3 billion Twitter accounts, which send 500 million tweets every day[11].
- **YouTube**[12] **and Vimeo**[13] are used for video hosting and viewing websites. YouTube has in present 1.9 billion monthly users and 400 hours of video are uploaded every minute[14].
- **Flickr**[15] is an image and video hosting website, which has an entire online community behind it. From this social network photos can be shared on other social networks like Facebook and Twitter. There are over 90 million monthly users which shared over 500 million images under Creative Commons license[16].
- **Instagram**[17] is an application that allows users to share photos and videos. Users can process their photos and can apply digital filters and special effects. In 2019, there were 1 billion active users and over 40 billion photos were sharing[18].
- **Snapchat**[19] is a mobile app that allows users to send photos and videos to their friends or add them to their "story". There are 203 million daily active users, which post on average 3.5 billion Daily Snaps[20].
- **LinkedIn Groups**[21] is a place where professionals with similar fields of interest can create groups and can share information on interest subjects. LinkedIn has over 610 million members[22] and the total number of LinkedIn Groups is over 2 million, where it appears 200 discussions per minute[23].

---

[10] https://twitter.com/home
[11] https://www.websitehostingrating.com/twitter-statistics/
[12] https://www.youtube.com/
[13] https://vimeo.com/
[14] https://www.brandwatch.com/blog/youtube-stats/
[15] https://www.flickr.com/
[16] https://expandedramblings.com/index.php/flickr-stats/
[17] https://www.instagram.com/
[18] https://www.brandwatch.com/blog/instagram-stats/
[19] https://www.snapchat.com/
[20] https://zephoria.com/top-10-valuable-snapchat-statistics/
[21] https://www.linkedin.com/
[22] https://99firms.com/blog/linkedin-statistics/
[23] https://expandedramblings.com/index.php/linkedin-business-page-and-group-statistics/5/

## II.2 Resources

### II.2.1 Textual Data

**Introduction**

Semantic Analysis in Social Media (SASM) refers to language processing of messages from social media, which we supplement with semantic information and meta-data from social networks (Farzindar and Inkpen, 2015). See Figure 1 below how SASM is based on collecting data from social networks, and how the results obtained by SASM can be used by single users or by decision-making groups that analyze this data.

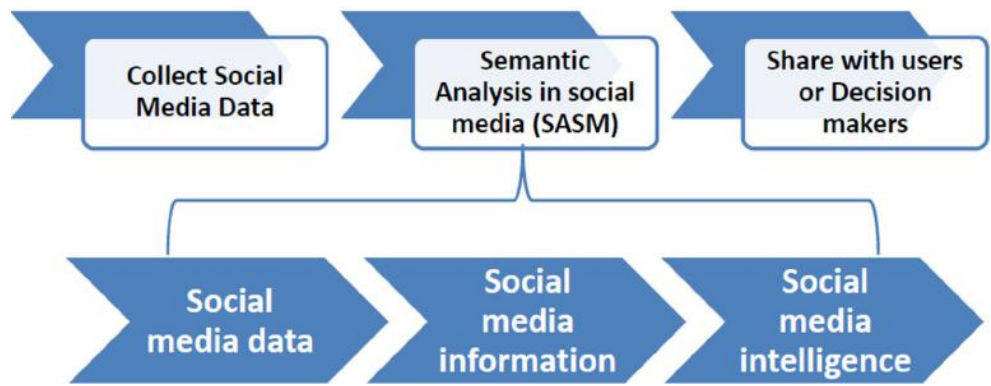

**Figure 1**: Exploitation of information from social networks (Inkpen, 2017)

We can obtain the properties of the data from social networks (Inkpen, 2017):

- They can be obtained for **free** (the Twitter API gives us access to some of the data they have for free, and to access them all you need to pay a monthly subscription);
- There is information that is posted in **real-time** on the social network (immediately after posting, the information is indexed and is offered as a result when they meet the search criteria);
- They often have **geospatial information** (we must take into account the fact that following GDPR rules, this information is no longer saved by default, but often users allow the use of this information, to easily transmit to what event it is found or easily search for information about nearby places);

- They have specified the **language** in which they are written (this allows us to use language processing tools specific to the language of the job);
- Have **emoticons** or **hashtags** that allow us to quickly identify the feelings of the one who wrote the post or the field of posting, making it easy for us to make connections with other previous posts or other events that have been written before;
- The text that appears is **unstructured**, is written by unprofessional people using abbreviations, acronyms, emoticons, punctuation marks often in excess, capital letters, repetitions of letters and words to underline something.

The language preprocessing, which applies to the tweets, is made with traditional natural language processing tools, which have been adapted to how to write specifically to social networks. The most used processes for natural language processing are performed with:

- Lemmatizer (identifies the root of the word);
- Tokenizer (splits the text into words);
- POS Tagger (identifies the parts of speech);
- NER (identifies and classifies the name entities);
- Chunkers and parsers (which group words between links and extract additional syntactic and semantic information);
- Language and dialect identifiers (when the language of posts is not specified).

To adapt these tools to social networks, it is necessary to create specific resources with training data taken from social networks. Also, for the processing of data from social networks, it is necessary to use specific tools that treat the special character of the posts:

- to *remove duplicates* (letters or words),
- *replace the abbreviations* with the multitude of words that represent it,
- *replace the capital letters* with the normal letters,
- if it is necessary to *replace even the emoticons* with the words that represent their significance.

Some of the most used tools for natural language processing are:

- **Stanford CoreNLP - Natural language software**[24] (Manning et al., 2014) (we can see an example of using it in Figure 2);
- **Apache OpenNLP**[25];
- **Natural Language Toolkit** (NLTK)[26];
- **GATE**[27] (Cunningham et al., 2013);
- **FreeLing**[28].

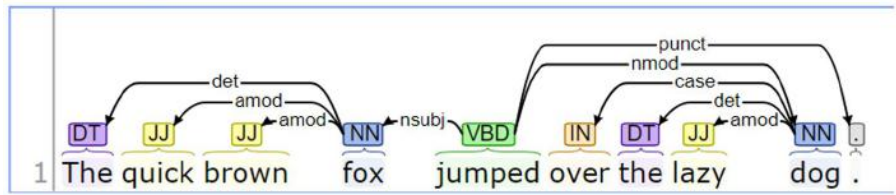

**Figure 2**: Example of using Stanford CoreNLP[29]

Most of them also offer support for social networks, and in addition, they are tools specially built to process information from social networks:

- **TweetNLP**[30] (an example is shown in Figure 3 below);
- **Twitter NLP Tools**[31].

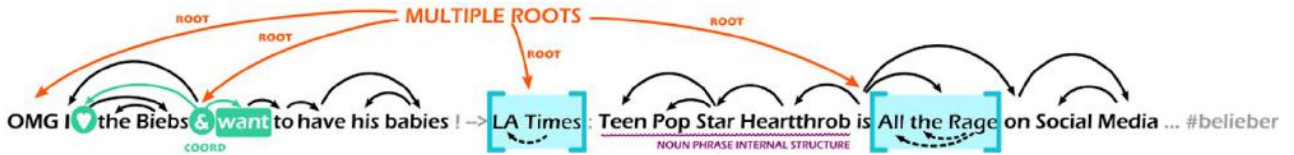

**Figure 3**: Example of using TweetNLP[32]

---

[24] https://stanfordnlp.github.io/CoreNLP/
[25] https://opennlp.apache.org/
[26] https://www.nltk.org/
[27] https://gate.ac.uk/
[28] http://nlp.lsi.upc.edu/freeling/index.php
[29] http://www.linguisticsweb.org/lib/exe/fetch.php?w=300&tok=894016&media=linguisticsweb:tutorials:linguistics_tutorials:automaticannotation:stanford-corenlp-example.png
[30] http://www.cs.cmu.edu/~ark/TweetNLP/
[31] https://github.com/aritter/twitter_nlp
[32] http://www.cs.cmu.edu/~ark/TweetNLP/deptree.jpg

**Identifying Similarities between Tweets**

In some situations (like *event detection* for example), we want to see news on Twitter, without duplications. To remove similar news from our visualization system, there were used few similarity algorithms which calculate the distance between two tweets (Minea and Iftene, 2017). For that the next four similarity distance algorithms were compared in terms of number of identified similarities: the *Levenshtein* (Levenshtein, 1966), the *Needleman-Wunsch* (Needleman and Wunsch, 1970), the *Jaro-Winkler* (Jaro, 1989) and (Winkler, 1990) and the *Smith-Waterman* (Smith and Waterman, 1981). Performed experiments show that the Smith-Waterman is what we are looking for, as it is able to find the similarities between a tweet and its retweet.

Also, to see which algorithm is faster the above similarity distances were applied on a set of 2,000 tweets. The faster algorithm was the Jaro-Winkler, 3 to 9 times faster than the others. Thus, when we need quality of the comparison and the time is not important, we use the Smith-Waterman algorithm, but when we need to show results in real-time, we use the Jaro-Winkler algorithm.

**Applications of Data Mining in Social Media**

Applications that use data from social networks are becoming more and more diverse. Areas of applicability include the *medical* field, *financial* applications, *voter prediction*, *security* applications, identification and action in the event of *natural disasters* or *disasters*, *user profiling*, *entertainment* applications, *real-time monitoring of data* in the social environment. Below we will see more details for some of these areas (Inkpen, 2017).

*Medical Domain*

This includes specialized platforms or groups on specific topics within the social networks, where they discuss with *pros and cons about vaccinations*, *mammograms*, the *positive/negative effect of certain treatments*, etc. Discussions are usually informal and care must be taken on the personal data of the users participating in the discussions, to protect their identity (*name, address, date of birth*, etc.).

Another increasingly exploited direction is related to the identification at an early stage of signs indicating *mental illness* (*depression, suicidal intentions, self-injury, anorexia*, etc.). During the CLEF labs[33], exercises were introduced in recent years, specific in the eRisk lab[34], which aims to identify such risks on the Internet. Usually, the organizers make available to the participants annotated data with information containing relevant information to identify a certain type of behavior. Based on these training data, which are labeled with useful information, it is required that labels be added to a collection of test data (with unlabeled data) to signal the targeted behaviors. Participants use techniques in the area of artificial intelligence, such as machine learning, neural networks, deep learning, etc. During this lab, we participated in the task of identification of anorexia (Cușmuliuc et al., 2019). We analyze different techniques of detecting early signs of anorexia in social media, their performance and how we can fine-tune them to improve the actual results.

*Financial Applications*

Behavioral economics studies the correlations that are made between the status of users in social networks and economic indicators, between financial news and fluctuations of the stock market. Studies in recent years have shown that data from social networks (such as Twitter[35], Sinaweibo[36], Seeking Alpha[37]) can be successfully used to identify users' moods, which can then be exploited in financial applications. Several experiments have been done successfully to predict market fluctuations for NASDAQ and other stock markets.

*Predicting the Intention to Vote*

In recent years, in the elections of any type, besides the classic electoral struggles, intense struggles also take place within the social networks, between those who support the participants in the vote. After filtering the posts according to the topic we want to pursue (which is usually done based on keywords), we have access to users' opinions and opinions. By classifying these opinions into positive and negative opinions, we can realize the support/opposition that someone or something has. Numerous studies have been conducted based on data collected during the Senate elections in the Netherlands (Tjong Kim Sang and Bos, 2012), the general elections in Ireland (Bermingham and Smeaton, 2011), and the American elections (Yaquba et al., 2017), etc. Also interesting is the study carried out by Gayo-Avello in 2013 related to the influence of social networks in elections in different areas of the world (Gayo-Avello, 2013).

---

[33] http://clef2019.clef-initiative.eu/index.php?page=Pages/labs_info.html
[34] http://erisk.irlab.org/
[35] https://twitter.com/
[36] https://www.weibo.com/
[37] https://seekingalpha.com/

In the article (Bender, 2017), the author shows how data from social networks can predict very well what will happen in the elections, and the detailed analysis of this information should be a priority for campaign leaders in the future.

*Security and Defense Applications*

The statistics on social networks are impressive: there are almost 3.5 billion active social media users, who produce daily on Facebook and Whatsapp only around 60 billion messages (text, images, video, audio, etc.) (Smith, 2019). With the volume of data being so high, people can only read part of this data, and their analysis is done largely automatically to detect criminal threats to public safety and security. Once detected by software applications, which use keyword lists, messages are read and analyzed by human users, who can decide whether the threats are real or not. Similarly, images are automatically analyzed, either based on the associated keywords and titles or using collections of annotated images. Then they are classified in images that have contained terrorist or not.

Detecting emotions, especially anger and hatred, in social media posts can also be alerted for possible terrorist messages. Especially when their intensity is significant, as shown in the studies in (Ghazi et al., 2010), (Keshtkar and Inkpen, 2012), (Iftene et al., 2017) and (Şuşnea and Iftene, 2018).

*Managing Crisis Situations*

Social networks benefit from the rapid reaction of users regardless of the time and regardless of the event reported by them. Discussions between users in a particular area on an issue related to *earthquake*, *tsunami*, or *fire*, or anything else, may signal the occurrence of an extreme phenomenon in that area. Numerous experiments have been done on the detection of events on Twitter: about *disasters* (Imran et al., 2013), about *earthquakes* (Robinson et al., 2013), about *fires* (Power et al., 2013), about *protests* (Iftene and Gînscă, 2012), etc.

*Creating Profiles for a User*

All user actions (*posts, likes, comments, searches, views*, etc.) are monitored, saved, analyzed within social networks and are used to create profiles for them. These profiles are accessed when we are offered advertisements for certain products, or when we are offered certain promotions at the accommodation, or when there are events taking place either near us or on topics that interest us, etc. (Şerban et al., 2016). Usually, a user is created several types of profiles: from the *medical* point of view, from the *gastronomic* point of view, from the *religious* point of view, from a *political* point of view, from the point of view of the *events* in which he participates, from the point of view of *locations* where it is active in the network, etc. All this contributes to the creation of a unique "fingerprint" for the user. More details on current approaches can be found in CeADAR project[38], which aims to uniquely identify a user on multiple social networks. To improve the quality of these components, or to improve the quality of the information that they have about a user or about the objective, etc., short questionnaires have started to be offered later, which the user may or may not complete.

*Entertainment Applications*

The researchers studied the activity of users on social networks for important events, such as the Oscars, or the 2012 Wimbledon final, but also regarding the movies that are broadcast at one time on Netflix, or regarding the forecast, or regarding the audience of a television station at any given time. The frequency of the posts, as well as the feelings that appear in them, can be a measure of the experiences of those attending these events. The statistics and the graphs realized can help to identify the key moments through which the actors or players pass during the course of the events. Netflix uses the popularity of Facebook to suggest to users a specific movie. Also, part of the success of Netflix is based on the attention they give to the activities of their social media users (Franck, 2017).

---

[38] http://www.ceadar.ie/

**Conclusions**

The fact that social networks have become a part of the lives of a large number of people, has led to the adaptation of the way in which today's applications are thought and built. They exploit information from these networks to see users' opinions about a product, about a political personality, about a party, about a company, about an event, etc. The fields of applicability are increasingly diverse and it is anticipated that in the future their area of applicability will grow even more. The researchers found a suitable place from which they can easily collect their information, based on which they can build resources, which are the basis of the algorithms they work with.

## II.2.2 Visual Data

**Introduction**

Recent years have seen rapid growth in the *digital image*, *audio* and *video* collections. However, we cannot access or use this information properly, if they are not organized so we can quickly search and find what we want. For example, the most used techniques by search engines to find images are:

1. *Image File Search* - The search engine detects the presence of an image by detecting specific tags , when inspecting web pages, or image-specific extensions, the most common are *.gif* (Graphics Image File), *.tif* (Tagged Image File) and *.jpg* (Joint Photographic Experts Group). Internet Graphic Hunter[39] is one of the search engines that uses this technique, and which, based on a *URL*, returns all the images present at that address.
2. *Keyword Search* - Each image has metadata associates (*title*, *keywords*, *description*, etc.) that include information about its content. Image annotation can be done manually, by the user, or automatically by means of automatic learning algorithms, which try to find the correspondence between the visual characteristics and their semantics. Numerous problems can arise in the case of this approach, which results both from the incorrect or incomplete annotation of the images, as well as from not using the right words in the case of the queries.

---

[39] http://internet- graphic-hunter.en.softonic.com/

3. *Search by Title* - The search engine can browse web pages whose titles indicate the presence of images corresponding to the subject is searched. This technique can work properly if the titles are a good match for the content of these pages, which is not always true. This method can be used if the images are indexed based on their titles, as is the case of the ARTCYCLOPEDIA[40] collection, which allows finding the pictures searched by specifying words found in their title.

4. *Manual Indexing* - There are search engines that employ specialized personnel browsing the web pages to find and index the images present. This work is time-consuming and, although effective, is used only in a limited search space. An example is The Library of Congress image archive[41], where each image has an *identifier*, a *title*, and a *description* on which to search.

5. *Content-Based Search* - In this case, the search is performed based on the semantics of an image, abstracting from the metadata associated with it. This content consists of *colors*, *shapes*, *textures* - virtually any kind of information that results from the image itself. This method is preferred over the others presented above, as both metadata-based search and manual annotation can result in the intoxication of images with no relevance to the initial query.

The system that we built in MUCKE project[42] (Iftene, 2017) allows searching images in a collection with over 90 million images, which was built using the main platforms that collect and index images: Google, Bing, Flickr, Picasa, Photobucket, Panoramio, etc. The architecture considered during the project development can be seen in Figure 4 below.

---

[40] http://www.artcyclopedia.com/
[41] http://www.loc.gov/index.html
[42] https://profs.info.uaic.ro/~mucke/

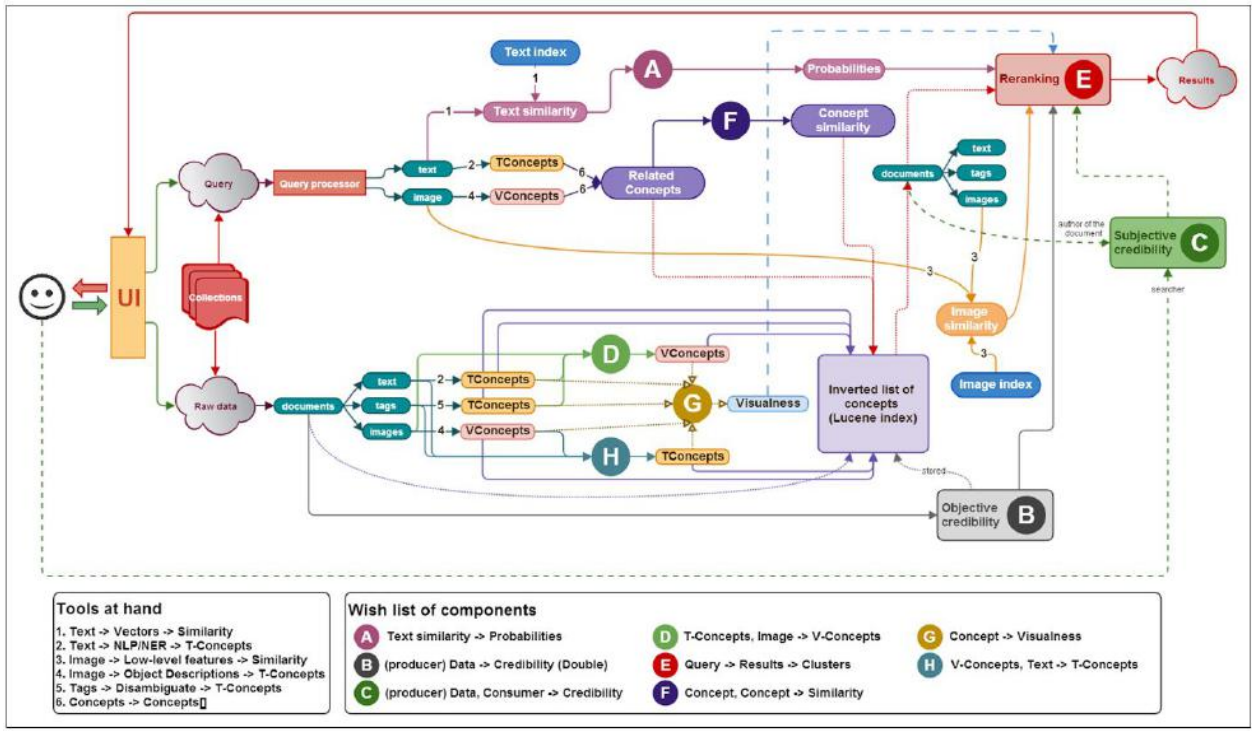

**Figure 4**: Data flow within the multimedia search system based on concepts and credibility (Bierig et al., 2014)

**Creating a Corpus of Images on Flickr**

Within the MUCKE project, we created an image corpus based on the images on the Flickr[43] platform. The data was collected using the public API provided by Flickr and the operation complied with the conditions set out in the Flickr Community Guidelines. The complexity of the project objectives required the involvement of various multimedia data sources to extract all the knowledge associated with the user. With over 10 billion photos being shared, Flicker is one of the most important image repositories and it has been the main source of visual information in MUCKE project. An example with images from the Flickr platform can be seen in Figure 5.

---

[43] https://www.flickr.com/

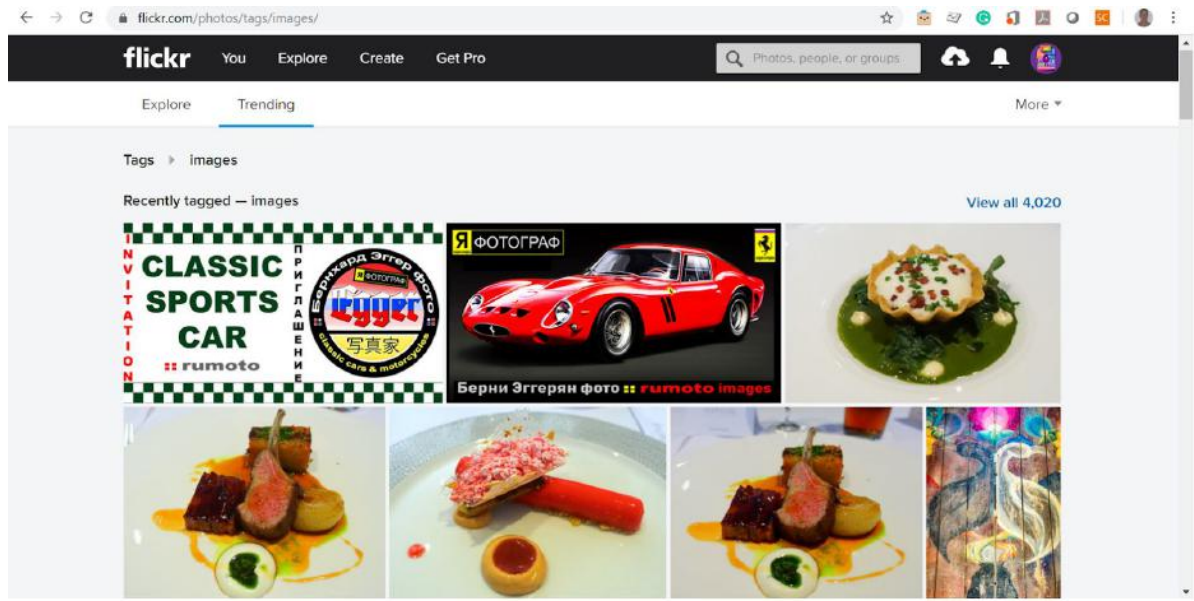

**Figure 5**: Example of images on the Flickr platform

Consortium members[44] with experience in searching and processing large data repositories have taken information (images and text metadata) for more than 1,000,000 Flickr users. Because Flickr contains a conglomerate between personal and social data, the focus was on the second category that was relevant for the tasks of image extraction. The group from UAIC[45] coordinated the entire data collection effort, but to increase the speed of the entire process, it was distributed among all the project partners.

Flickr offers a filtering system that allows its users to mention both the types of photos they have uploaded and the images they want to access. Flickr offers a complex API[46], which can be accessed through both REST and SOAP, with extensive documentation and API kits for every modern programming language.

One of the main features that Flickr presents is how it organizes the images that users upload (Alboaie et al., 2013). They can tag their photos, which will allow other users to easily find images related to a particular domain. Flickr is one of the first applications that implemented the visual representation "tag cloud" and is also considered a good example of the effective use of a folksonomy. In addition to tags, an image posted to Flickr has associated metadata such as *title,*

---

[44] https://profs.info.uaic.ro/~mucke/partners.html
[45] "Alexandru Ioan Cuza" University
[46] http://www.flickr.com/services/api

*author, the date the photo was taken, date uploaded to Flickr, number of views*, etc. There is also the possibility to access the EXIF data of some images, which refers to the *camera, light, aperture, focus, orientation, color spectrum*.

Flickr offers public and private image hosting, with the possibility of releasing images under Creative Common Use Licenses[47] or labeling them with "all rights reserved". The images we downloaded in the MUCKE project were public and appear under the Creative Commons license.

**The MUCKE Corpus**

The Mucke corpus contains around 90 million entries with metadata and images based on *Wikipedia concepts* (Spitkovsky and Chang, 2012), which are often used to annotate Flickr images. Wikipedia concepts are sorted by the number of Flickr images that correspond to a concept, divided by the number of links returned by Wikipedia, to penalize very common concepts. A sample from the Flickr metadata file is below:

```xml
<?xml version="1.0" encoding="utf-8" ?>
<rsp stat="ok">
<photos page="1" pages="2" perpage="500" total="896">
      <photo id="123456789" owner="1122334455@N66" secret="77889900"
server="1234" farm="5" title="Hercule Poirot's Chrismas, Agatha Christie"
ispublic="0" isfriend="1" isfamily="1" datetaken="2000-11-22 20:12:34"
datetakengranularity="2" tags="collins agathachristie herculepoirot
fontanabooks alexisorloff romanspoliciers vintagedetectives
vintageagathachristiebookcovers" dateupload="123454321" views="123" />
```

The following parameters were extracted using the public Flickr API:

- *photo id* - unique Flickr image registration number;
- *owner* - unique registration number of the Flickr account that uploaded the image;
- *title* - the title of the image taken from Flickr;
- *datetaken* - image creation date;
- *tags* - tags added by Flickr user;
- *dateupload* - upload date of Flickr of the image;
- *views* - number of unique image views.

---

[47] http://www.flickr.com/creativecommons

**Internal Sharing Mechanism**

The system architecture was based on Lucene[48] application (for the text retrieval part) (McCandless et al., 2010), combined with LIRe[49] (for the image retrieval part) (Lux and Marques, 2013), (Şerban et al., 2013). LIRe is a simple but efficient open-source library built over Lucene, which offers a simple way to find images from the content. LIRe creates a Lucene image index and provides the mechanism for searching in this index, but also for navigating and filtering of the results. Being based on an integrated text search engine, it is easy to integrate into applications without the need for a database server. Moreover, LIRe can be used for up to millions of images due to the approximate indexing mode with hash tables.

LIRe is built using Lucene, an open-source text search engine. As in the retrieval text process, images must first be indexed to be retrieved later. Field documents, which have name and value, are organized in the form of an index.

The system uses a modular architecture, which will allow the dynamic integration of new modules or algorithms to achieve better results in the future.

**Extraction of Visual Properties and Indexing**

Using LIRe, the following properties of raster images can be extracted, indexed and searched (Lux and Marques, 2013), (Lupu et al., 2013), (Alboaie et al., 2013):

- **Color Histograms in RGB** (Red-Green-Blue) and **HSV** (Hue-Saturation-Value) spaces. These histograms contain the color distribution in an image;
- **MPEG-7 Descriptors scalable color, color layout, and edge histogram**. MPEG-7 contains features that allow structural and detailed descriptions of audio-video information;
- **Tamura Properties** for *roughness*, *contrast,* and *directionality* are used in LIRe;
- **Color and Edge Directivity Descriptor** (CEDD) contains the color and edge directivity descriptor for an image;

---

[48] Lucene: http://lucene.apache.org/
[49] LIRe: http://www.lire-project.net/

- **Fuzzy Color and Texture Histogram** (FCTH) contains the fuzzy color and texture histogram. This property also combines color and texture information into a single histogram;
- **Joint Composite Descriptor** (JCD) represents the common composite descriptor. JCD was designed for natural color images and it can be obtained from the combination of CEDD and FCTH;
- **Auto Color Correlation Feature** represents the property for automatic color correlation.

To create an index and search for its content, follow these steps (Alboaie et al., 2013):

1. A *document* is created for each image from a collection. It can contain both text fields and visual features for images (from the ones mentioned above);
2. To perform searches, you must first create a *query document*. This document must contain the search criteria composed by textual or visual fields.

The search result using this query is a list of documents with scores attached. These scores represents the degree of match between an index and query document (1 is the best score, and 0 the worst).

## II.3 Contributions

In this section, we will see a few directions where we start to develop applications that exploit information from social networks. Part of this work was done with students during the software engineering classes at bachelor level and during the advanced software engineering classes at master level. Many ideas were developed later like bachelor or master theses, and after that, students were involved active in writing scientific articles for different conferences.

### II.3.1 Social Media Credibility

**Introduction**

Since the 1990s, the popularity of social networks has increased significantly, thereby impacting business (Edosomwan et al., 2011). For news consumption, these social networks respond very well on two plans to users: (1) low costs, easy access, rapid dissemination of information, and (2) widespread of low quality news (often intentionally to mislead readers) (Shu et al., 2017), known straight "fake news". The rapid spread of fake news has become a natural activity, with a high negative potential for users and societies they are part of. The news is also *marked with opinions* (see Facebook, 2004), *retransmitted* (retweets on Twitter, share on Facebook) without having to check many times whether it is true or false news.

The problem that arises, *how we identify whether or not news/ users are trustworthy or not*? The proliferation of misleading information in social media (Perez-Rosas et al., 2017) is increasing the need for computational approach able to analyze the reliability of online content centered on credibility term. In what follows, we focus on the automatic identification of the user's credibility and credibility of tweet issues that thoroughly clarify what fake news means. Our contribution is twofold. First, we introduce two elements for the task of fake news recognition, covering a fertile area for spreading untrustworthy information to an impressive number of people in real-time. Second, we conduct a set of methods to build accurate classifiers of authors of tweets and messages posted by them on Twitter. This study is challenging, given the huge amount of data collection that is incomplete, unstructured and noisy.

Currently, we do not have a comprehensive automatic method of figuring out in real-time for evaluating the credibility of online news. In this work, *fake news* focuses on classifying the credibility of a tweet post (Wu and Liu, 2018), in close contact with the authors of tweets.

The *credibility of news* spread through social media networks become very attractive, presenting unique features and challenges that make any kind of algorithms ineffective (Ishida and Kuraya, 2018). In recent years, many studies have begun to address the issue of identifying fake news (non-credible information). The literature includes articles about the credibility of *TV and YouTube video information* (Ciampaglia et al., 2015), (Clark, 2009), *Twitter* (Castillo et al., 2011), (Chu et al., 2010), (Cook et al., 2012), (Iozzio, 2012), (Atodiresei et al., 2018), (Cuşmuliuc et al., 2018), *Facebook* (Allcott and Gentzkow, 2017), (Chen et al., 2015). Furthermore, a few sites train and help users who want to identify fake news (TenQuestionsForFakeNews[50]). There is still a lot of talk about the influence of Twitter on the US elections in 2016 (Bovet and Makse, 2019) and the fact that there is a lot of fake news on Twitter (Brummette et al., 2018).

---

[50] https://teachingcivics.org/lesson/ten-questions-for-fake-news-detection/

Even if we witness a permanent struggle between fake news and real facts is currently playing on social media, the technology is fighting back to identify them and notify the online consumer. In (Conroy et al., 2015) the notion of detecting fake news is "*defined as the task of classifying news across a continuum of veracity with an associated measure of certainty*". Although designing a system for detecting fake news is not a simple matter, the most promising directions for conceiving the most efficient system were almost the same after 2015:

(1) **Linguistic approaches** – are based on involuntary "leaks" of speakers, and existing methods are trying to catch such anomalies (Mihalcea and Strapparava, 2009).
   - *Representation of data* - typically uses statistics on *n*-grams, which are analyzed to identify fake information (Hadeer, 2017), (Hadeer et al., 2017).
   - *Advanced linguistic structures* - sentences are transformed into more advanced forms of information representation (such as parsing trees), which then analyze probabilities attached to identify anomalies (Conroy et al., 2015), (Perez-Rosas et al., 2017).
   - *Semantic analysis* - it analyzes semantically the contents of a user's statements, constructs pairs of the attribute form: descriptor and calculates compatibility scores (Shu et al., 2017).
   - *Rhetorical Structures and Utterance Analysis* - relations between the linguistic elements are built, which help determine the proximity to the centers of truth or deception (Popoola, 2017), (Rubin and Lukoianova, 2014), (Rubin et al., 2016).
   - *Classifiers* - SVM classifiers or Naïve Bayesian-type classifiers are used to predict future clutter-based fraud and distances (Rubin et al., 2016), (Sing et al., 2017).
   - *Deep learning* - use neural networks that identify fast fake news (Bajaj, 2017), (Sneha et al., 2017).
(2) **Social networking approaches**
   - *Linked data* - knowledge networks are exploited to identify the lie (Conroy et al., 2015), (Idehen, 2017).

- *The behavior of users on social networks* - the fact that users are forced to authenticate when using the social network, provides increased confidence in the data that appears here (Shu et al., 2017), (Shu et al., 2019).

In the last few years, the **hybrid approach** (combining machine learning in computational linguistics with social networking approaches) seems very promising. Compared to previous approaches, which displays all the tweets collected over a certain period, we can manage to display credible or not credible tweets, credible or not credible users' tweets, and statistics on countries and continents in real-time.

**Data and Methods**

We describe the data set (tweets) and different formulas to calculate the user's credibility and credibility of tweets. In general, a similar approach involves the application of four general processes: *sampling, synchronic analysis, lexical framing, and diachronic analysis* (Berente et al., 2018). In our case, we consider three general processes: sampling, lexical framing, and statistical analysis.

**Data Set**

To compute credibility scores for tweets, we needed conclusive data. For that, firstly, we collected more than 2,500 tweets (from 50 users with at least a few thousands of followers). To build an annotated dataset, we collaborated with 5 experts for data annotation. Each annotator worked independently on others and his/her task was to classify tweets and user's Twitter (*credible* and *not credible*).

For tweets considered credible, the annotators assigned 0 and for not credible tweets, 1. All *ambiguous tweets* (tweets on which annotators have held controversial discussions whether it is credible or not) were eliminated. After that, we remained with 2,270 tweets (1,248 not credible and 1,022 credible). Retweets were considered not credible because we were not able to efficiently retrieve additional information used when computing the credibility score (Twitter API impose some limits that were too low to be usable within the terms of response time, effectiveness and usability).

For each tweet, we collected the following types of information: (1) *retweetsNo* - the number of times the current tweet was retweeted; (2) *favoritesNo* - the number of times the current tweet was marked as favorite; (3) *creationDate* - the date this tweet was posted; (4) *wordsNo* - the number of words in the current tweet (excluding stopwords); (5) *relevantWordsRatio* - ratio between the number of words within the text that are not stopwords nor punctuations and the total number of words; (6) *charactersNo* - the number of characters in the current tweet.

Additionally, for each user we collected the following information: (1) *The most recent 40 tweets* posted by that user; (2) *hasLocation* - true, if user filled the location field, false otherwise; (3) *hasDescription* - true, if user filled the description field, false otherwise; (4) *hasGeo* - true, if user turned on geolocation, false otherwise; (5) *isVerified* - true, if user was verified by Twitter, false otherwise; (6) *creationDate* - the date when the account was created; (7) *followersNo* - the number of followers.

For our experiments we selected different entities from: **politics** (*Donald Trump, Barack Obama, Hillary Clinton,* etc.), **business** (*Tim Cook, Bill Gates*, etc.), *companies* (*Google, Microsoft*), **organizations** (*Discovery, NASA*, etc.), **television** (*CNN, NatGeo*, etc.), **music** (*Eminem, Justin Timberlake, Miley Cyrus,* etc.), **sport** (*Maria Sharapova, Simona Halep*), **other** (*Android, Kim Kardashian, Dalai Lama,* etc.).

**Methods**

*Using Formulas to Calculate Credibility for User and Tweet*

In the initial approach, we composed a specific formula which proved to be relevant for credibility score. To compute the tweet's credibility, we consider the following formula:

$$TweetCredibility_{score} = w_R \times T_R + w_F \times T_F + w_W \times T_W + w_S \times T_S$$

where $T_R$ represents the *retweets score* (the number of retweets divided by the number of reachable followers of the author, we considered that a tweet reaches 3% of the followers base just by posting it), $T_F$ represents the *favorites score* (number of this the tweet was marked as favorite divided by number of reachable followers), $T_W$ represents the *ratio of relevant words* contained by tweet's text, $T_S$ represents the *sentiment score* (cumulative sentiment score for tweet's text computed using the Stanford Sentiment Analysis component; very negative words weighted 0.75, negative and very positive words weighted 0.50, positive words weighted 0.25, neutral words weighted 0.00).

To identify the best distribution for the weights from tweet' credibility formula, we performed more experiments on our annotated tweets and in the end we came up with the following weights: $w_R = 0.1$, $w_F = 0.3$, $w_W = 0.5$, $w_S = 0.1$ for which we have the best precision.

To compute the **user's credibility**, we considered the following parameters: (1) the *location*, (2) the *URL*, (3) the *description*, (4) if he *is verified or not*, (5) the *geolocation*, (6) the *creation date*, and (7) the *most recent 20 tweets* tweeted by this user.

Below is the formula that covers 50 Twitter users included in our database. We saved the result for later comparison to find the best values for weights that respected the associated credibility to these users and their tweets.

$$UserCredbiliyscore = w_L \times U_L + w_U \times U_U + w_D \times U_D + w_V \times U_V + w_G \times U_G + w_C \times U_C + w_{A20} \times U_{A20}$$

where $U_L$ is 1 *if the user has location set*, else value is 0, $U_U$ is 1 *if the user has URL set*, else value is 0, $U_D$ is 1 *if the user has description set*, else value is 0, $U_V$ is 1 *if the user's account is verified*, else value is 0, $U_G$ is 1 *if the user has geolocation enabled*, else value is 0, $U_C$ is the *division between the number of months* from when the account was created until the date the user's credibility is computed and the number of months from 15 July 2006 (the day Twitter went public) until the date the user's credibility is computed, $U_{A20}$ is the *average credibility of the last 20 tweets* of the current user.

To identify the best distribution for the weights from users' credibility formula, we came up with the following weights: $w_L = 0.01$, $w_U = 0.01$, $w_D = 0.03$, $w_V = 0.1$, $w_G = 0.08$, $w_C = 0.07$, $w_{A20} = 0.7$. As we can see the most significant value for the users' credibility is the $U_{A20}$ (the average credibility of the last 20 tweets of this user).

*Using Neural Networks*

To compare the results to both manually annotated scores and to the ones obtained by the manually created formula we used a neural network model (a mathematical function whose weights are refined iteration after iteration). After training this neural network, any further inputs are processed using the formula with the adjusted weights resulting in correct or incorrect decisions depending on how well trained is the model.

Our neuronal network model uses **five input neurons** for: (1) the *retweet score*, (2) the *favorite score*, (3) the *relevant words ratio*, (4) the *number of hashtags* and (5) the *number of hashtags characters*; and **one output neuron** that produces a value between 0 and 1. A value above 0.6 means the tweet is credible, else it is not credible. We decided to use a single hidden layer with 13 neurons because our model does one thing - computes the credibility of tweet. Furthermore, we use 13 neurons after doing some testing and not go higher to avoid overfitting. The number of neurons in the hidden layer is picked after the next formula that helps determine the upper bound of hidden neurons such that the training will not result in overfitting:

$$N_h = \frac{N_S}{\alpha \times (N_i + N_o)}$$

where $N_S$ is the number of samples in the training dataset, $N_i$ is the number of input neurons, $N_O$ is the number of output neurons, α is an arbitrary scaling factor usually between 2 and 10. In an attempt to get the best results possible out of the available data we considered to train several versions of the model with different inputs and various combinations.

Below, we discussed each different configuration. In the end, we benchmarked all configurations and chose the one that provided the closest scores to the ones manually annotated.

*Configurations*

We created the following variations of the neural network model:

- *Basic* (C1) - the basic configuration for which we consider the retweets score, the favorites score, and the relevant words ratio;

- *BasicWithNoRetweets* (C2) - at basic configuration we completely exclude retweets from the training dataset;
- *BasicWithSentiment* (C3) - additionally to the basic configuration, we add the sentiment score of the tweet;
- *BasicWithSentimentHashTagLength* (C4) - at the basic configuration, we add the sentiment score and the hashtag length;
- *BasicWithSentimentHashTagCount* (C5) - at the basic configuration, we add the sentiment score and the hashtag count;
- *BasicWithHashtagsCountAndHashtagsLength* (C6) - a combination of the previous two models.

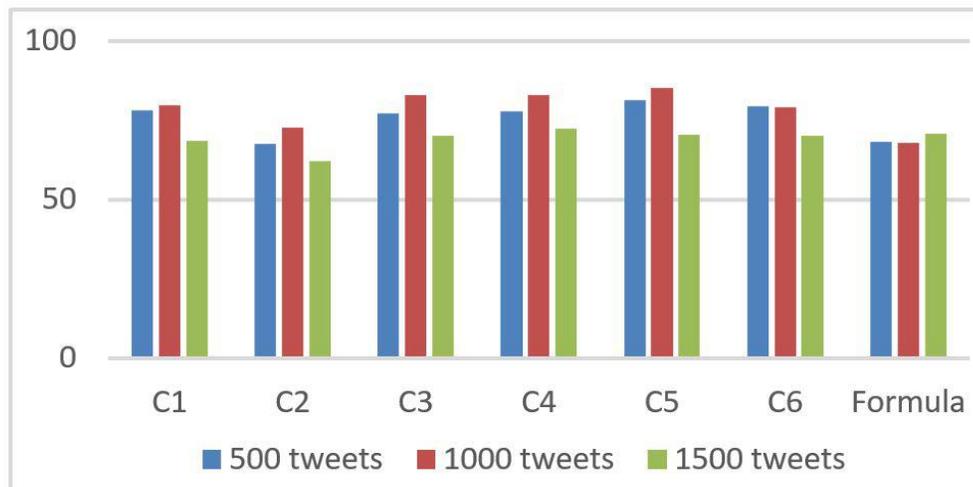

**Figure 6:** The results for considered configurations

Basically, we constructed a base configuration over which we added or removed elements to find relevant connections between the content of the tweets and credibility score. After training the presented models on 500, 1,000 and 1,500 tweets, we calculated the accuracy using one-third of the training data from the rest of the tweets from the dataset. The results for the above six configurations are presented in Figure 6, where we add also the results for the system based on the formula.

We observed that by removing the retweets from the basic configuration, we get noticeable lower results on accuracy (with around 10%). Information about sentiments and hashtags (length and count) helps us get better results. The highest rate of success that we obtained is 85.2% for configuration 5 with 1,000 tweets considered for training. An interesting aspect is that the increase in the number of tweets for the training data of more than 1,000 does not improve our results anymore; in fact, it leads to a decrease in quality. We also tweaked the number of training iterations and we settled to about 100,000 training iterations for an optimal balance between training time and success rate. After 100,000 iterations the success rate increase is insignificant, moreover, it starts decreasing after a certain point.

**Experiments**

From the entire set of data presented above, we describe some use cases following the evolution of the user's credibility and tweet credibility in time.

*Tweets Monitoring*

One of the components of our system allows us to specify a tweet id and for it, we can monitor the evolution of his credibility over time. Next, we will see different types of behavior for credible tweets caught by this component (with scores close to 1), but we have similar behavior for not credible tweets (with scores close to 0).

**Use case 1 - Constant Behavior**

For this use case, we monitored one of the tweets of user @realdDonaldTrump with id: 946731576687235072 with the following text: *The Democrats have been told, and fully understand, that there can be no DACA without the desperately needed WALL at the Southern Border and an END to the horrible Chain Migration & ridiculous Lottery System of Immigration, etc. We must protect our Country at all cost!*

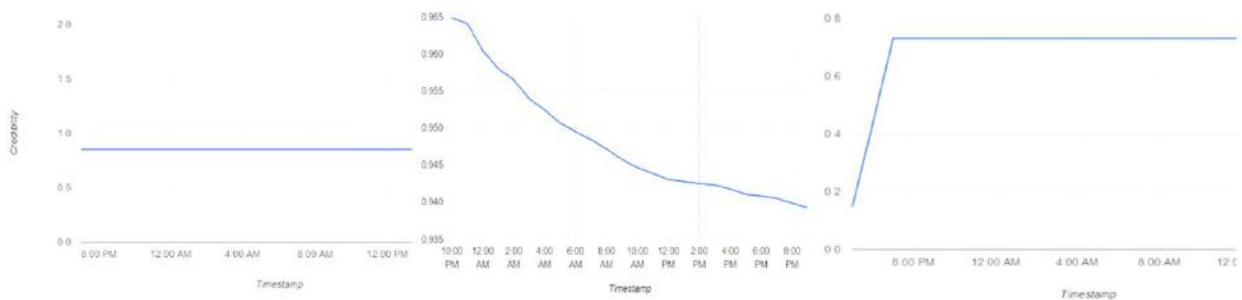

**Figure 7:** Evolution of tweet credibility in time: constant (left), variable decreasing (middle) and constant growing (right)

Monitoring this tweet posted by Donald Trump, we obtained linear credibility of 0.85, meaning that this tweet is a highly credible one (see Figure 7, left). The reason why this tweet has constant credibility over the hours it was monitored by us is because that moment in time was far away from the creation date of tweet. The conclusion would be that the interest in it has diminished drastically in the last period.

**Use case 2 - Variable Decreasing Behavior**

We take a tweet from the user @NASA, the tweet with id: 1112414759419371527 and the text: *It's gettin' hot in here! "A Engineers recently conducted a static hot-fire test of our @NASA_Orion spacecraft to ensure it's ready for missions to explore the Moon. Watch us turn up the heat: https://go.nasa.gov/2FEXwlx*.

In this case, our system captures the decreasing of the credibility of tweets, which fluctuates from almost 0.965 to 0.935, mainly because it was monitored after the interest in this news has diminished (see Figure 7, middle).

**Use case 3 - Constant Growing Behavior**

Here, we choose another tweet from the user @realdDonaldTrump, the tweet with id: 947461470924820480 and the text: *Why would smart voters want to put Democrats in Congress in 2018 Election when their policies will totally kill the great wealth created during the months since the Election. People are much better off now not to mention ISIS, VA, Judges, Strong Border, 2nd A, Tax Cuts & more?*

In this case, we see that the increase of the credibility is more abrupt at the beginning (Figure 7, right). The reason for this to happen would be that the system monitored the tweet's last hours before becoming irrelevant or followers focusing over new tweets.

**Not Credible Tweets**

A part from not credible tweets are below:

- *Just a moment prior to being told to "piss off" in classic style. #makingfriendsinmanchester @... https://t.co/AV9QvT26v8* (score 0.000046);

- *The patient voice in cancer research @sysbioire #patientsinvolved https://t.co/qb7OP6nB0A* (score 0.00034);
- *IT HURTS NOW???? BUT WHAT DOESN'T KILL YOU MAKES YOU STRONGER??IN THE END??? #brokenfamily… https://t.co/avuxg2L2zS* (score 0.000125);
- *Everyone knows what I look like, not even one of them knows me?? #everybodyhatesme… https://t.co/Ad37jfCIjA* (score 0.001889).

From our analysis, the system classifies as not credible tweets, short posts without meaning, with many punctuation signs, posted by users without many followers without retweets and without interest from other Twitter users.

*Users Monitoring*

Another component of our system allows us to monitor a user and to see the evolution of his credibility in time. Next, we will illustrate different types of behavior for different users, according to their activity on Twitter.

**Use Case 1 - Donald Trump**

We monitored the credibility of Donald Trump over 10 hours and we observed some fluctuations (Figure 8, left). An important drop of credibility happened around 2:00 AM, but the credibility of the USA president returned close to its previous scores. This was caused by one of his tweets that causes divided opinions through his supporters. After a while, we can observe linear credibility, meaning that Donald Trump had no activity between those hours and/or his followers' base was probably inactive during that period. Around 3:00 PM, there is a remarkable increase in credibility due to another tweet posted by the president, which led to a sudden increase in credibility, followed by a constant drop.

**Use Case 2 - Justin Bieber**

Monitoring Justin Bieber we can see a continuous decrease of credibility over 10 hours. Even so, the difference of credibility from the beginning of the monitoring until the end of it illustrates an insignificant decline (from 0.4667 to 0.4666) (Figure 8, right). The reason could be that he posted new tweets, but it took some time until these tweets got relevant up to the point the credibility of the user stabilized. On the other hand, for Donald Trump, the scores decreased and then increased, but the difference of credibility from the beginning of the monitoring until the end of it is higher (from 0.41 to 0.53).

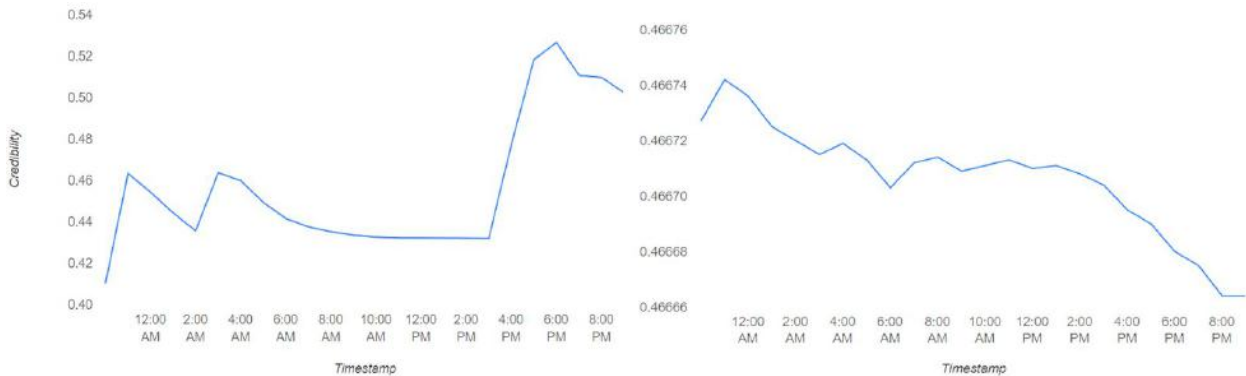

**Figure 8:** Evolution of user credibility in time: for Donald Trump (left), for Justin Bieber (right)

*Use Cases Conclusions*

We observed that the credibility of tweets fluctuates after they are posted. The credibility stabilizes after a few hours - when a tweet becomes outdated. A tweet that does not become viral or is not posted by a user with many followers may have a less volatile credibility score or does not fluctuate at all. In the case of monitoring users, their credibility fluctuations are more noticeable than in the case of tweets, because the most recent 20 tweets are accounted for when calculating the credibility score every 60 minutes from the moment of monitoring. However, a user's Twitter may also have low or no credibility fluctuations, depending on his/her activity (low or his/her image is not high). The biggest changes in both credibility situations occur in the early hours of posting. Another observation would be that if we increase the number of training tweets, we will not get significantly better results, because messages often have similar features.

**Statistics and Interpretation**

In this section, we will present and we will analyze the results obtained by our system on 50 selected users and 3,004 tweets (collected in the period March-April 2019) using the most suitable neural network presented in the previous section. The credible/not credible label was assigned for the tweets using the NN described earlier in this article. 1,344 tweets were labeled as not credible and 1,660 tweets were labeled as credible.

*Users Credibility*

In Figure 9, we can see the overall credibility of all users.

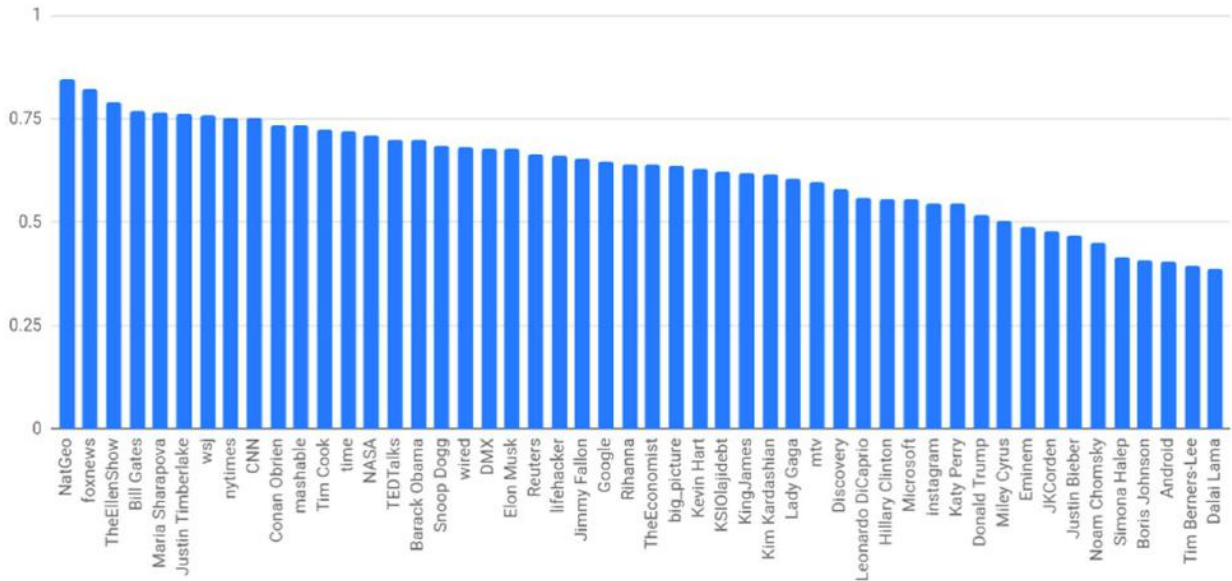

**Figure 9:** Statistics in monitored users.

We can remark: (1) in *politics* Barack Obama is more credible like Hillary Clinton and Donald Trump; (2) in *business*, Bill Gates has a highest value of credibility and he is more credible than Tim Cook, which has also a good value of credibility; (3) for *companies*, Google is more credible than Microsoft, but both have medium values of credibility; (4) for *organizations*, Discovery and NASA have the lowest values of credibility; (5) for *televisions*, NatGeo and Foxnews have close and very good values for credibility; (6) in *music*, Justin Timberlake has the highest value in comparison with Snoop Dogg, Rihanna, Lady Gaga, Eminem and we can deduce that he is the most in vogue artist out of those who have been monitored; (7) in *sport*, Sharapova has a better credibility in comparison with Simona Halep, and this is due to the fact she returned after a pause in which she was suspended, and Simona lost 1st place in the WTA rankings and her activity and followers are more active on Twitter. It is interesting how Android, Tim Berners-Lee, and the Dalai Lama have the lowest value of credibility, due to the low activity on Twitter in the last period.

We also note from Figure 9 that there are many ambiguous situations with the credibility of around 0.6, which makes the decision as a user is credible or not difficult to take. But we note that the information provided by our system is very useful when we want to compare two Twitter users to figure out which one is more credible.

What is interesting is the evolution of these users' credibility over time: when we started in 2018 to collect information about some of these users, (1) *Donald Trump* had a much greater credibility compared to other politicians, which means that lately his credibility has been affected by his posts on Twitter as well as by his political activity, (2) *Simona Halep* had more credibility than Maria Sharapova, who was suspended at that time, but the loss of her first position and Maria's return to the circuit made the hierarchy change. Our plans for the future aim to make clearer when the credibility of a user increases or decreases and try to justify these changes. The reasons we have identified so far for decreased credibility are poor network activity, or posting a controversial tweet, or moving attention to someone else who may have better results in the same domain of activity, etc. Reasons for increasing credibility are continued work within the network, notable results achieved in the domain of activity, posts on Twitter supported by network followers, etc.

*Credibility by Continents and Countries*

In the following figures, we can see the number of tweets (both credible and not credible) by continents and by countries. As can be deduced from Figure 10, in the right, the zones with most tweets are North America and Europe and from these continents comes the most number of not credible tweets (Figure 10, in the left). In terms of percentage, these continents have also the highest number of not credible tweets, and their number is around 5% greater than in Oceania, South America, and Asia.

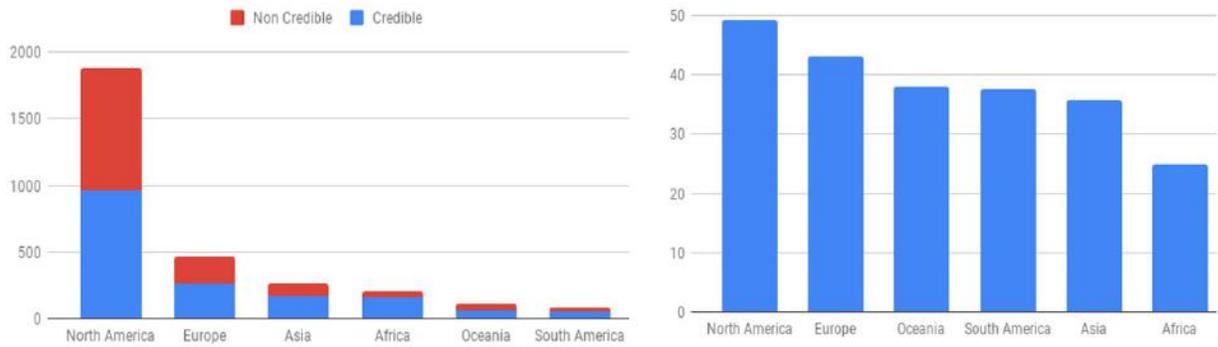

**Figure 10:** Statistics by continents, by number of credible/not credible tweets (left) and by the percentage of credible/not credible tweets (right)

If we go deeper for a country statistics, we get the results presented in Figure 11 (countries that had less than 25 tweets analyzed were excluded from the statistics).

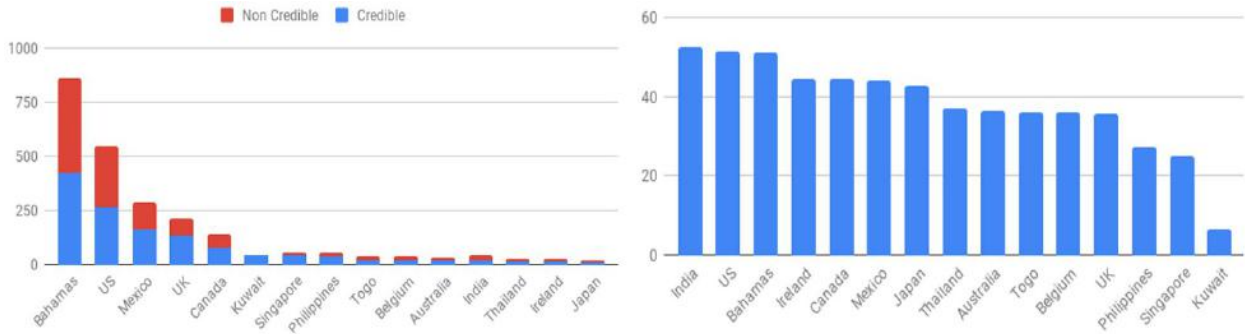

**Figure 11:** Statistics by countries, by number of credible/not credible tweets (left) and by the percentage of credible/not credible tweets (right)

As can be deduced from Figure 11, the countries with most tweets are Bahamas, US, Mexico, UK and Canada and from these countries come the most number of not credible tweets (Figure 11, in the left). In terms of percentage, countries with the highest number of not credible tweets are India, US, Bahamas, Ireland, Canada, Mexico, and Japan, and their number is around 10-20% greater than in Thailand, Australia, Togo, Belgium, and the UK and with around 40% greater than Kuwait.

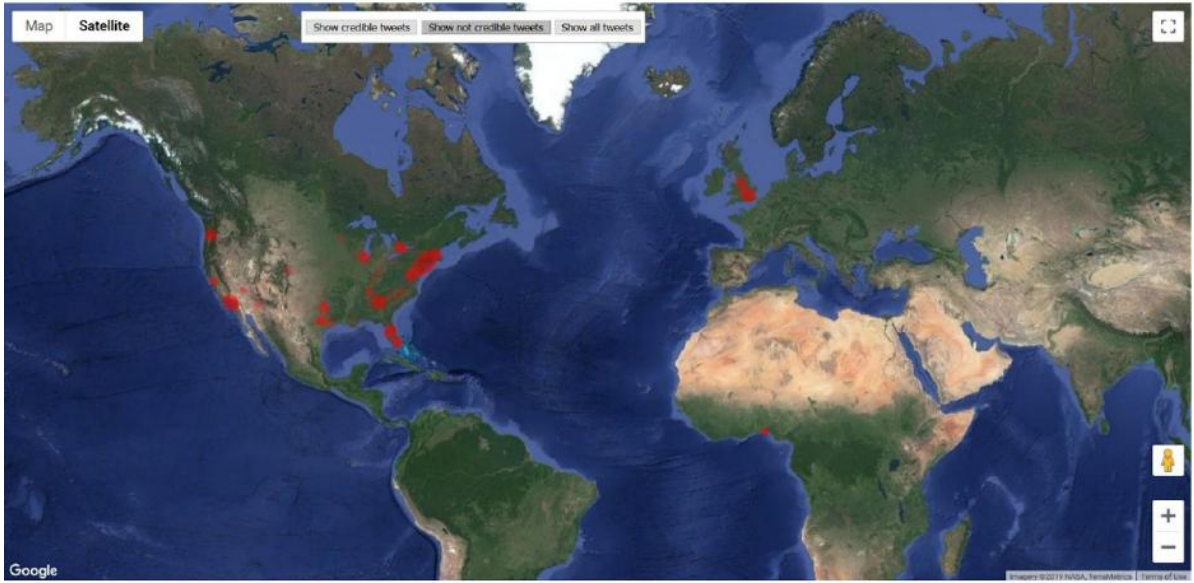

**Figure 12:** Heatmap of not credible tweets

Our application allows the user to display these tweets over a heatmap to see the distribution of credible and not credible tweets. Figure 12 contains a visual representation of the 1,344 not credible tweets (the red areas).

*Error Analysis*

The small number of relevant tweets compared to the total number of tweets that can be collected from Twitter exists because many tweets don't have geolocation information. This information is very important for our system because the statistics for a user based on continents or countries or the heatmap need mandatory this information. The biggest number of irrelevant tweets comes from the Bahamas, a small country, where most messages are by type advertising or job announcements, automatically added to Twitter by bots, containing no information besides user's location or advertisement. These tweets are so frequent that they make up almost 30% of tweets collected by us. For the future, we need to pay more attention to how we collect information for our system to avoid such tweets coming into our database.

Regarding the quality of the neural network approach, we still investigate ways to improve it. The main problem here is related to the fact that we want to keep a balance between the speed of the system, which works in real-time, and the quality of the results. When we want to see in real-time evolution of the credibility for users and we need for that to collect more tweets from Twitter and then for everyone we need crawling, processing, classifying, updating statistics and maps, every delay may affect the quality of the user experience.

Also, we want to come with more relevant data in our training dataset and we investigate what kind of data can be useful for our system. Another problem comes from the component that assigns a tweet to countries or continents when the tweet is posted on Twitter in a region which is very close to a border between the two countries. We intend to assign a country parameter to every Twitter user, which represents the most common country from which he posts the last 10 tweets. This value will be used instead of the geolocation of the tweet in limit situations when the user is near a border. Our experiments show until now that this value can be used with success.

**Similar Work**

Similar work was done in (Atodiresei et al., 2018) and in (Cuşmuliuc et al., 2018).

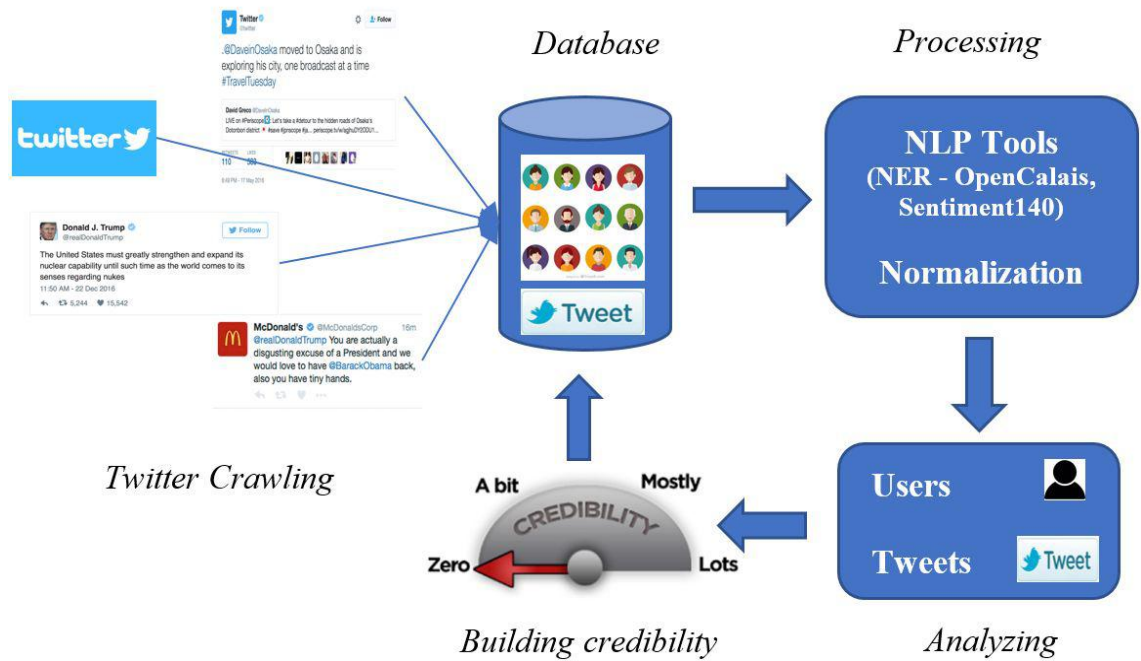

**Figure 13**: System architecture (Atodiresei et al., 2018)

The general architecture of the system from (Atodiresei et al., 2018) is presented in Figure 13. The main components are (1) the *Twitter crawler component*, which takes tweets from Twitter and save them to our computers; (2) the *Processing module*, which calculates the credibility of a new tweet; (3) the *Analyzing module*, where we combine the analysis of the current tweet with existing information from our database; (4) the *Building credibility module*, where we calculate the credibility for users.

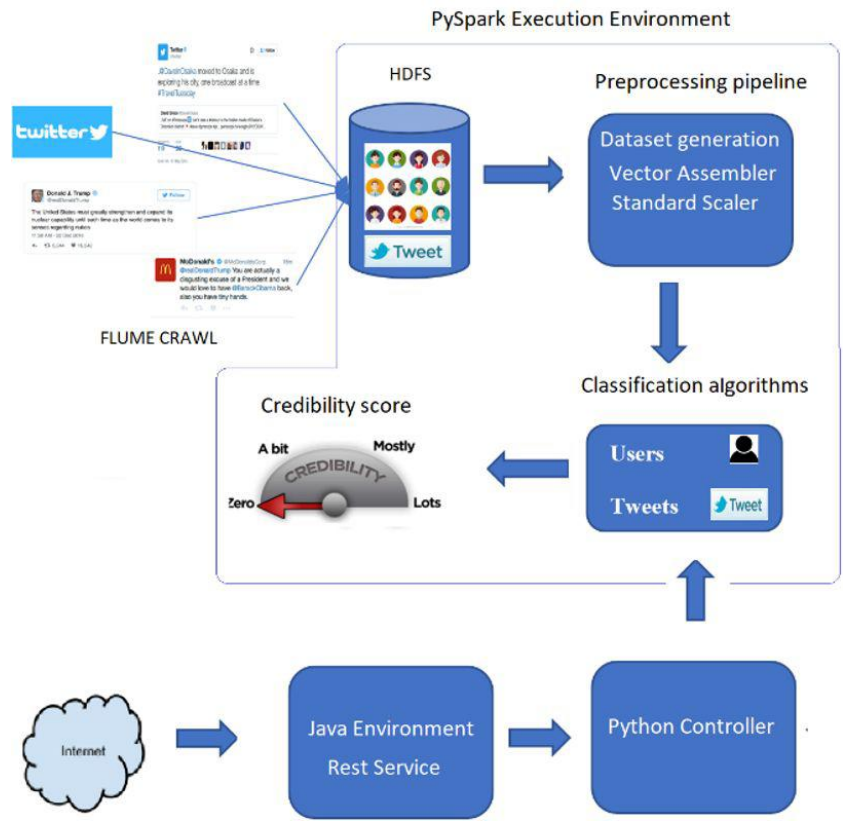

**Figure 14**: System architecture (Cuşmuliuc et al., 2018)

In (Cuşmuliuc et al., 2018), we analyze different techniques of detecting fake news, their performance and how we can do fine-tuning to improve the actual results (the system architecture is presented in Figure 14). For that, we use three classifiers like Support Vector Machine, Random Forest, Naïve Bayes. After performed experiments, we have found that Random Forest has the maximum accuracy followed by SVM.

**Conclusions**

What we have presented in this study provides a way of classifying users and the information they post in credibility classes. To classify tweets and Twitter users, we trained a neural network model using a collection of tweets that were manually annotated by experts. Also, in our experiments, we consider entities from different fields, such as politics, business, companies, organizations, television, music, sports, and others (Iftene et al., 2017), (Atodiresei et al., 2018).

The current and next work comes with a new proposal to display in real-time statistics and information related to credible and not credible tweets on Google Maps using heatmap. Compared to previous approaches, which displays all the tweets collected over a certain period, we can manage to display credible or not credible tweets, statistics on countries and on continents in real-time.

## II.3.2 Sentiment Analysis

**Introduction**

This section presents a system created to take data from Twitter and to classify it into *positive*, *negative* or *neutral* (Iftene et al., 2017). The results of this processing can be seen in a dynamic way on Google Maps, where the users can apply different filters (select the language, the date or the type of opinion of the tweets).

Tweets from Twitter began to be the input for many of the applications developed in recent years, especially for applications that want to know opinions of users (Borruto, 2015), (Jenders et al., 2013), (Liang and Dai, 2013), (Luo et al, 2013), (Maynard and Funck, 2011), (Pak and Paroubek, 2010), (Iftene et al., 2017) (in particular for products or companies, for presidential candidates, for local elections or for an event, etc.).

A system that processes data from Twitter has three classical modules: (1) a *module for search and local storage for tweets*, (2) a *module for data processing* (usually natural language processing services), (3) a *module to display the results*.

**Opinion Mining on Twitter**

In (Kharde and Sonawane, 2016), (Iftene, 2016), the authors present the main techniques used to identify opinions in existing data on Twitter. Their study shows how the existing techniques process these tweets to categorize them into *positives*, *negatives* or *neutrals*. Existing techniques based on machine learning algorithms (Naïve Bayes, Max Entropy and SVM), or lexicons based techniques are presented, analyzed and compared. In (Go et al., 2009) and (Pak and Paroubek, 2010) there are proposed models for classifying tweets using emoticons. (Barbosa and Feng, 2010), (Gamallo and Garcia, 2014), (Liang and Dai, 2013), (Parikh and Movassate, 2009) and (Xia et al., 2011) implemented models based on maximum entropy or Naïve Bayes to classify tweets. Features spaces contained retweets, hashtags, links, punctuation marks in combination with features like words polarity and POS (Part-of-speech). In (Bifet 2010) the author experimented using multinomial Naïve Bayes, stochastic gradient and Hoeffding tree. In (Agarwal et al., 2011) he experimented with models based on unigrams based on features and based on trees. (Davidov and Rappoport, 2010) proposed an approach using hashtags from Twitter, punctuation marks, words, n-grams and templates for various features, which are then combined into a single vector of features used to classify tweets. They used the *k*-NN (*k-Nearest Neighbor*) algorithm to label opinions by building a vector with features for each example of test data. (Turney, 2002) used methods based on "bag-of-words" perform opinion mining - the relationships between words are not taken into account, and the document is represented as a collection of words. Kamps et al. 2004 used a lexical database like WordNet (Fellbaum, 1998) to determine the emotional content of a word in a multi-dimensional space. (Luo et al., 2013) brought into focus the difficulties that can be encountered when we want to classify tweets.

**Real-Time Analysis of Twitter Data**

One of the problems encountered when operating with information on Twitter is related to scalability of the application you want to develop (Cuesta et al., 2014), (Karanasou et al., 2016) and (Sheela, 2016). Existing techniques involve the use of *Hadoop machines* (Bingwei et al., 2013), (Jimmy and Kolcz, 2012) and (Mohit et al., 2014), *HPC platforms* (High Performance Computing) (Jiang et al., 2012), and hardware components included *EC2* (Amazon Elastic Compute Cloud). Big Data analysis techniques like Map-Reduce are used as well (Michal and Romanowski, 2015). Machine learning techniques were implemented on these super-machines to perform supervised classification, in the context of having to deal with a large amount of data (millions of tweets). Many existing approaches have focused both on processing speed and the quality of the results obtained by them, currently, the accuracy of results is between 74% and 82%, depending on the used technique and input size.

**Proposed Solution**

For developing our tweet classification application we used data from Twitter, more specifically tweets collected between March 4, 2017, and April 3, 2017. From the collected tweets we decided to mark and analyze more carefully the tweets that are related to the *refugee crisis* and the *terrorist attacks*. To filter these tweets, we prepared a list of keywords grouped into six sections: (1) NATO - contains a list of the 28 NATO member countries, (2) the European Union - the list of the six-member countries of the EU, but not members of NATO, (3) Fight and attack - contains the lexicon of the words (i.e. words usually used to express the concepts of fighting and attacking), (4) Civilian - contains the lexicon of the word (5) Peace - contains the lexicon of the word and (6) European refugee crisis - contains the lexicon of the word. Also, we considered the top three origin countries of the immigrants (Syria, Afghanistan, and Iraq). In total, the list contains 103 words for each of the six target languages (English, German, French, Greek, Turkish and Italian). We initially started with English, German, French languages, but, considering the refugee topic we selected for our analysis, we also included Greek, Turkish and Italian, as the languages of the most common transit countries.

*Ensuring scalability*

Twitter provides live data including details about the user who posts the tweet, the language of tweet, the location, etc. The connection between Twitter and our application is made using a module that constantly collects tweets. Received data is filtered in two steps (1) **firstly**, we keep only those in English, German, French, Greek, Turkish or Italian that contain information related to the geolocation of the user who made the posting (we collected this way 66,975 tweets), (2) **secondly**, we filter the tweets that contain words from the list with information related to the refugee crisis and the terrorist attacks (we thus marked 2,702 tweets). These 2,702 tweets were then annotated with opinions (negative/positive/neutral). To ensure scalability, we used threads to asynchronously process the tweets that are going to be stored in the database: (1) identify the opinion of the tweet and (2) grouping similar tweets. After marking a tweet as processed, it was ready to be visualized on Google Map.

*Grouping Similar Tweets*

Redistribution of information on Twitter is getting more and more popular, it is a simple retweet or copy/pasting a tweet and adding hashtags, emoticons or links. The purpose of this step was to group similar tweets, to ensure displaying in a diversified way at the GUI interface. For this, we perform experiments with four distance similarity algorithms: Jaro-Winkler, Levenshtein, Needleman-Wunsch, and Smith-Waterman, presented in the previous section.

*Opinion Mining*

The collected tweets were classified using Sentiment140 API[51] into positive, negative and neutral (Șușnea and Iftene, 2018). Because, Sentiment140 API works only with Spanish and English languages, we translate with API Yandex[52] the tweets from native language into English. After translation the tweet is sent to Sentiment140 API[53] and then it is persistently stored in our MongoDB database for later usage (offline analysis). Once annotated, they will be sent to a module that is responsible for distributing them to the web clients interfaces. Also, annotated tweets will be stored to the database to enable offline analysis, when needed. The web interface simply takes the information from the module that distributes it and presents it to users in a graphical way.

*Display Tweets on Google Maps*

Tweets are grouped into clusters, represented by circles or hexagons distributed on the global map. The color of a cluster will be intuitive and related to its contain: *white* for neutral clusters (containing the same number of positive and negative tweets or only neutral tweets), a color between *green* and *red* (green representing positive tweets and red negative tweets). On the right side are displayed the latest tweets without duplicates (*green* is for the tweets positive annotated, and *red* is for the tweets negative annotated).

---

[51] http://help.sentiment140.com/api
[52] https://www.yandex.com/
[53] http://help.sentiment140.com/api

**Results**

*Statistics*

First of all, it is important to note that we only had access to 1% of Twitter's Public Stream consisting of sample real-time tweets. Between 4 March and 3 April, we collected a total of 66,975 tweets that have geolocation attached. Among all languages that we monitored English and Turkish are the most popular (full languages distribution can be seen in Table 1).

| Language | Number of tweets | Positive / Negative tweets |
| --- | --- | --- |
| English | 28,208 | 7,052 / 1,399 |
| German | 1,115 | 31 / 2 |
| French | 2,305 | 53 / 4 |
| Italian | 2,231 | 40 / 5 |
| Turkish | 33,017 | 343 / 19 |
| Greek | 99 | 2 / 1 |
| **Total** | **66,975** | **7,521 / 1,430** |

**Table 1**: Tweets count by native language

*Results analysis*

Analyzing tweets from our database we instantly observe that on the second day (22nd of March 2017 - date of the Westminster Attack in London) tweets count is significantly higher than the day before (see Figure 15). A higher number of tweets means a higher number of keywords occurrences which means higher chances of tweets being related to the events we are interested in. Also, the content of tweets is related to this attack: 22 April: "*UK Terrorist Attack: President Buhari Sympathises With Britons*", 23 April: "*British police raid after deadly Westminster attack*", "*Another terrorist attack, this time in my hometown of London! We are mad, pissed, sad, horrified*", "*Four killed in UK parliament terrorist attack*".

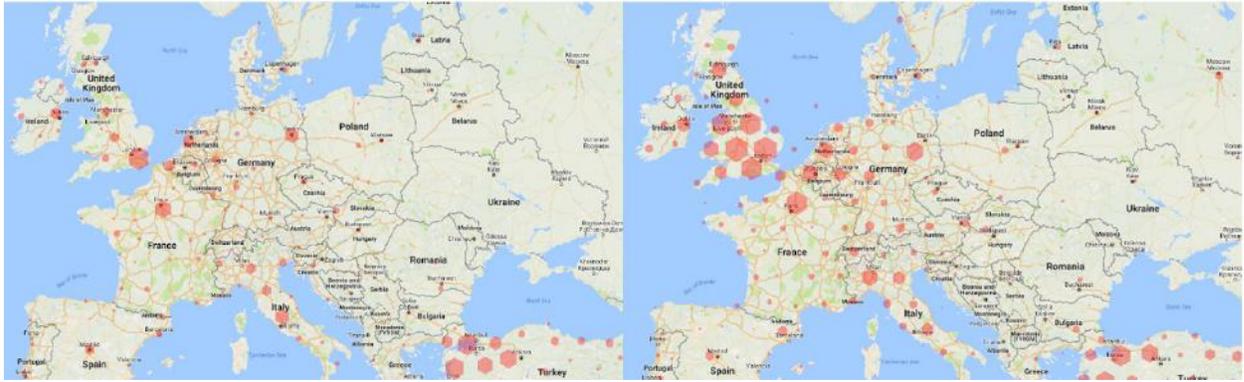

**Figure 15**: Tweets distribution for 21 March (left) and 22 March (right) (Iftene et al., 2017)

Also, we notice a big increase in usage of keyword "*attack*" starting with the 22nd of March 2017. What is interesting is that the day following the attack presents a bigger increase in usage of "*attack*" keyword than attack's day. This happens mostly because of mass media starting to cover the incident. Two days after the incident, there is a clear frequency decrease of keyword "*attack*" and no occurrences at all for "*terrorist*" and "*raid*".

*Error Analysis*

The small number of relevant tweets compared to the total number of tweets collected is due to massive amounts of tweets, that are automatically posted (*weather forecasts, check-in, jobs announcements, advising*, etc.). This kind of tweets is irrelevant for our analysis so there is no point to rely on their content. For the future, we need to pay more attention to how we collect information for our system to avoid such tweets coming into our database. The biggest number of irrelevant tweets comes from Turkey - mostly check-in posts or messages of the type advertising automatically added to twitter containing no information besides the user's location or advertisement. These tweets are so frequent that they make up almost 50% of tweets collected by us.

Another problem is related to the fact that most tweets are classified as *neutral* by Sentiment140. But many of these tweets contain words or hashtags that would allow us to classify them as **negative** (*silly, death, fuck, kill, bad, lose, fucking, block, incident, protest*, etc.) or **positive** (*great, champion, good, inspiring, thank*, etc.). Due to this problem, we created a list with triggers for negative and positive opinions and we completed the result of Sentiment140 service with an additional checking in case the identified opinion is neutral. For example, on the 8th of April 2017, we collected 4,354 tweets on English and classified them with Sentiment140 service, 3,704 tweets are neutral, 569 positive and 83 negative. After additional checking, we classified 363 neutral tweets into positive tweets and 72 from neutral into negative.

Furthermore, our analysis was made only on tweets that contain geolocation information because we have to know the location where the data is coming from to place it on the map. Hence we could have more relevant results if more tweets would contain this data.

**Conclusions**

Social networks can provide a lot of information, especially during the night or when there are special natural phenomena (like earthquakes, floods, fires, snow, chills or extreme heat), or when there are presidential elections, or when there are attacks with many casualties that stir up panic. In such cases, the amount of data can increase very fast, and from here appears the need to provide scalability and showing results in real-time. To analyze data on Twitter we used a dictionary with specific terms to monitor the refugee crisis and the terrorist attacks, with specific techniques from natural language processing: POS, removing stop-words, identification of opinions, etc. We also saw how big data analysis techniques such as map-reduce allow us to obtain useful information from the data gathered.

## II.3.3 Image Retrieval

**Introduction**

An image is said to convey far more than a thousand words. Through images we have the opportunity to convey ideas, feelings or experiences lived, in a very simple way, but at the same time very expressive. It has been a long time since the first photographic image was made in 1826[54], and since then we have witnessed a continuous evolution of technology. In Figure 16 we have "View from the Window of Le Gras", a heliographic image and the oldest photograph that survived the passage of time. The image was created by Nicéphore Niépce in 1826 or 1827 at

---

[54] https://www.thevintagenews.com/2016/07/23/43278-3/

Saint-Loup-de-Varennes, France, and shows parts of his buildings and estate, Le Gras, as seen from his window (Blazeski, 2018).

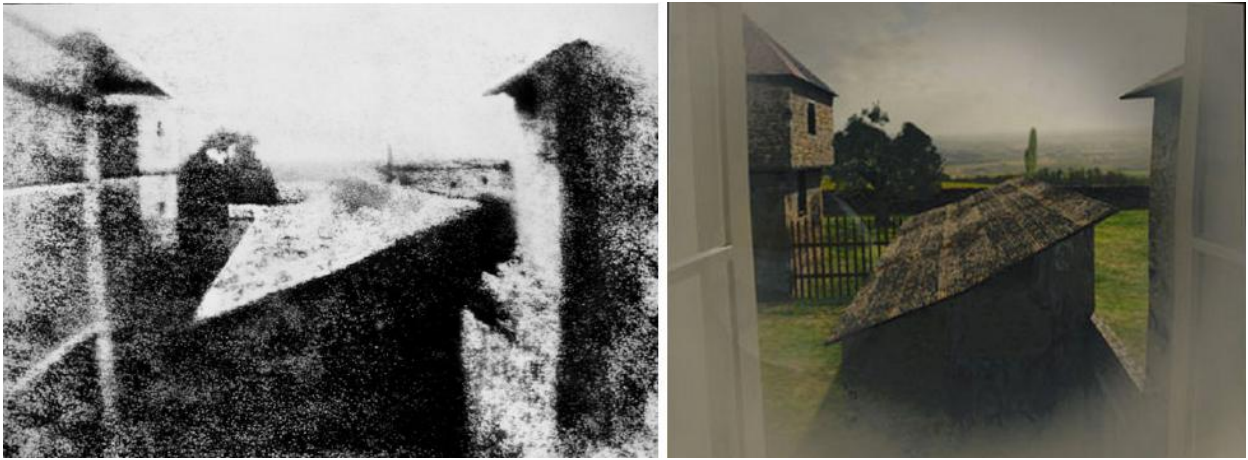

**Figure 16**: The first photo "View from the Window at Le Gras"[55] (left) and a current color photo[56] (right)

Today you can take pictures and you can shoot anywhere, anytime, using ever more diverse and advanced tools: from mobile phones, smartphones, tablets, video cameras, webcams, to the most sophisticated cameras and film or flying drones. Some interesting statistics about what's happening on Facebook[57]: it uploads over 300 million images daily, and every 60 seconds it adds over 510,000 comments, changes 293,000 user statuses, and uploads over 136,000 photos.

Increasing amounts of visual content are created day by day. These resources are becoming available in a very short time to everyone in the world through the Internet. If we look back at the evolution of the Web, we can see how it went from the current Web version 1.0 to the year 2000, "a version that first allowed the reading of information", to the Web 2.0 version until 2005, "a version of mass reading and writing", in the Web version 3.0 of 2010 "oriented on the Internet porting of personal information", until the Web version 4.0 which is expected to appear in the next period (see Figure 17 for details). In this context, intelligent image search is a topical and interesting topic. In the past, static web pages were created, mostly text-based, and search engines focused very little on finding relevant images in connection with a user query. At present, things have changed a lot. Finding images has become just as important as finding textual resources - or maybe even more important if we take into account those mentioned above.

---

[55] https://upload.wikimedia.org/wikipedia/commons/thumb/5/5c/View_from_the_Window_at_Le_Gras%2C_Joseph _Nic%C3%A9phore_Ni%C3%A9pce.jpg/1280px-View_from_the_Window_at_Le_Gras%2C_Joseph_Nic%C3%A9 phore_Ni%C3%A9pce.jpg
[56] https://66.media.tumblr.com/cb28b29f2e3211ba257781bcfddbe9f2/tumblr_nv1jmjqs6l1upgwl7o5_540.jpg
[57] https://zephoria.com/top-15-valuable-facebook-statistics/

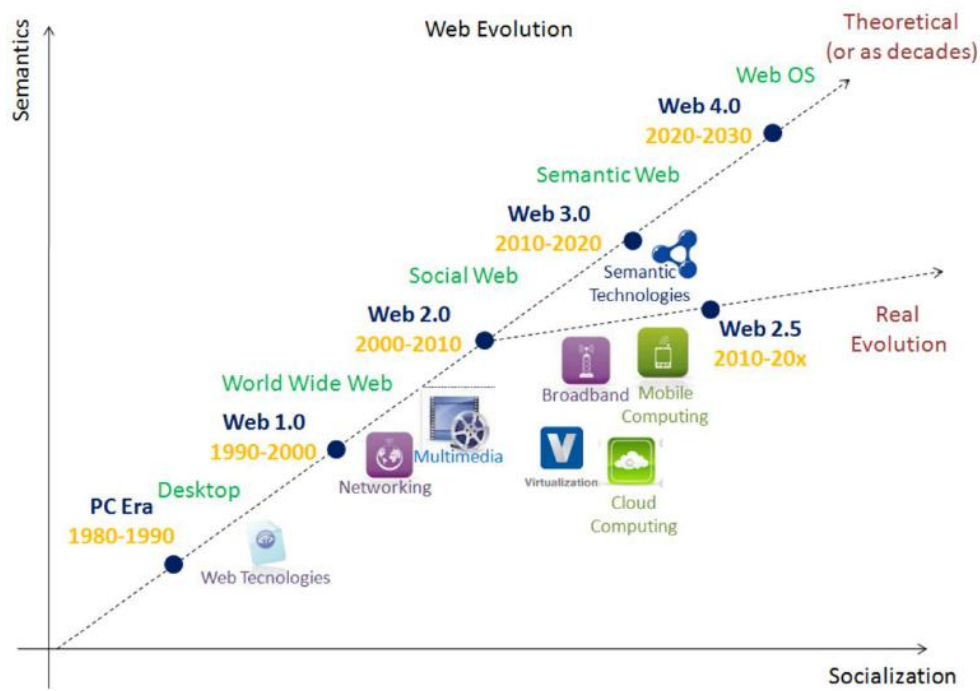

**Figure 17**: Web evolution (Madurai, 2018)

The question that arises now is: *What makes image retrieval so complicated*? The idea of starting, sharing the photos we take with other people sounds great in theory, and the possibilities offered today by the Internet are endless and include specialized applications for taking pictures on the web (like Flickr or Picasa) or through social networks (such as Facebook or Twitter) or simply by posting directly from your computer, which is connected to the Internet. It's great that we have so many possibilities, but *what happens after this content reaches the Internet*? *How easy can it be to find and access afterward*? Due to the heterogeneous nature of all these resources of interest, from which a user can choose what he likes most, it is very difficult to look for all the resources that might contain something that might interest him. In addition to this, problems may occur due to the language in which the annotations associated with an image are made, the distance to the resource, the transfer speed or security aspects.

Another problem comes from the fact that to look for something, we only express ourselves through words. Word search is getting better and better when searching in textual data collections, since natural language processing has evolved lately, coming with a wide range of specific algorithms and techniques. Things change, however, when it comes to image collections, primarily because we have the skills to reach visual representations with textual descriptions. The *name* of an image, one of the most used features in the search process, is not far enough to describe its content. To make searches as accurate as possible, we should add other *metadata*, such as *keywords* related to the content, *characteristics* of how the picture was made, *description of the content*, etc. However, it is difficult to estimate how many would be the most relevant metadata to be able to search efficiently in large image collections.

In this context, it is understood the usefulness of creating an image search engine that can put together several collections of images, offering a common search interface. The purpose is to create an application that can use both external image resources and locally stored resources on our server, along with the metadata associated with each image. This engine would allow a user to issue a search request and then get relevant results in response. The server module should provide the answers depending on the type of device from which the request came, giving specific answers to a desktop computer, a laptop, a tablet, a palmtop or a mobile phone.

**Image Search Engines and Hosting Sites**

There are several image search engines, which use different indexing techniques to return eloquent results to user queries, as well as sites that host and enable image sharing, viewing, and retrieval, of which we recall:

- **Wikimedia Commons**[58] hosts over 55 million (July 2019) free licensed images, is considered the largest free image domain. Commons is the main repository of multimedia files of the Wikipedia encyclopedia, and they are structured according to *type* (images, audio, video), *subject* (nature, society-culture, science, engineering), *location* (earth, space), *authors, licenses* and *sources*. Wikimedia Commons also provides a REST API that is accessed via HTTP (GET or POST) requests to their URL[59] from which a response is expected, which includes information from the MediaWiki database, in the format specified in the query (XML, JSON, etc.).

---

[58] http://commons.wikimedia.org/wiki/Main_page
[59] http://en.wikipedia.org/w/api.php

- **Google Images**[60] is a service provided by Google, which was introduced in July 2001, and offers users the ability to search for images on the web. Image search can be done both by keywords and by the content of an image. Keywords are based on the *file name*, the link text that connects to the image, or the text that is associated with an image. Image search uses automatic image processing techniques to match the user-uploaded image with the other images in the Google Image Index or other collections. Google also provides an API, which offers up to 100 free queries per day and up to 10,000 queries per day, at a fee of $ 5 per 1,000 queries[61]. To use this API, a developer must create a Google account and register an application that can associate multiple search engines. Following this step, the developer will get the information needed for a query: the *API key* that is associated with an application and the *identifier* for each search engine that it wants to use.
- **Bing Images**[62] allows users to efficiently search for images most relevant to the specified topic. Images are displayed on a single page and are added gradually as you scroll down. Advanced filtering allows you to refine the search based on the properties associated with the images: size, aspect ratio, color hue, image type, etc. To access the API provided by Bing, a developer must create an account on Windows Azure MarketPlace (also called a Microsoft account), based on which he will receive a *customer key* (Customer ID) and a *primary key* for the account (Primary Account Key). All queries are secure, from a security point of view, and authentication is done through an Oauth type 2 mechanism. There are several subscriptions offered by Bing, the free one being 5,000 free queries per month.
- **Yahoo! Image Search**[63] is a search engine owned by Yahoo! Incorporation which provides a specialized image search service. Since June 2007, Yahoo! Search has introduced an API called "Build Your Own Search Service" or BOSS, which gives developers the ability to create their search engines based on the Yahoo indexing platform. Unfortunately, all the services they provide are for a fee.

---

[60] http://images.google.com
[61] https://developers.google.com/custom-search/v1/overview
[62] http://bing.com/images
[63] http://images.search.yahoo.com

- **Flickr**[64] is a website that hosts images and video files, created by Ludicorp in 2004 and purchased by Yahoo! one year later. Some statistics from July 2019 show that Flickr has a total of over 90 million registered members, of which over 75 million as photographers, with a maximum of 25 million images uploaded each day, with a total of over 10 billions of images shared by users[65]. The API provided by them allows access to multimedia files without the need to create an account, which is only required when uploading images to their platform. The Flickr API allows developers to obtain EXIF information for a photo. A search can be done based on keywords but can be slow if EFIX metadata is required.
- **Picasa**[66] is a platform for hosting, organizing and editing digital photos, which was originally owned by Lifescape and later purchased by Google in 2004. For good organization, Picasa offers the ability to import files, add tags, create collections, as well as tracking and facial recognition features. Search can be done by *image name*, associated *tags*, *folder name*, and other metadata. The API they provide allows integration with Picasa albums, giving users the ability to create, edit or delete albums, upload images and publish comments to them, but also search and access metadata associated with the results obtained. Its use is possible following the creation of a Picasa account, and authentication can be done through the Oauth, AuthSub or ClientLogin protocols. The API can be accessed using both a REST service and the Java client library.
- **Photobucket**[67] is a free alternative for hosting images, video files, creating presentations and sharing photos. It is mainly used for creating personal albums that can be accessed remotely or displayed on sites such as eBay, MySpace, Facebook, LiveJournal, Open Diary. Photobucket also provides a developer API for uploading images and videos, obtaining all the recent multimedia files of a particular user or group of users, searching for materials based on specified criteria, and retrieving all metadata associated with an image (*URL*, *thumbnail*, *title*, *description,* etc.). The communication is done through a REST service, based on a unique key, obtained from the registration, and authentication by the standard OAuth protocol.

---

[64] http://flickr.com
[65] https://expandedramblings.com/index.php/flickr-stats/
[66] http://picasa.google.com/
[67] http://beta.photobucket.com/

- **Panoramio**[68] was a website that made it possible for photographers to organize their images based on where they were taken and can view them both on Google Earth and Google Maps. Unlike other search engines, Panoramio focused more on exploring and illustrating places around the world (*cities* or *nature*), searching based on *latitude* and *longitude* coordinates. With the help of the API provided by them, developers could access both images and associated metadata, the communication is done through a REST service, the answer is returned in the standard JSON format. Since 2007, the site has been purchased by Google.
- **ImageNet**[69] is an image database organized according to the WordNet hierarchy (only nouns are used). For each node in the hierarchy exists hundreds and thousands of images (on average 500 images per node). ImageNet offers access to the word class and the synonym of a synset[70], to the URL of the images, to the actual images (for educational or research projects), to SIFT (Scale-Invariant Feature Transform) features and to the annotations associated with the boxes of objects, boxes are annotated and Verified with Amazon Mechanical Turk[71].
- **Picsearch**[72] is a Swedish company that develops and provides image search services for large and important websites. Picsearch's clients include several large companies (e.g. Lycos), regional portals in Germany, Turkey or Arab countries, as well as entertainment, sports or e-commerce sites. The services they provide include *face recognition*, *color filtering*, *size*, *content*, *animation*, etc. An interesting thing about Picsearch is that it indexes images from the web using a crawler, known as Psbot. It uses a branched load technique that reduces the voltage that could occur on indexed servers.
- **Cydral**[73] is an add-on for Mozilla that offers filtering features for finding images, illustrations, and icons across the Internet. It can find the place where a particular image appears in the webspace, but it can also return relevant images based on keywords or by selecting an image.

---

[68] http://www.panoramio.com/
[69] http://www.image-net.org
[70] Sets of cognitive synonyms
[71] https://www.mturk.com/mturk/welcome
[72] http://www.picsearch.com
[73] https://addons.mozilla.org/en-us/firefox/addon/cydral-image-search-engine

- **Incogna**[74] is an image search engine that organizes its files according to their content. By using parallel processing, all existing forms are searched for in an image and a large-scale visual search index is created with them. Unlike the other search engines presented above (excluding Google Images), their technology does not require the image to present associated metadata, although it can also be used successfully for text indexing.
- **Getty Images**[75] is an American agency that hosts and delivers images, targeting three branches: *advertising and graphic design*, *media* (print and online advertising), as well as *marketing and communication*. They provide an API for accessing images from their database only to those who own a company account. This account can be obtained following a request in which is presented the reason why the access to these resources is desired and it is possible to charge some fees for saving the images.

**MUCKE Corpus**

Image retrieval based on the associated metadata has been used from the beginning due to its simplicity and low computing cost. Images are automatically or manually annotated by keywords that are then used in retrieval process (Alboaie et al., 2013).

The MUCKE corpus based on Flickr images provides a set of metadata associated with each image, which contains information about the respective resource. The most important data offered by Flickr are the fields related to the *owner*, *title*, *important calendar data* (upload or creation), *location* (GPS coordinates), *tags* associated with the image by the user. These fields store data received from the user (*tags*, *title*) or automatically identified information (GPS location, calendar data) and are a key component of image retrieval. All of these metadata are textually processed. In the first stage, anaphoric resolution and retrieval of documents are done using Lucene. In the second stage, it is intended to use semantic processing, recognition of named entities, etc.

---

[74] http://incogna.com
[75] http://www.gettyimages.com/

Anyway, textual annotations don't offer information about the visual characteristics of the image and depend on the subjectivity of the annotator. This leads to the need to integrate both content descriptions and metadata for better image information management.

**The Skeleton of the Prototype**

The main steps made within the prototype for the internal sharing mechanism will be described below:

1. The u*ser enters the query*: the query mode based on keywords (textual or content-based);
2. The server receives the *keywords* and searches the NoSQL database for tables containing them:

    2.1 If *yes*, the database is queried and skipped to step 6;

    2.2 If *not*, the keyword is added to the SQL database;

3. The *server requests the Flickr API* and receives a collection of results;
4. A *clustering* mechanism is applied depending on the query results from step 1;
5. The *results* are saved in the NoSQL database;
6. The *results* are transmitted to the client to be displayed.

Our collection is a collection of Flickr images. In the prototype, a search page allows the user to enter keywords for searching (Figure 18). At beginning, keywords are searched in image titles and descriptions. Then, based on image content, a cluster with similar images is created (see Figure 19).

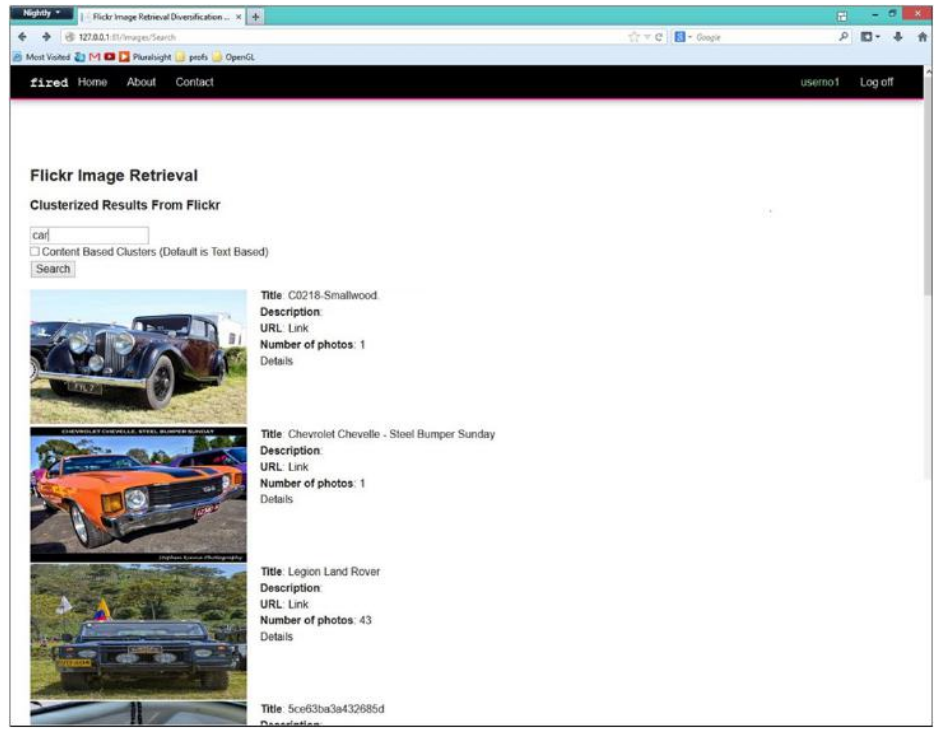

**Figure 18:** The UAIC image retrieval prototype on Flickr

Content-based clustering - In Figure 19 you can view a group from the search above.

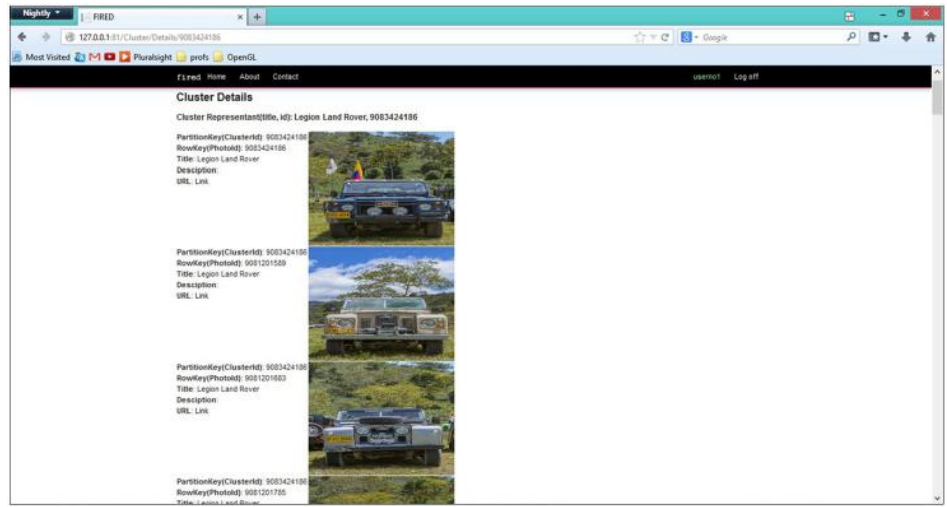

**Figure 19:** Group based on the image prototype

One of the main advantages of this architecture is modularity. Thus, it is possible to add new components for image or text processing at any time, to improve the overall quality of the system. For the future, we intend to add more components to this architecture and to consider also a credibility component in the search process (Gînscă et al., 2015).

**Diversification in Image Retrieval with YAGO**

For the next experiments, we consider around 30.000 images from entire MUCKE collection. Our aim was to process existing data at textual and image levels and after the retrieval step to offer the results in a diversified way (Iftene and Alboaie, 2014).

To offer the results in a diversified way, we identify the relevant keywords in the query. For that, standard tools were used for POS-tagging (Simionescu, 2011), lemma identification[76] and named entity identification (Gînscă et al., 2015). Next, a query expansion module that makes use of the YAGO ontology[77] is used (Rebele et al., 2016) (see Figure 20).

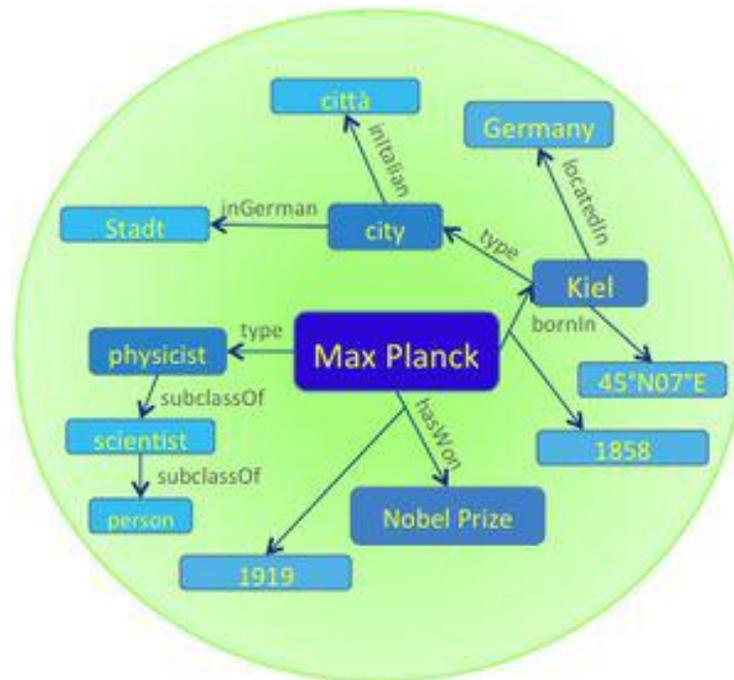

**Figure 20**: Sample from YAGO ontology[78]

YAGO ontology has knowledge about the world (Hoffart et al., 2013) organized in an intuitive way (see Figure 20). The information from YAGO is extracted from semantic resources like Wikipedia[79], WordNet[80] and GeoNames[81]. It is structured in elements called **entities** (*persons*,

---

[76] http://nlptools.infoiasi.ro/WebPosRo/
[77] https://www.mpi-inf.mpg.de/departments/databases-and-informationsystems/research/yago-naga/yago/
[78] https://www.mpi-inf.mpg.de/fileadmin/_processed_/a/5/csm_yago-graph_b488494451.png
[79] http://en.wikipedia.org/wiki/Main_Page

*prizes*, etc.) and **facts** about these entities (*where is a city located*?, etc.). For example, starting from the query "*tennis player on court*", where two entities ("*tennis player*" and "*court*") are identified, for "*tennis player*" are searched in YAGO available instances (Iftene, 2017) and we found "*Roger Federer*", "*Novak Djokovic*", "*Serena Williams*", etc. Therefore, instead of doing a single search with the initial query, we get to the situation where we perform several searches with the new queries, following which, in the end, we obtain the final result by combining the partial results obtained. Figures 21 and 22 show the results obtained with query "*tennis player on court*" in our application and Google. We can see how in our results appears both concepts ("*tennis player*" and "*court*"), while in the Google are cases of results without first concept.

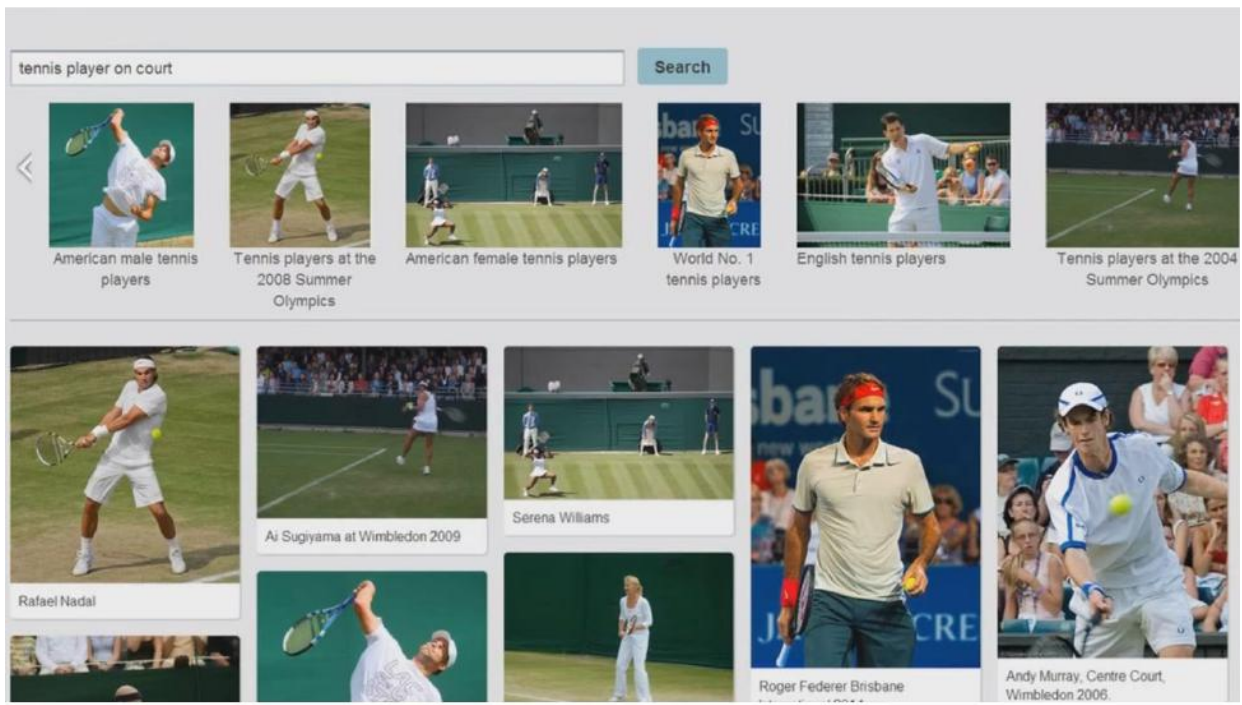

**Figure 21**: Results for query "*tennis player on court*" in our application (Iftene and Baboi, 2016)

---

[80] http://wordnet.princeton.edu/
[81] http://www.geonames.org/

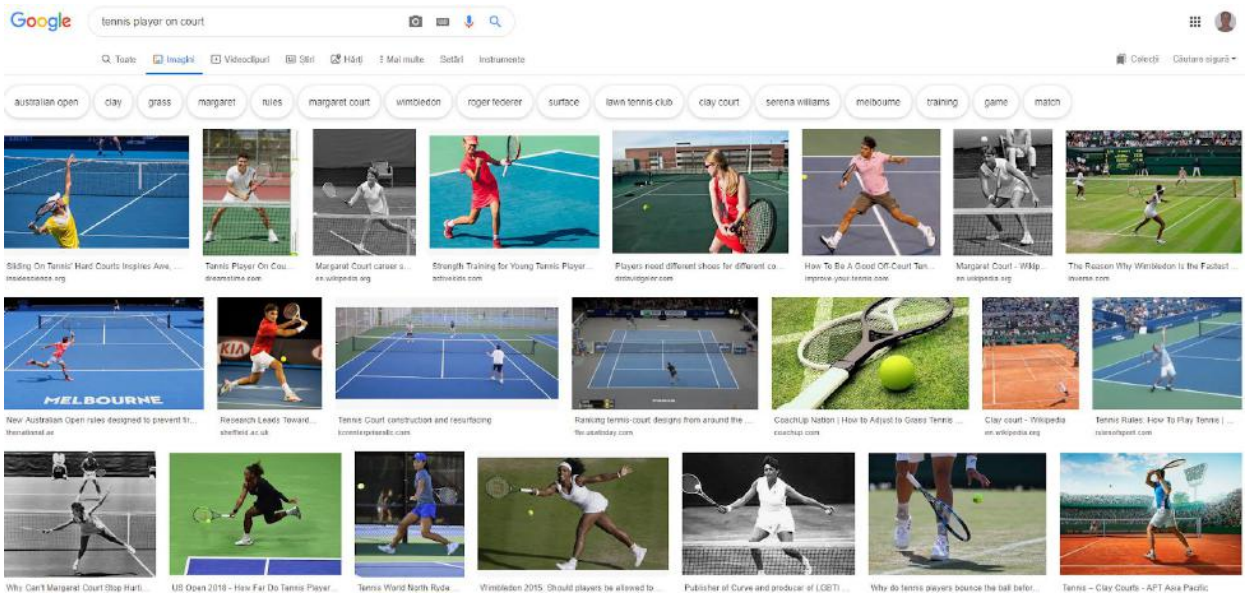

**Figure 22**: Results for query "*tennis player on court*" in Google application

**Exploiting Information from Social Networks to Improve the Quality of Image Searches**

To improve the search quality, there is a need for annotated corpora with the help of human annotators, which will then be used in the search process. Creating these resources is often a costly operation (as a time, but also from a financial point of view). For these reasons, the use of existing information on social networks, where users voluntarily contribute to the annotation of images (by adding comments, by adding emoticons, etc.), has become increasingly used lately.

On the one hand, social media can be used to add new tags to the images in the comments associated with them, elements that can be used later in their classification and search (McAuley and Leskovec, 2012). On the other hand, the quality of a component that identifies images that hurt religious, political, or instigating violence, can be improved by automatically increasing the resources used by it (Goel et al., 2016). As a social network records all user's activities (posts, comments, likes, etc.) a profile can be created for it, which can be used successfully later when it searches (Ntalianis and Doulamis, 2017). The results can be further filtered according to this profile, thus offering personalized results and therefore more relevant to it. Interesting is how based on the images and comments associated with them on Instagram can determine the health status of the population in a certain geographical area (Garimella et al., 2017).

**Conclusions**

Today, big companies like Flickr, Google, Bing, Yahoo and Microsoft are interested in hosting multimedia content and allowing users to access it. In the MUCKE project we created a collection of images from the Flickr network and the Wikipedia encyclopedia. To improve the quality of search operations, we used natural language processing techniques and tools (POS taggers, tools for lemmatization, for anaphora resolution, for identification of named entities) and semantic resources (such as YAGO, Wikipedia, WordNet, Freebase and GeoNames).

## II.3.4 Graphic View of Information on Twitter

**Introduction**

Data on Twitter has started to be input data for many of the applications developed in recent years. The purposes of these applications range from *identifying user emotions* (in relation to certain products or companies, in connection with candidates to local or presidential elections, in connection with an event, etc.), *to identifying extreme natural phenomena* (earthquakes, fires, hurricanes, etc.), *to identify the satisfaction status of the people* (regarding the medical services they benefit from, or regarding the services offered by the local administration, etc.), *to identify the depressive states to certain persons or groups of persons*, etc. A system that processes data on Twitter has three classic modules:

1. *local search and save of tweets* (in databases or in the form of XML or JSON),
2. *processing of saved data* (usually with language services), which eliminates stop-words, identifies the part of speech, identifies name entities, identifies the positive, negative, neutral opinions of those who express them, etc.,
3. *displaying the results obtained* (either statistics or geographically distributed information), when this is possible. Here are some ways to view the results obtained after processing.

Displaying opinions and feelings identified in Twitter posts using Google Maps (Hao et al., 2011) (Figure 23, in the left) is very similar to displaying friendship links between Facebook users (Ma, 2012) (Figure 23, in the right).

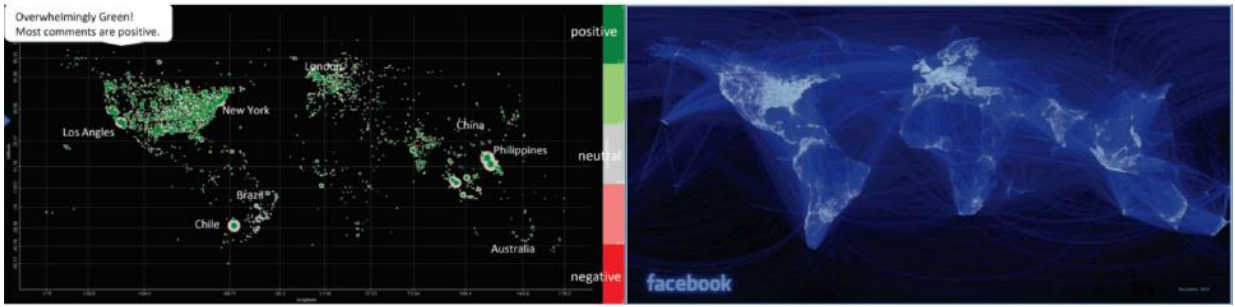

**Figure 23**: Displaying feelings that appear in tweets (left), Friendship links between Facebook users (right) (Hao et al., 2011)

In the paper (Malik et al., 2013) the authors graphically suggest using the TopicFlow application, the most discussed topics on Twitter, using the time axis (Figure 24).

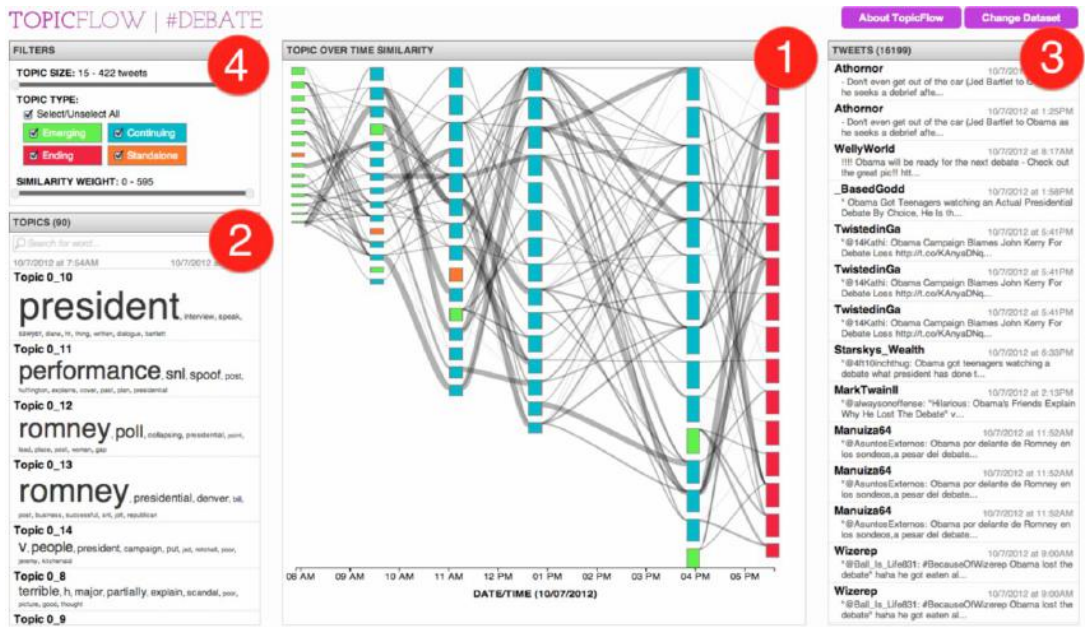

**Figure 24**: TopicFlow has 4 areas: (1) TopicFlow diagram, (2) summary list of topics, (3) a list of tweets from the data used, and (4) filtering modes (Malik et al., 2013)

Also interesting are two other approaches: (1) the approach that uses hash-tags in tweets to make statistics (Stojanovski et al., 2014) (Figure 25, left) and (2) how to display feelings from one's posts by states in the US (Engel, 2016) (Figure 25, right).

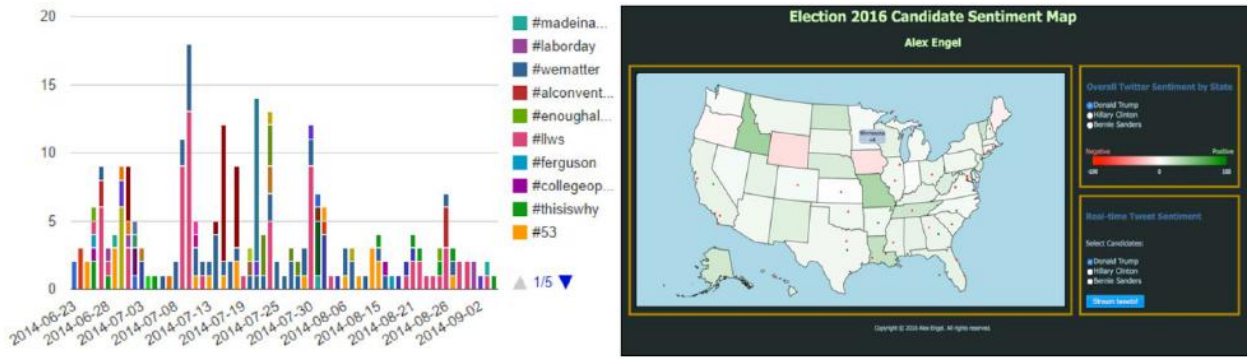

**Figure 25**: Hash-tag statistics (Stojanovski et al., 2014) (left), Centralized display by states at US presidential elections in 2016 (Engel, 2016) (right)

**Sentiments on Twitter**

For the application that we developed together with the master's students, we used as an input data source the Twitter platform. This provides real-time continuous data flow, along with details about the location of the users posting this information. The connection between Twitter and our application is made using a module, which constantly receives data from it. The data received are filtered (first we keep only those in English or Spanish, which have information about the geolocation of the user, who made the posting) and then sent to the way that annotates them (according to the opinions of the users who posted them).

For annotation with feelings, we used Sentiment140 API, which uses a template especially trained on tweets (in a semi-supervised method), to classify them. Once annotated, they will be sent to a module that is responsible for distributing them to connected web clients. Also, the annotated stations will be stored to allow us to do offline analysis when needed. The web interface simply takes this information from the module that distributes it and presents it to the users in a graphical way.

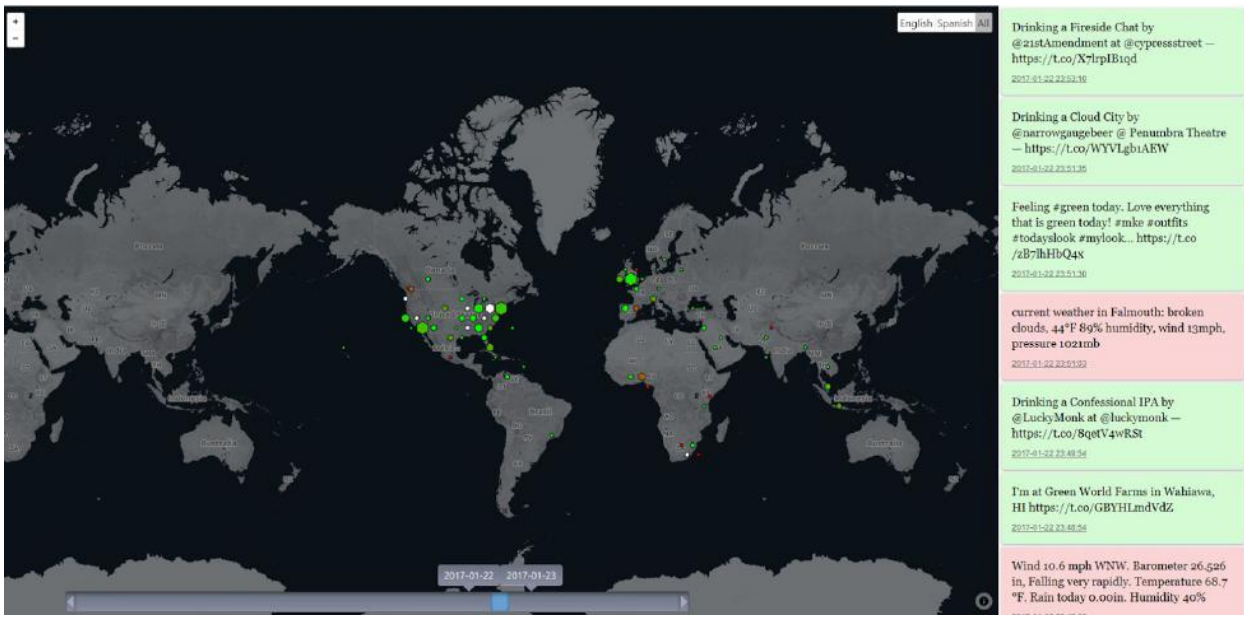

**Figure 26**: View clusters on Google Maps

As can be seen in Figure 26, tweets are grouped into clusters, represented by circles, distributed on the globe map. The color of a cluster will intuitively represent what it contains: *white for neutral clusters* (containing the same number of positive and negative tweets or just neutral tweets), *a color between green and red* (green representing positive tweets and red tweets negative negatives).

When one cluster grows, it affects the other clusters on the map, which will shrink in proportion, so that the largest clusters will be the most visible. If the user uses the zoom option in a particular region (see Figure 27 left), where a larger cluster is located, it will be divided into smaller clusters, until we see only clusters composed of a single tweet (see Figure 27 right).

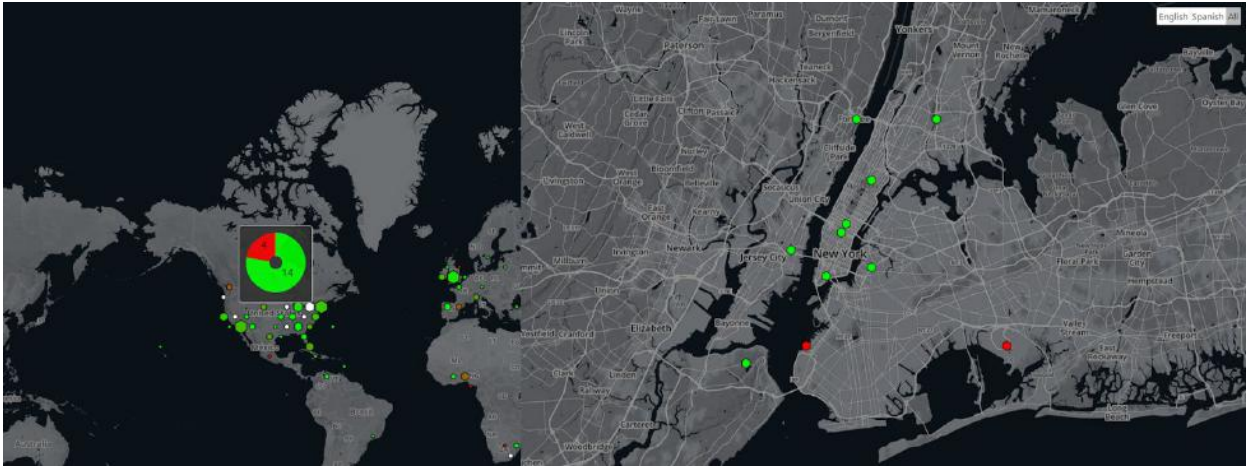

**Figure 27**: View details of a larger cluster (left), View map using zoom-in (right) (Iftene et al., 2017)

To see how our application works, we let the application run for more than 24 hours (between December 31, 2016, and January 31, 2017). As shown in Figure 20 at the bottom, we can select the period for which we want to view the information on the map, the minimum is one day (implicitly the current day is used), and the maximum is 7 days (we do not have to process a large amount of data).

**Conclusions**

In the activity we did, we came up with a new proposal to display information on Google Maps. Compared to the previous approaches, which display all the tweets collected in a certain period, we can manage the period we display the tweets from 1 day to 7 days. Interesting fact is when we use the period of a day and move from the oldest date to the newest date and we can see over time how Twitter posts across the globe evolve. Also, we can manage to display credible or not credible tweets, statistics on countries and continents in real-time

Also in Figure 12, we present a new proposal to display in real-time statistics and information related to credible and not credible tweets on Google Maps using heatmap.

## II.4 Participation in Evaluation Campaigns

Above algorithms have already been tested in **CLEF evaluation campaigns**[82], such as ImageCLEF (PlantIdentification, Tuberculosis, Scalable Concept Image Annotation Challenge), CheckThat!, QA4MRE, GeoLife, Machine Reading Evaluation, etc.

The UAIC group was involved in **2013** in the Plant Identification task at Image CLEF, where it was ranked 5th out of 12 participating groups (Şerban et al., 2013). Because, the proposed architecture was modular, it allows us to analyze, combine and improve the results after applying methods such as *image retrieval using LIRe*, *metadata clustering,* and *Naïve Bayes classification*. The UAIC group also participated in the QA4MRE task at CLEF (Iftene et al., 2013), where the system contains special modules for natural language processing as *data pre-processing*, *data indexing*, *identification of relevant snippets*, *textual entailment*, *anaphora resolution* and *identification of the correct answer*,.

In **2015** and **2016**, we participated at Image CLEF in the task of annotating images with concepts (Scalable Concept Image Annotation Challenge) (Calfa et al., 2015), (Cristea et al., 2016). In 2016, we used an *ontology* created in 2015 with relations between *concepts* (represented by words, synonyms, hyponyms and hypernyms).

In **2019**, we were involved in GeoLife challenge (Atodiresei and Iftene, 2019), where the participants need to predict plant species given their location. Also, our group participated in Check That!, where we need to evaluate the worthiness of a political claim in a debate. The proposed method to achieve the goal of the task was to represent each claim by a *feature vector* and feed it to a machine learning classification algorithms to classify if the claim is check-worthy or not (Coca et al., 2019). At ImageCLEF 2019, Tuberculosis task our group submitted a solution that addresses the problem of tuberculosis' severity prediction in low-resource environments by attempting to minimize the information required from the CT scan using a regularized variant of the SAMME.R algorithm (Tabarcea et al., 2019).

---

[82] http://www.clef-initiative.eu/

In addition to participating in the CLEF evaluation campaigns, we also participated in **SemEval evaluation campaigns**[83]: Contextual Emotion Detection in Text, Discovering Humorous Tweets, Discover Word Similarity and Identification of Sentiments in Tweets.

In **2016**, the UAIC team was involved in Task 4 related to the identification of sentiments in tweets (Florean et al., 2016) and (Ciubotariu et al., 2016). The systems sentiment analyzers consist of several freely available resources and tools, additionally enhanced with a classifier proposed by us trained on the SemEval-2016 training data.

In **2017**, the UAIC team participated in SemEval 2017 Task 6 related to discovering humorous in tweets with a system combining Naïve Bayes and neural networks (Flescan-Lovin-Arseni et al., 2017). Also, UAIC participates in the SemEval-2017 Task 2 "Multilingual and Cross-lingual Semantic Word Similarity", which deals with identification of semantic similarity between two words. The system was built using neural networks (Rotari et al., 2017).

In SemEval **2019**, in Task 3, EmoContext: Contextual Emotion Detection in Text (Chatterjee et al., 2019), the organizers ask participants to classify users messages in one of four classes: *Happy*, *Sad*, *Angry* or *Others*. In (Simionescu et al., 2019), we present the architecture of a system designed by us for this task.

**Conclusions**

Many parts of these systems were developed or used resources which were built during the MUCKE project. The obtained results demonstrate the quality and robustness of the architecture proposed in the project, but also, we demonstrate the extensibility of the architecture, because many times the researches have led to the completion of the MUCKE architecture with new components, these extend the main functionalities or improve the existing ones.

The students preferred to get involved in such competitions because the organizers provide the resources necessary to evaluate the systems proposed by the participants. In this way, most of the effort made by the students was related to the implementation and testing of various algorithms, instead of being allocated to the creation and validation of resources. When they didn't find what they wanted, they created additional resources and ontologies just to improve their results to a certain extent.

---

[83] http://alt.qcri.org/semeval2019/index.php?id=tasks

## II.5 Conclusions

In this section, we presented how we can use social networks such as Twitter and Flickr to create our resources and applications. These resources were used in the MUCKE project where we created an image retrieval system. The image retrieval system comes with new implemented concepts like **credibility** (of users and resources), **diversification of results** (with semantic resources like YAGO), **scalability** (due to a large number of images). On the one hand, in the MUCKE project, the credibility of resources and users is seen based on the correlation between the concepts existing in the image and the concepts existing in the textual description of an image. Based on this correlation score, we can calculate the credibility of a resource and we can reorder the results after the information retrieval step using this additional information.

On the other hand, in Twitter, the credibility of news and users is calculated based on the impact that the posts have (number of followers, retweets, comments, etc.), based on the current activities of those who post them, based on how they write a post (capital letters, use of hash-tags and emoticons, etc.), etc. How we show, it is very hard to identify what is fake and what is true in the Twitter network, but we can order based on the credibility score, the following entities (personalities from politics, sports, music, etc., organizations, political parties, televisions, etc.)

It is interesting how sentiment analysis of tweets and the different ways of displaying them, can indicate the new events that take place on the globe and the areas where much is discussed about them. Our proposal to display the information on Google Maps takes into account the location, the calendar date, the language, the positive/negative feelings, the number of postings in a certain area, the existence of certain keywords in the post.

## II.6 Future Work

We believe that social media mining will become a key component used to find users' opinions about a subject, to plan marketing strategies, to decide whether a customer will buy a product or not, to help in case of emergency, etc. That is why in future any application, no matter how small it is, it should acknowledge this kind of information and exploit it.

In the future, we have in mind to find a more efficient way to classify tweets credible and not credible, while having real-time processing of data. Also, we have in mind to build flexible architectures based on microservices, which are scalable and fault tolerant (Baboi et al., 2019). We believe that the credibility of information will become a key component used to find true news on the Internet. That is why in future any application that will use data from social networks, no matter how small it is, it should make a difference between credible and not credible data.

## **II.7 References**

Simionescu, C., Stoleru, I., Lucaci, D., Bălan, G., Bute, I., **Iftene, A.** (2019) *UAIC at SemEval-2019 Task 3: Extracting Much from Little*. In Proceedings of the 13th International Workshop on Semantic Evaluation, Association for Computational Linguistics (ACL), pp. 355-359.

Simionescu, R. (2011) *Hybrid POS Tagger*. In Proceedings of Language Resources and Tools with Industrial Applications, Workshop (Eurolan 2011 Summer School).

Singh, V., Dasgupta, R., Sonagra, D., Raman, K., Ghosh, I. (2017) *Automated Fake News Detection Using Linguistic Analysis and Machine Learning*. In Proceedings of Conference SBP-BRiMS, DOI: 10.13140/RG.2.2.16825.67687.

Smith, K. (2019) *126 Amazing Social Media Statistics and Facts*. Brandwatch. Marketing. https://www.brandwatch.com/blog/amazing-social-media-statistics-and-facts/#section-2.

Smith, T. F., Waterman, M. S. (1981) *Identification of Common Molecular Subsequences*. In Journal of Molecular Biology, vol. 147, pp. 195-197.

Sneha, S., Nigel, F., Shrisha, R. (2017) *3HAN: A Deep Neural Network for Fake News Detection*. In 24th International Conference on Neural Information Processing (ICONIP 2017), Springer International Publishing AG 2017, part II, LNCS vol. 10635, pp. 1-10, https://doi.org/10.1007/978-3-319-70096-0_59.

Spitkovsky, V. I., Chang, A. X. (2012) *A Cross-Lingual Dictionary for English Wikipedia Concepts*. In Proceedings of the Eighth International Conference on Language Resources and Evaluation (LREC-2012), pp. 3168-3175, Istanbul, Turkey.

Stojanovski, D., Dimitrovski, I., Madjarov, G. (2014) *TweetViz: Twitter data visualization*. In Proceedings of the Data Mining and Data Warehouses.

Şuşnea, E., **Iftene, A.** (2018) *The Significance of Online Monitoring Activities for the Social Media Intelligence (SOCMINT)*. In Proceedings of the Conference on Mathematical Foundations of Informatics MFOI'2018, pp. 230-240, July 2-6, Chisinau, Republic of Moldova.

Tabarcea, A., Roşca, V., **Iftene, A**. (2019) *ImageCLEFmed Tuberculosis 2019: Predicting CT Scans Severity Scores using Stage-Wise Boosting in Low-Resource Environments*. In Working Notes of CLEF 2019 - Conference and Labs of the Evaluation Forum, Lugano, Switzerland, September 9-12.

Tjong Kim Sang, E., Bos, J. (2012) *Predicting the 2011 Dutch Senate Election Results with Twitter*. In 13th Conference of the European Chapter of the Association for Computational Linguistics, April 23-27. Avignon, France.

Turney, P. D. (2002) *Thumbs up or thumbs down? Semantic orientation applied to unsupervised classification of reviews*. In Proceedings of the 40th annual meeting on association for computational linguistics, pp. 417–424, Association for Computational Linguistics.

Winkler, W. E. (1990) *String Comparator Metrics and Enhanced Decision Rules in the Fellegi-Sunter Model of Record Linkage*. In Proceedings of the Section on Survey Research Methods. American Statistical Association, pp. 354-359.
76

# III. Technological Trends

## III.1 Augmented Reality

### III.1.1 Context

**Introduction**

The term **augmented reality** (AR) first appeared in the 1950s, when Morton Heilig, a moving film cameraman, "*believed that cinema as an art should be able to attract the viewer to screen activity*" (Alkhamisi and Monowar, 2013). In 1962, Heilig built a model of this idea, which he called in 1955 as the "*Cinema of the Future*", known as *Senorma*, which existed before digital computing (Carmigniani et al., 2011). Then, Ivan Sutherland invented the head-mounted in 1966 (Krevelen and Poelman, 2010), (Carmigniani et al., 2011). In 1968, he comes with first AR system (Yuen et al., 2011). In 1975, Myron Krueger set up an artificial reality lab where users can easily interact with virtual elements (Carmigniani et al., 2011), (Carmigniani and Furht, 2011).

In the early 1990s, AR became an important topic of study for many researchers. In 1997, Ronald Azuma comes with the first study in AR and propose the notion of AR as "*the assembly of the real and the virtual environment, while both are recorded in 3D and interactive in real-time*" (Carmigniani et al., 2011), (Ford and Höllerer, 2008). In 2000, Bruce Thomas invented the first AR mobile game and presented it during the International Symposium on Laptops (Carmigniani et al., 2011), (Carmigniani and Furht, 2011). Then, several AR applications were designed especially for mobile applications. The Wikitude AR Travel Guide was created in 2008 (Carmigniani et al., 2011). In 2008, Gartner Inc. expected AR to be among the top 10 technologies in the period 2008-2012 (Yuen et al., 2011). Also, it is clear that the number of AR-accessible applications has grown sharply and expanded to include not only location-based search applications, but also social media, eLearning, games, training, entertainment and medical applications (Ford and Höllerer, 2008). Figure 28 shows the evolution of AR throughout history.

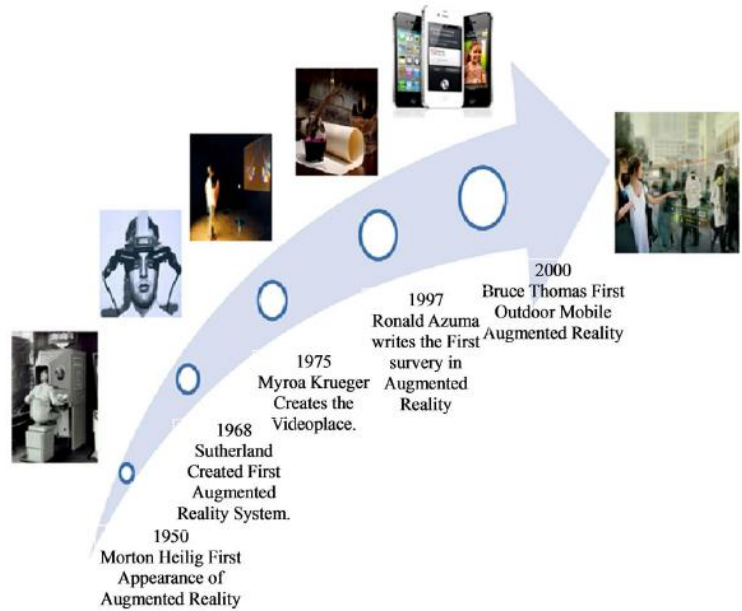

**Figure 28**: History of Augmented Reality (Alkhamisi and Monowar, 2013)

**Definitions**

**Augmented reality** (AR) is *a variant of Virtual Environments (VE) or virtual realities*, as it is called. VE technologies immerse complete a user within an artificial environment (Azuma, 1997).

To limit AR to specific technologies, in the paper (Azuma, 1997), the author defines an AR system as *a system having the following three characteristics:*

1. *It combines the real and the virtual environment;*
2. *Real-time interactive;*
3. *Works in 3-D.*

Augmented reality enhances the user's perception and helps him understand and interact with the real world. Virtual objects display useful information for the user and helps him to execute concrete tasks in the real world. AR is an example of what Fred Brooks calls intelligence amplification (AI): *using the computer as a tool to make a task easier for a human to perform* (Brooks, 1996). Several types of applications using AR have been explored: *medical domain, maintenance and repairs, annotations, robot route planning, entertainment, navigation and orientation of military aircraft*.

In 2010, in the paper (Krevelen and Poelman, 2010), the authors present the notion of the **reality-virtuality continuum**, initially defined by (Milgram and Kishino, 1994), in *which AR is part of the domain-general of combined reality*. Here, virtual and augmented environments are combined and the user can see real objects combined with virtual ones. When analyzing not only artificiality but also user transport, (Benford et al., 1998) classifies AR as being separate from both VR and teleoperation. According to (Azuma, 1997) and (Azuma et. Al, 2001), an AR system:

- *combines real and virtual objects* in a real environment;
- *aligns real and virtual objects* with each other; and
- *runs interactively*, in three dimensions and in real-time.

In (Carmigniani and Furht, 2011) the authors define **augmented reality** (AR) as *a direct or indirect real-time visualization of a real-world physical environment that has been enhanced/ augmented by adding computer-generated virtual information*.

**AR Devices**

The main devices used by AR are the special screens used for display, input and tracking devices and computers (Carmigniani and Furht, 2011).

- **Screens used for Display** - there are three main types: head-mounted displays (HMD), portable devices and space screens.
- **Input Devices** - there are several types: some use gloves, others wireless bracelets, for smartphones, even the phone, others interpret the user's gestures, others interpret the user's touch screen, etc.
- **Tracking Devices** - consist of digital cameras with or without optical sensors, GPS, accelerometers, compass, wireless sensors, etc.
- **Computers** - an AR system requires a powerful processor and enough RAM to process the images taken by the camera. So far, laptops have been used, and the growth in performance of smartphones and tablets has made the evolution of these systems faster and faster. Fixed complex systems still use classic computers with powerful graphics cards.

**AR Interfaces**

There are four main modes of interaction in AR applications: *tangible AR interfaces*, *collaborative AR interfaces*, *hybrid AR interfaces* and *emerging multimodal interfaces* (Carmigniani and Furht, 2011).

- **Tangible AR Interfaces** - allow direct interaction with the real world using real objects for this. For example, VOMAR application (Kato et al., 2000) allows a person to select and rearrange the furniture in a room using an AR application.
- **Collaborative AR Interfaces** - permit to use at the same time multiple displays to support collaborative activities.
- **Hybrid AR Interfaces** - combine different interfaces and offer the possibility to interact through many devices (ISMAR, 2002).
- **Multimodal AR Interfaces** - use different types of inputs (speech, touch, gestures or gaze). These types of interfaces have recently emerged.

**General Architecture of an AR System**

The general architecture of a system using augmented reality is shown below in Figure 29 (Glockner et al., 2014).

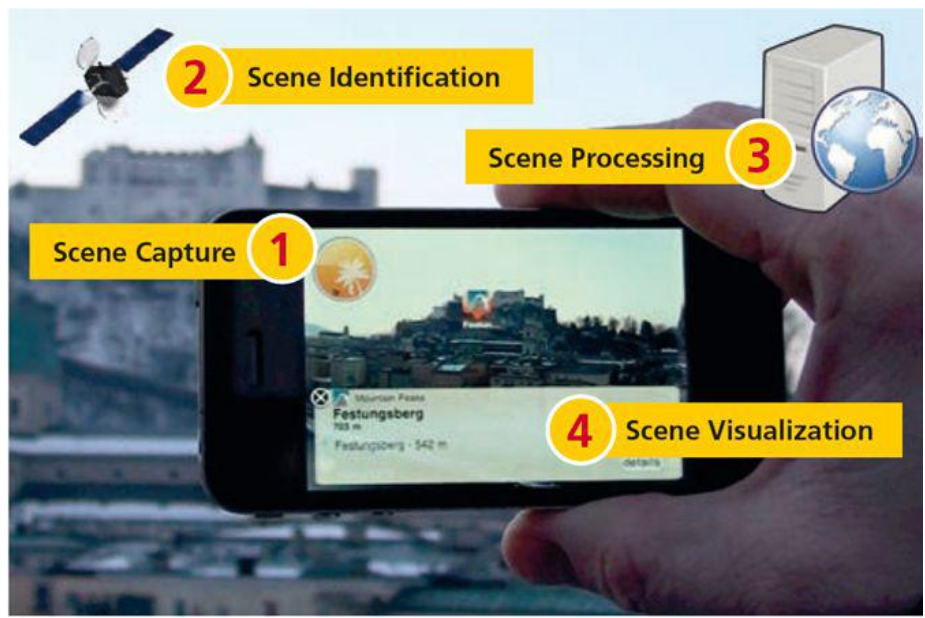

**Figure 29:** The basic functionality of a system based on augmented reality (Glockner et al., 2014)

The four tasks performed by the AR system are *scene capture*, *identifying scenes for choosing accurate information to stimulate them*, *scene processing* and *augmented scene visualization* (Cai et al., 2012) (Lopez et al., 2010). These tasks are described in detail as follows (Alkhamisi and Monowar, 2013):

*Capture the Scene*

In general, the devices used in the scene capture are physical devices that recognize the reality that should be augmented. There are two types of devices used for scene capture:

- **Camcorder devices**: such devices capture reality differently than other devices used to view augmented reality (for example, video cameras and smartphones) (Lopez et al., 2010).
- **Visualization Devices**: These devices capture reality and provide an image of it with augmented information (e.g., head-mounted devices) (Lopez et al., 2010).

*Scene Identification*

Scene identification is also considered one of the main actions taken in augmented reality. There are two types of scene identification techniques with the following characteristics:

- **Marker-Based** approach uses markers that are in the form of visual labels contained in the actual scene of the entire AR system (Lopez et al., 2010).
- **Marker Not Required**: AR systems that do not use markers use scene identification devices. As an AR browser, it uses tags to help users analyze and navigate through data in a digital environment associated with the real environment.

*Scene Processing*

After calculating the spot of a specific marker in the real space, depending on the internal and external parameters of the camera, the system looks for the virtual model corresponding to each marker in 3D. This virtual model will be used in the next step of processing, related to the viewing of scene.

*Viewing the scene*

Finally, the system generates the image of the 3D object and the actual projected space and passes on the stage image that mixes reality and virtualization when using the marker and presents the digital information when not used as a marker scene identification technique.

**Applications**

In (Azuma, 1997) and (Glockner et al., 2014) authors present the most important domains where were built applications based on AR. An overview of all these domains can be seen in Figure 30 from below.

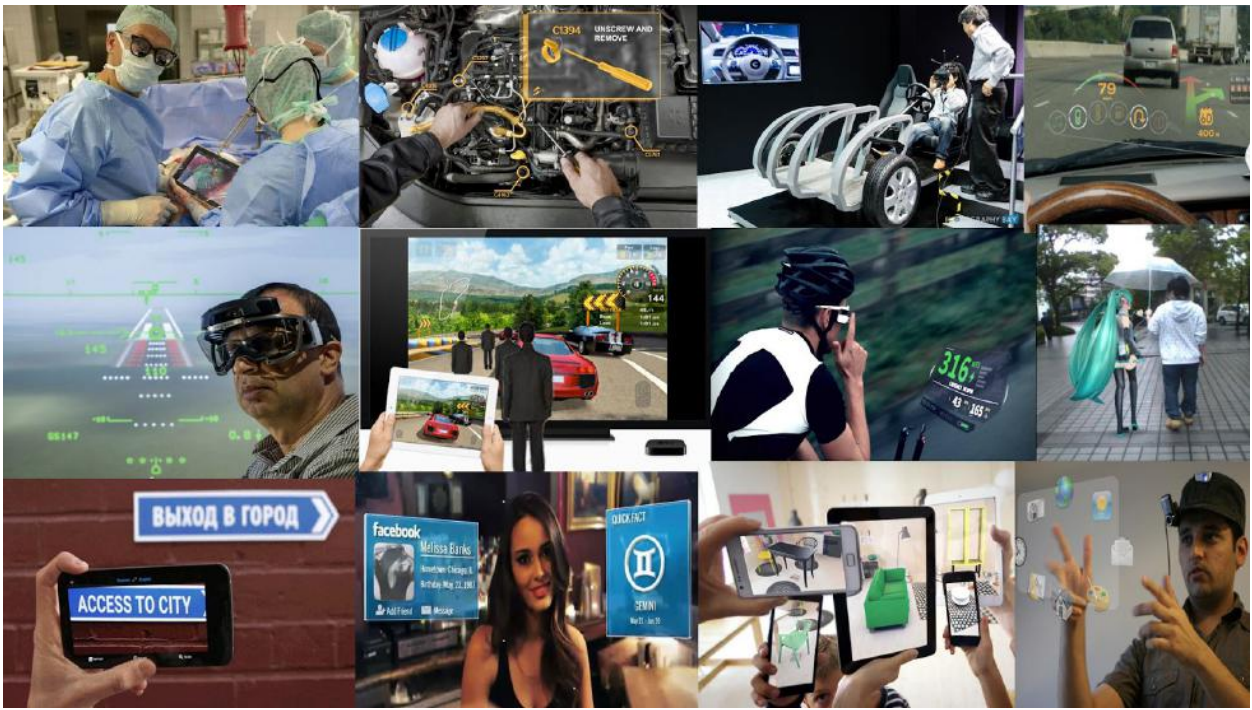

**Figure 30**: The most important domains in which AR applications were created

- **Medical Domain -** Doctors could use AR as a visual and training aid for surgery (Azuma, 1997). It is possible to collect real-time data of a patient, using non-invasive sensors such as magnetic resonance imaging (MRI), computerized tomography (CT) or ultrasound imaging. Thus, AR technology could provide an increased internal vision of the patient, without the need for larger incisions. Currently, the Liver Explorer[84] application of

---

[84] https://www.mevis.fraunhofer.de/en/press-and-scicom/institute-news/fraunhofer-mevis-liver-explorere-app-at-the-apple-special-event.html

Fraunhofer MEVIS is used. This app offers real-time AR guides and healthcare. The camera shoots the liver and, using AR, overlays the surgical planning data on the organ. Also, the software can react in real-time (for example, updating the surgical plan according to the speed of the blood).

- **Production and Maintenance** - The instructions could be easier to understand if they were available, not as text and image manuals, but as 3D drawings that overlap the actual equipment, showing the tasks step-by-step and how to be done and the way to do them. These 3D images can be animated, making the instructions even more explicit. In the automotive field, those from Volkswagen developed the MARTA[85] (Mobile Augmented Reality Technical Assistance) system. This system helps when a car is not working properly, helping the user to perform repair and maintenance of the vehicle. The application recognizes the vehicle parts by recognizing objects and describes text and images all the steps that need to be executed in real-time. This application runs on various mobile devices.

- **3D Collaborative Modeling** - Canon's MREAL[86] (Mixed Reality System) supports the design process, allowing the 3D models generated by the computer to be combined with real-world objects. MREAL allows users to work in a collaborative way and simultaneously over a model. The application does this by creating a 3D model of both existing and new components, which it then combines. For example, an existing car seat can be integrated into a new car. Because MREAL offers a mixed reality, users can sit in the (real) seat and see around him the AR environment (not real).

- **Annotation and Visualization** - AR could be used to annotate objects and environments with additional information. Applications that use public information require the availability of public databases that they can use. For example, a portable display might provide information about the contents of library shelves while the user walks in an AR environment (Fitzmaurice, 1993). At ECRC (European Computer-Industry Research Center), a user can indicate certain parts of an engine model, and the AR system displays the name of the selected part (Rose et al., 1995). Wikitude[87] and Metaio Junaio[88] are two

---

[85] https://www.volkswagenag.com/en/group/research/virtual-technologies.html
[86] http://www.canon.com/technology/future/mixedreality.html
[87] https://www.wikitude.com/

AR browsers that provide context-sensitive information about recognized objectives that are around the user.

- **Planning a Robot Route** - Teleoperation of a robot is a difficult problem, because usually the robot is far away, and we have delays in communication. In this situation, instead of controlling the real robot, it would be preferable to control an AR version of the robot. Once the plan is tested and verified in artificial environment, the user can tell the real robot to execute the specified plan. This avoids problems caused by delays due to communication. During the testing we can predict the effects of environmental manipulation, thus serving as a preview tool to help the user accomplish the list of tasks. The ARGOS system has shown that such solutions work better than traditional solutions (Milgram et al., 1995).
- **Auto Industry - Assistance during the Trip** - Currently BMW[89] displays on the car windscreen information that helps on the move, such as speed, direction of travel based on the recognition of objects in the real environment, parking assistance, information taken from sensors, warnings related to traffic and possible collisions, etc.
- **Military Domain** - From many years ago, pilots of military aircraft and helicopters used headphones with HUDs (Head-Up Displays) and HMS (Helmet-Mounted Sights) displays during the landing and take-off (Wanstall, 1989). A military application currently used is the Q-Warrior helmet[90]. AR elements have the role of providing the military with situational attention, identification of friends or enemies, night vision and an increased ability to remotely coordinate small units of soldiers equipped with similar equipment.
- **Entertainment** - At SIGGRAPH '95, several exhibitors showed "Virtual Crowds" that combine actors with virtual environments, in real-time and in 3D. The actors stand in front of a large blue screen, while a computer-controlled mobile camera records the scene. Because the location of the camera is monitored and the actor's movements are known in advance, it is possible to digitally combine the actor into a 3D virtual scene. Of course, this as a way to reduce production costs: creating and storing scenes is practically cheaper than constantly building new physical scenes from scratch.

---

[88] https://www.facebook.com/junaio-326372139104/
[89] http://www.motoringfile.com/2011/10/12/bmw-group-developing-augmented-reality-windshield-displays/
[90] https://www.youtube.com/watch?v=3ZD1ywnvdKM

- **Recreational Activities** - Recon Jet[91] is an AR system already available for recreational activities. The head-level display (glasses or lenses) connects to third-party sensors, such as Bluetooth and WiFi, and provides real-time route and weather information, enables social media access, and provides real-time performance information. For example, a runner would want to know speed, distance to the finish line, altitude difference and heart rate. With these capabilities, Recon Jet gives us a picture of future developments in the wearable AR field, which could monitor, in addition to the vital signs of a person and the work environment, in dangerous environments.
- **Virtual Friends** - A Japanese hacker used an available 3D model and motion sensors to have an "AR meeting" with a famous Japanese cartoon star (Hatsune Miku[92]). In his video, he shows how he goes with Miku to a real park and how Miku recognizes and reacts to objects in the real world (for example, sitting on a real chair). This software even allows interaction with the virtual pop star (e.g., the tie or the head). This app gives us the idea that soon, people can be accompanied by a virtual presence that could provide assistance when needed (for example, in the medical or engineering fields or as a human interface for daily activities, such as managing a personal calendar, notes and contacts).
- **Translation** - another promising area of application in AR is translation domain. An existing application is Word Lens[93] software that runs on almost any smartphone and can translate text from one language into another. The user points their device towards a piece of text written in a foreign language, and it will display the translated text in the user's native language. The translated text will be written with the same font and on the same background as in the real world.
- **Facial Recognition** - Accessing information from our applications can be made easy by combining face detection with AR. One application that use facial recognition is the Infinity AR[94] app. The idea is to analyze a face and identify it by searching it on social networks in profile pictures (eg Facebook), and once we have found the user access the data in his profile. Obviously such an application could also be used by the people of the law as they search for certain people who have problems with the law and use this application to identify them faster.

---

[91] https://www.reconinstruments.com/products/jet/
[92] https://www.youtube.com/watch?v=9jpWiTVR0GA
[93] http://bigthink.com/design-for-good/word-lens-real-time-translation-via-augmented-reality
[94] http://www.infinityar.com/

- **Interior Design** - One suggestive examples is included in the Ikea catalog that uses AR[95]. The AR app allows users to use smartphones and tablets to view, select, place 3D models of Ikea furniture in their homes. Once the user is satisfied with the chosen furniture, he can place the order online and he will receive the order sent home as soon as possible.

- **Use of Gestures in AR** - The new generation of applications is becoming more intelligent and allows us to interact with them in increasingly natural ways. Gesture communication is simple and intuitive at the same time, and the combination with AR creates an attractive environment for those who use these applications. An example of a gesture interface system that uses AR is SixthSense[96], developed by MIT. The system recognizes the user's gestures, which he captures with a video camera and translates them using techniques based on "computer vision".

- **Internet of Things** - Lately, AR-based interfaces have begun to enter the Internet of Things area, allowing us to control with them cars, audio and video equipment, appliances or heating systems. A very good example in this area is the Revolv AR[97] system, which is under development. By combining AR with Google Glasses, the system allows the user to interact with digital devices in the home (lighting system, air conditioning, heating system, etc.).

**Conclusions**

Augmented reality with a consistent history since the 1950s, has become increasingly popular in recent years. In this first part, we saw the basics of this field and the elements that underlie a system based on AR. The number of areas of applicability has increased from year to year, but at the same time it has diversified. Classic fields, such as medicine, engineering, military, 3D collaborative modeling, robotics, have begun to diversify in recent years through entertainment, social networks, interior and exterior design, translation, security, car assistance during travel, etc. In the next period, we envisage a greater involvement in the area of transmitting orders and actions through gestures, either in the entertainment applications or in the sensor management applications in a smart home.

---

[95] htttps://www.youtube.com/watch?v=vDNzTasuYEw
[96] http://www.pranavmistry.com/projects/sixthsense/
[97] https://www.f6s.com/revolvar

## III.1.2 AR in eLearning

**Introduction**

Research from (Agape et al., 2015) showed that 65% of the population is visual and 80% of what they experience is much better retained. As the authors of (Johnson et al., 2010) state, augmented reality has immense potential, to provide a useful context to the learning process, allowing learning and discovery experiences connected to real information in the surrounding world.

Visual techniques based on augmented reality have been used in education since ancient times (Munnerley et al., 2014): increasing the walls of the caves with suggestive images to help younger generations in hunting and survival; The Heilig patent for Sensorama presents the necessity of teaching people in the army, industries and schools (Heilig, 1962); Ivan Sutherland saw the utility of enlarged images in fields such as physics to more easily explain atomic concepts (Sutherland, 1965); and the activity of Caudell and Mizel in augmented reality from Boeing was aimed at teaching the workers to perform normal operations in the assembly of airplanes (Caudell and Mizell, 1992).

AR in education has become much easier since smartphones and tablets became widely available, and the combination with Internet access has opened up new possibilities for teachers and students today. In education, augmented reality was initially used in industry, medicine, auto service, architecture, urban and environmental education (Forsyth, 2011). The focus is on providing helpful information, which is missing in the real world, but which helps the student to better understand the concepts they are learning. A classic example is related to the learning of anatomy, which requires a certain type of laboratory with certain minimal equipment (most often very expensive), which can be replaced by an AR application, much cheaper and more accessible anywhere and anytime (Blum et al., 2012).

With the help of AR, the learning is removed from the traditional spaces (classrooms, laboratories, seminar rooms) and can be carried out where the student is (home, in the park, in the bus, etc.) Already in the current applications students are no longer passive. In the learning process, they become important contributors to the learning materials courses, projects, themes (Billinghurst and Dünser, 2012).

AR could allow (Wu et al., 2013):

1. learning content in *3D perspectives*,
2. *ubiquitous, collaborative and contextual learning*,
3. learners' senses of *presence, immediate character, and immersion*,
4. *visualization* of the invisible,
5. *formal and informal learning*.

AR can be used in addition to the elements of traditional pedagogy and can come with more advanced methods, such as SOLO (Structure of observed learning outcome) taxonomy (Biggs and Collis, 1982). More studies have been done to see how AR and VR students can benefit from the learning process. (McLellan, 1994) and (Gardner, 1985) are of the opinion that AR addresses all types of intelligence (musical, visual, verbal, logical, bodily, etc.). Some researchers claim that these intelligences do not exist independently (Pashler et al., 2008) and learning should not be limited to just one or several, but to all.

AR offers several special educational opportunities:

1. *Mobility* (can be accessed anywhere).
2. *Visualization* (which can be manipulated by the viewer).
3. Students *can build and generate content* (and can contribute in an active way to their education).
4. *Alternative perspectives* (open new way to see the world and to improve their knowledge).
5. *Comparison* and differentiation between several perspectives (can create interesting debates between participants).
6. *Integration* of multiple perspectives (solicit participants, developing their synthesis capabilities).

**AR Applications**

If in the first part of this section, we have presented a history of the use of AR in education, in the second part we will see some applications developed with the help of students from the Faculty of Computer Science, such as bachelor theses or dissertation theses.

*Development of Communication and Teamwork Skills*

Currently, the school and employers are starting to focus on developing teamwork skills for students or employees. This is because we always need to work on large projects where communication and collaboration are key elements, on which the chances of success of the project may depend. For this, it is necessary for each of us to get to know our colleagues better, to improve their communication and teamwork skills. In the case of autistic children, the communication can be indirect, through games or with the help of images, and the improvement of this type of communication can lead to the improvement of direct communication. With the aim of creating applications that enrich the communication we have created two applications that work in the network, (1) the *first* one for autistic children, where the indirect communication during the game where they have tasks that they can solve together bring them additional bonuses and facilities, and (2) the *second* aims to synchronize the actions of the players through direct verbal and visual communication.

**Game 1 - Supporting Ships in the Air[98]**

In this application, we have several ships that fly to perform their various missions and which must be fed while flying (Iftene and Trandabăţ, 2018). When the missions of the ships are completed, they move on to the next level, where the difficulty of the game increases (either there are more ships, or there is a shorter time in which they must be fed, etc.). During the game, when they synchronize their activities they receive bonuses, and moving to the next level is easier. At the final levels, it is obligatory to work in the team otherwise the levels cannot be finalized and thus cannot reach the end of the game. In Figure 31 below we have a sequence during the game.

---

[98] https://www.youtube.com/watch?v=DrdcZBIg0WE&list=PLddW60TN_y-WsMfQoEcz-O-6QdZJ8wZ3y&index=3&t=0s

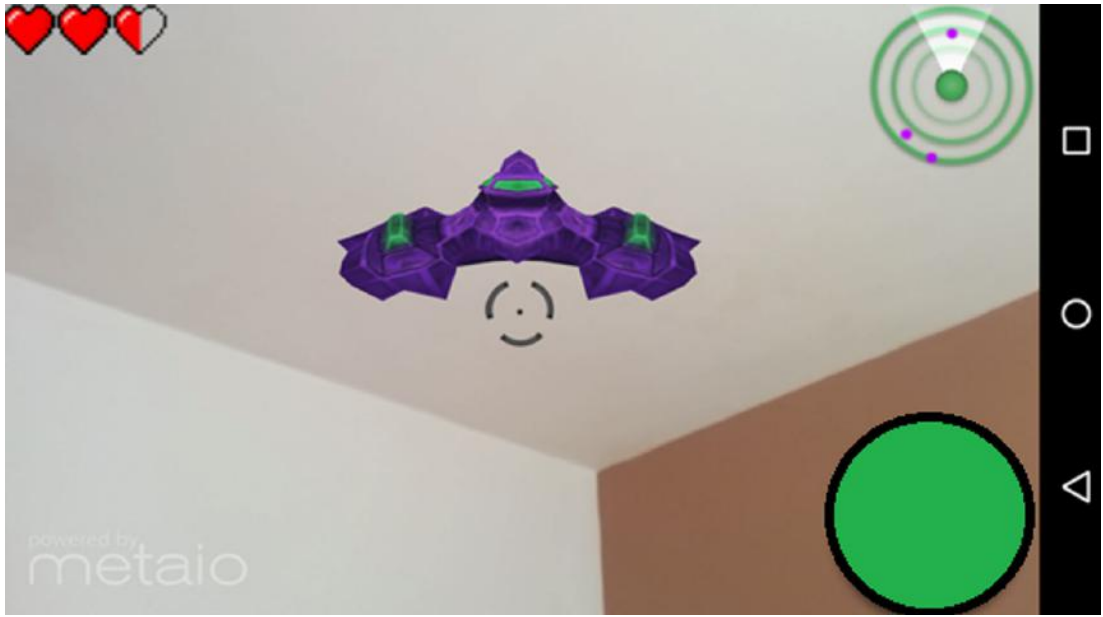

**Figure 31**: An image from the application with one of the 3 ships

The application is of client-server type, the smartphones of the players representing the client, these communicating with the server that has the situation of all the ships of the current level. The application on the *client's* phone was implemented with the help of Metaio[99] and works on Android devices. The *server* was implemented in Java and manages the data needed to run the game in good conditions. Thus for the current level, it has the position of ships, ensures communication between players, and between players and server, updates scores at the player level and global statistics level, etc.

In order to find out the opinion of the users we have performed *usability tests*, and following the feedback received we have constantly improved the application from one version to another. Many functionalities of the application have been added following discussions with those involved in these tests (the conditions for switching from one level to another, increasing the difficulty level to require communication and collaboration, global statistics and scores, etc.).

To better understand how to build the application and what its effects are on children with autism, we called on an expert who guided us during the construction of the application and who in the end evaluated its effects on these children. His observations were positive, observing during the use of the application that the communication through images improved after each mission accomplished. This encourages us to continue with the development of the project in this direction, hoping for positive effects in the future.

---

[99] https://en.wikipedia.org/wiki/Metaio

**Game 2 - SMAUG - Sphero Multiplayer Augmented Game[100]**

The application uses the Sphero 2.0[101] device that can be controlled by several players at the same time (Pînzariu and Iftene, 2016). Players coordinate from the tablet or mobile phones the hardware device with the control of the direction and speed. To achieve the objectives it is necessary for the players to communicate and coordinate in the actions they undertake (collecting diamonds, solving puzzles, completing augmented routes, etc.). We can see in Figure 32 how the players collaborate in a game session.

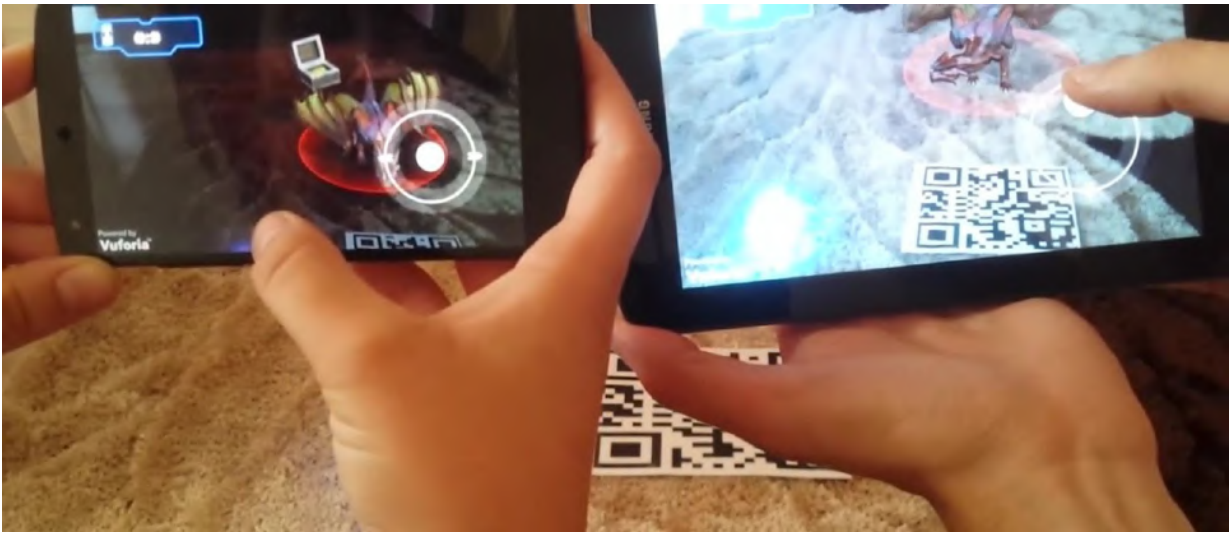

**Figure 32**: SMAUG - the multiplayer option (Pînzariu and Iftene, 2016)

The *client* on the mobile device controls Sphero (via Bluetooth) and communicates with a *server* (via TCP) that has information common to all players. The *server* can be started on any mobile device that calls this option, following which the other players connect to it to be able to use the multiplayer option. The AR component was made using the Vuforia[102] platform.

The usability tests we conducted identified the following: (1) the *negative elements*: the single-player component was observations related to the ambiguity of the interface, and the multi-player component were reported problems of performance and stability, (2) and the *positive elements*: the Sphero device can be controlled very easily, and the interaction during the game is very attractive.

---

[100] https://www.youtube.com/watch?v=gC13u2zfWgE&list=PLddW60TN_y-WsMfQoEcz-O-6QdZJ8wZ3y&index=4&t=1s
[101] https://www.sphero.com/products/sphero
[102] https://developer.vuforia.com/

*AR Applications for Primary and Secondary School*

Applications for primary classes help students more easily memorize information from biology, geography and chemistry lessons.

**ARBio - Using Augmented Reality in Biology**[103]

The ARBio application was built for primary school children but can be used by anyone who wants to learn biology (Iftene and Trandabăţ, 2018). The augmented reality component recognizes markers representing 2D animals and displays their 3D models. Besides the 3D model, the user has access to audio files with animal sounds, but also suggestive video files and useful textual information from Wikipedia. In Figure 33 we can see two suggestive images from this application.

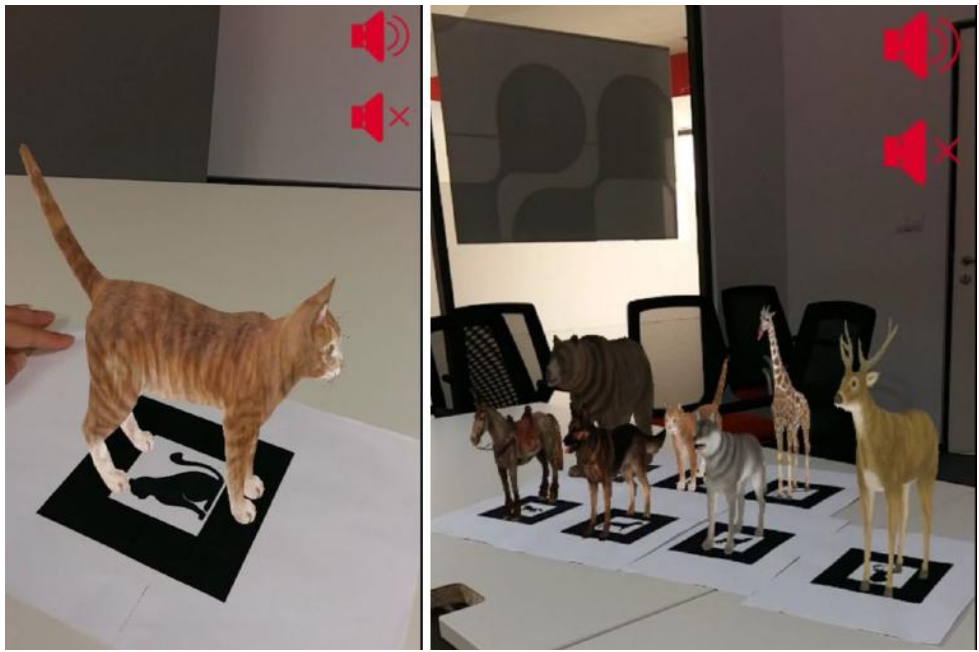

**Figure 33**: ARBio - visualization of a 3D model associated to a 2D model (left) comparison between 3D models of animals recognized by the application (right)

---
[103] https://www.youtube.com/watch?v=ldO33rPhKXc&list=PLddW60TN_y-WsMfQoEcz-O-6QdZJ8wZ3y&index=1

The application has two main modules (1) the module based on augmented reality and (2) the module that allows the visualization of information from Wikipedia. The augmented reality module was made using Artoolkit, which communicates with components made in C ++ (which integrates the AR module in the application on the phone) and Java (for the interaction with 3D models, with 2D markers and with sound files). 3D models were created with the help of Autodesk Maya[104] utility.

Following the usability tests we have reached the following conclusions: (1) the *positive aspects*: the association between the 2D marker, the 3D model, the associated sounds and the details on Wikipedia is very intuitive and helps students to learn more about this subject, (2) the *negative things* students reported several times that the application crashed when multiple markers were used and multiple audio files were started simultaneously.

**GeoAR - Learn Geography with Augmented Reality[105]**

The application helps students from secondary school to learn the geography of Europe (countries, capitals, flags and neighbors) (Chitaniuc and Iftene, 2018). Using augmented reality, the application has a learning component where information is displayed using map markers of countries (see Figure 34). After the new information is searched with the help of the AR component, the application allows evaluating the accumulated knowledge using games, which have different degrees of difficulty from one level to another, creating a real competition between students. Thus they get to learn without realizing it, increasing their desire to accumulate extra points based on the notions learned in addition. For this, the application allows redirects to Wikipedia which contains more detailed information about countries (flag, population, surface, etc.).

---

[104] https://www.autodesk.com/products/maya/overview
[105] https://www.youtube.com/watch?v=Dp58Gf6zyXg&list=PLddW60TN_y-WsMfQoEcz-O-6QdZJ8wZ3y&index=4

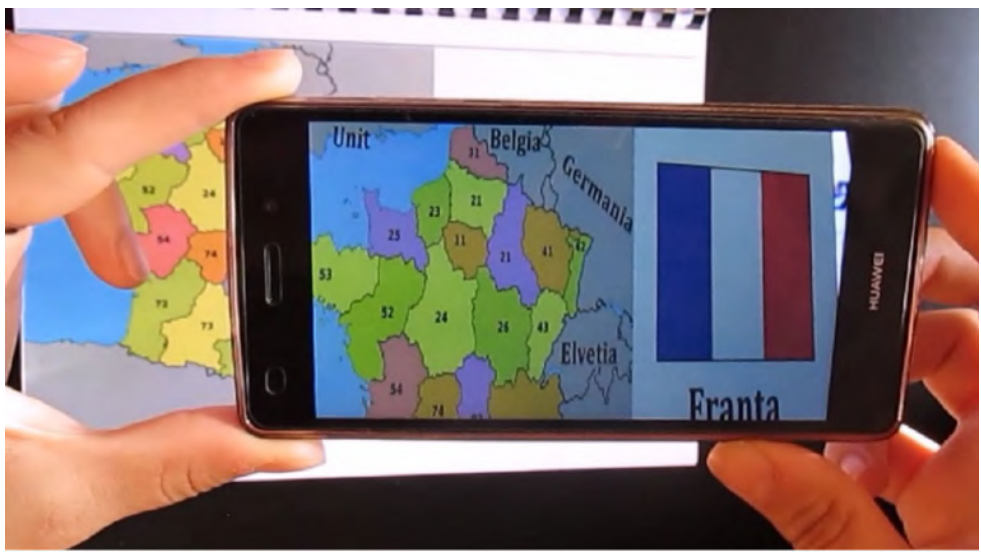

**Figure 34**: GeoAR - Augmented reality component (Chițaniuc and Iftene, 2018)

*The Learn Europe module* - contains the augmented reality component and was developed with the help of Unity[106] and Vuforia. *Test your knowledge module* - it was thought of as a game, the questions and answers being read from a configuration file. *Learn More Module - A*llows the user to redirect to Wikipedia when the user requests it.

**AR Chemistry Learn**

The AR Chemistry Learn app was built for educational purposes and aims to help those who want to learn chemistry (Macariu and Iftene, 2018). Since the age of 14, children are beginning to be familiar with this subject, and some concepts seem difficult to understand. Thus, the application is based on multicolored and explanatory visual elements that help them and they aim to improve their learning process. The AR Chemistry Learn app can be downloaded to the phone or tablet and is easy and intuitive to use.

To create the application, the Unity platform was used, with the basics of the game and the Vuforia kit for text and image recognition, which easily integrates with Unity. With the Adobe Illustrator[107], the cardboards were made (see Figure 35 for the Hydrogen cardboard). To have an educational purpose, the cardboard contains the full name of the substance, the chemical formula and Mendeleev's periodic table, the specific element being colored.

---

[106] https://unity.com/
[107] https://www.adobe.com/ro/products/illustrator.html

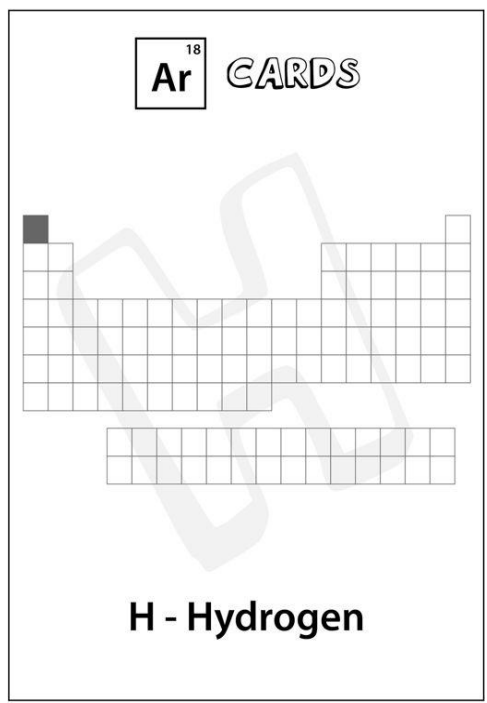

**Figure 35**: Image Target - cardboard for Hydrogen (Macariu and Iftene, 2018)

The main modules of the application are *Learn with the manual*, *Learn with the cardboards*, *Test your knowledge*, and *Add a substance*.

*Learn with the manual* - In addition to augmented reality, this module is based on text recognition. Thus, for certain words that represent names of substances, the desired information is displayed. The name of the module is suggestive, as its purpose is to help high school students understand and learn faster through an interactive way. Because there may be many unfamiliar and difficult words in a textbook, the application proposes another way of learning. It can be read from anywhere, manual, a simple sheet or directly from the phone. For example, if the student wants to find out about *methane*, he needs to position the phone's camera on the word and the compound will appear above it in 3D (see Figure 36). If the word appears several times in the frame, the 3D form will be displayed for each compound. The second step in this module is to press the 3D substance. A window will pop up where information is dynamically brought from www.Wikipedia.com, if there is an Internet connection (see Figure 36).

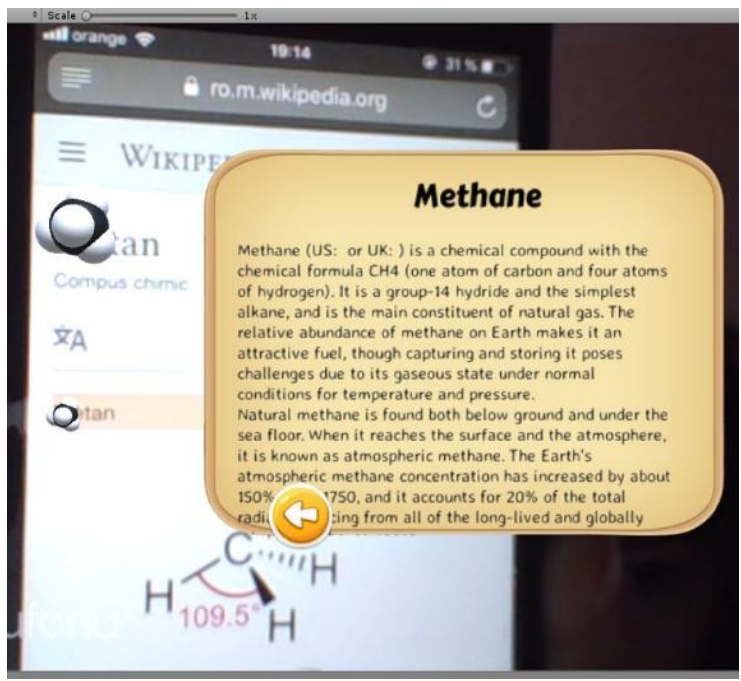

**Figure 36**: Additional information when pressing the 3D substance (Macariu and Iftene, 2018)

*Learn with cardboards* - Once the image/card has been recognized, the specific elements appear in Game View. Each substance corresponds to a molecule in the form of a sphere with its specific weight and color and a plane in which the specific data of the molecule are noted: the *atomic number*, the *name of the substance*, the *atomic weight*, and the *chemical symbol*. When two substances are close enough, a force of attraction is executed on the geometric bodies, and the cardboard changes its writing with the specific composition (see Figure 37).

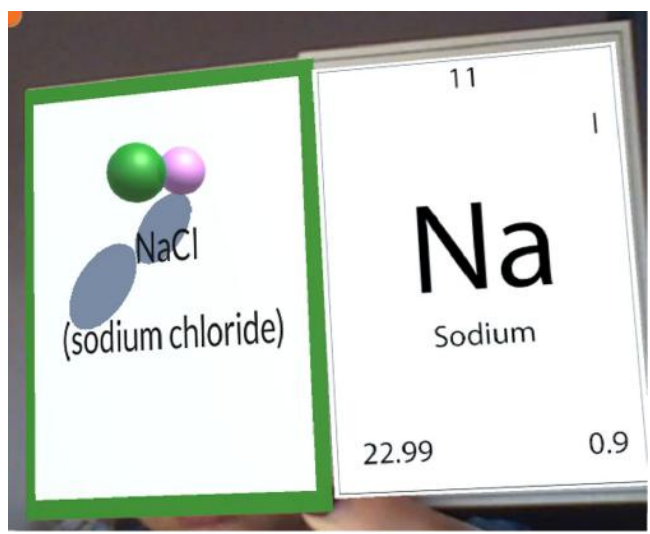

**Figure 37**: Combination of two substances (Macariu and Iftene, 2018)

*Test your knowledge* - This module, from the Test menu, wants to test the knowledge of building the compounds learned in the previous module. At the top it is specified what needs to be built, without giving further guidance on the cards to be brought, and next to it is the "Next" button. In the lower-left part, the score record is kept, how many are right or wrong.

*Adds a substance* - This module allows the user to enter a compound created by him into the application. When you open the "Add a substance" menu, you can see at the bottom a text that guides the user to use the application correctly. Once the elements you want to combine have been brought into the frame, press the "Ok" button on the bottom right. The details on the card on the left are missing because the name of the substance has not yet been given, to be written after the "Ok" button is pressed.

**Conclusions**

In the next period, we will witness a widespread use of AR in education. The purpose of AR in education will be to help the classical learning methods and not to replace them. Sounds, pictures, movies come to help the student better understand the new concepts they are learning.

As we have seen in the applications presented, the fields of *biology*, *geography*, *chemistry*, *collaborative work* are very well suited to applications based on AR. For the evaluation part, attractive games can be thought of, which eliminates the stress of oral or written exams. In addition to the applications presented in this section, we have also created applications to help students learn computer science (Păduraru and Iftene, 2017) or learn general culture elements (Pantea et al., 2019).

## III.1.3 AR in Gastronomy

**Introduction**

Since the 1970s, the amount of food and beverage consumed around the globe at daily meals increases yearly, as shown by the studies carried out in the USA (Duffey et al., 2011), (Piernas and Popkin, 2011), in Australia (Collins et al., 2014) and in Ireland (O'Brien et al., 2015). Among the factors identified as contributing to the consumption of larger portions of foods, we can include: (a) the perception of "*money value*", (b) *increasing sizes of pre-packaged products* such as food, beverages, pots and cutlery, (c) *continued exposure to larger portions* due to the food environment we live in and (d) *lack of awareness or understanding* of the recommended size for serving (Livingstone and Pourshahidi, 2014), (O'Brien et al., 2015) and (Steenhuis and Vermeer, 2009).

Consuming larger sizes of food portions is associated with an increase in the level of received energy, which can "*override the regulation of energy balance and can have persistent effects that could lead to obesity*," according to (Rolls, 2014). Although there is no clear link between large portions and obesity, as discussed in (Livingstone and Pourshahidi, 2014), a recent meta-analysis of 58 trials has demonstrated a small to moderate effect between the association of larger food portions and packs and increased energy intake (Hollands et al., 2015).

In this context, new technologies have started to be used more often to help people eat healthier or conform to a specific diet. The server application (Rollo et al., 2017) was built to test the augmented reality (AR) effect on guiding food serving. The experiments performed by the authors have revealed that users tend to overestimate the amount of food needed for a sufficient portion. When using an AR application which pre-visualized the selected food on a plate, the user has improved the accuracy and consistency of estimating food sizes, which demonstrates that AR can be used to support users in choosing more adequate food portions.

Another example is the application presented in (Bayua et al., 2013), where the authors provide information about the number of calories in a dish through an augmented reality application which scans the food using a phone's camera. The aim of this application is to help users regulating their diet plan, especially diabetes patients who need to have strict control of the number of calories in their blood. Their method involves scanning food, displaying associated objects in a 3D format, and viewing calories in graphical form. The visual graph is provided as a gauge, and the information changes according to the scanned 3D objects to help the user visualize the number of calories. Authors' tests have shown that the nutrition information generated with their application provides useful information on carbohydrates, proteins, and fats, information which is very important for users who follow a diet plan.

With this kind of applications, augmented reality starts to prove its importance as a helper in supporting users with dietary restrictions by instantly visualizing calories and food sizes to adjust the amount to what they really need.

**AREasyCooking**

Inspired by the recent interest in augmented reality for the nutrition area, we advance the challenge and propose a complete mobile application based on several new technologies, aiming to help users prepare home-cooked food (Chițaniuc et al., 2018).

There were two main questions we wanted to answer during the development of this application: (1) *Can we quickly introduce available ingredients in an application and then rapidly find a recipe that can be made with these ingredients*? (2) *While watching a movie with step by step instructions for preparing a recipe, can we find variants that allow us to control the movie without using the hands, which are busy preparing the recipe*?

The architecture of our application, combining augmented reality with eye and voice controls is described in Figure 38. Starting with a list of available ingredients, either introduced by the user or automatically identified by the application, a search is performed over a database of recipes to only suggest recipes including the available ingredients. Once a recipe is selected, the user can use eye and voice control to follow the indications while preparing the dish.

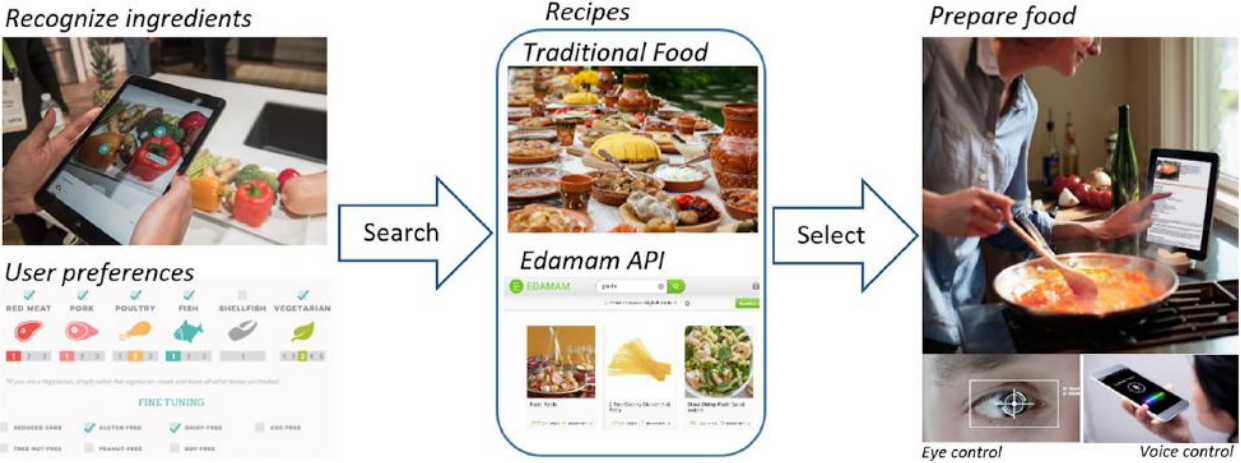

**Figure 38:** Architecture of the AREasyCooking System

*Configuration Option*

Before running AREasyCooking, a configuration step is needed. Thus, in the Configurations option of the application, the user can specify four lists of ingredients: (1) *preferences*, (2) *allergies* and *diet*, (3) *universal* and (4) *preferred source*. The fist list offers checkboxes where the user can select/deselect ingredients to only keep his preferences (*red meat, pork, poultry, fish, shellfish,* and *vegetarian*). The next list, allergies and diet, allow the user to manually add any ingredients that should be avoided in the returned recipes. In the third list, the universal one, there are again boxes, checked by default, for a set of common ingredients, found in most kitchens. This list contains soft ingredients like *salt, sugar, pepper, flour, oil, spices, mustard, mayonnaise, ketchup, water,* etc. Additionally, the user can add new elements like *eggs, milk, rice, potatoes, onions, garlic*, etc. or deselect specific soft ingredients. The necessity of this list came from the observation that, in most recipes, soft ingredients account for around 50% of all ingredients. The last list contains the possible sources for the recipes, which are Martha Steward, Jamie Oliver, traditional food, or other. The user can change the weights for the source of the recipes, or leave the predefined order. This ordering helps ranking the returned recipes, in case there are more than 20 results.

*Recognize Ingredients*

Every recipe starts with ingredients. Our application allows users to select among a variety of recipes using various filters based on preferred and available ingredients. The configuration option considered the preferences of the user and added a predefined list of soft ingredients. For the rest of ingredients, manually adding into an application the available ingredients can be uncomfortable for someone who wants to start cooking fast. Therefore, besides the possibility of manually introducing ingredients, our application also uses two other novel methods: augmented reality and a bar code reader. Thus, the identification of available ingredients and the time needed for their introduction into the application is significantly shortened.

*Augmented Reality*

While searching for tools that can help with automatic ingredient recognition, we found Clarifai API[108]. The food model recognizes more than 1,000 food items down to the ingredient level. This feature was adapted in our application. The use case is pretty simple and starts with the user taking a photo of the ingredients in the refrigerator. The image is further sent and analyzed by the Clarifai API. A report is generated, containing the identified ingredients. Each ingredient has a certain percentage and the user can validate the existence of the specific ingredient in the fridge, along with the desire to use the ingredient in the dish to be prepared. These details are shown to the user on the camera preview, as detailed in Figure 39.

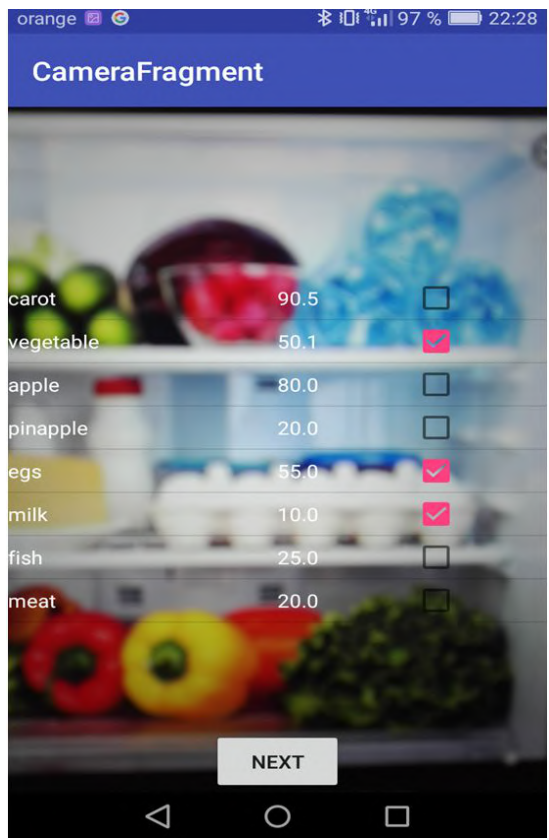

**Figure 39:** Augmented reality component used for the recognition of ingredients (Chiţaniuc et al., 2018)

Augmented reality can be used to present the ingredients in an interactive form. However, we have chosen to only show them in a static way and not pinned to a specific spot, so when the camera is moved the ingredients still remain on the telephone display even though the camera is not directed towards them anymore.

---

[108] https://www.clarifai.com/developer

The API predicts ingredients from the image with a good accuracy, yet some items cannot be correctly recognized. Another downside is the fact that the API does not recognize the quantity of the ingredients. Therefore, we have investigated also a second manner of facilitating the introduction of the ingredients.

*Barcode Reader*

Besides vegetables and fruits, many ingredients can be wrapped in cardboard or metal boxes that are labeled with barcodes. Therefore, we decided to use a barcode reader connected to a Raspberry Pi. This method of introducing ingredients has a definite advantage over the augmented reality method, having extremely high accuracy. Additionally, it also provides quantity-related information. However, the disadvantage of this method is that the hardware is not completely embedded into the phone, and extra space for the Raspberry Pi and the bar code reader is needed.

*Collection of Recipes*

After the ingredients are set, recipes are selected from a database to contain the available ingredients. We consider two sources of recipes for our application: one was obtained using the Edamam Recipe API[109] and one was built by with specific traditional East-European food.

The Edamam Recipe API makes an efficient search throughout over 1.5 million of English recipes stored in a semantically organized database and can be filtered by a list of ingredients, by calories, diet or allergy preferences. The REST API call returns a list of recipes, identified by a title, a summary, the total number of calories and a picture. The parameters for the search are the ingredients selected by the user. Subsequently, the results are filtered using dietary preferences from the user profile. The results of the search include the list of ingredients, with corresponding quantities and the instructions to prepare the chosen recipe.

---

[109] https://developer.edamam.com/

The recipes are usually extracted from popular culinary websites, including one of the famous cooks such as Martha Stewart[110] or Jamie Oliver[111]. Beside the information offered by the API, we added to our application the possibility to search for a YouTube movie relevant for the preparation of the specific food.

*Filtering and Sorting Food Recipes*

The AREasyCooking application allows the user to define a list of filters in his profile. These filers will be used while searching in the database of recipes for the ones that best fit the user's needs. Thus, the user can specify allergic substances which should not appear in the ingredients of the returned recipes. Additionally, if the user is on a specific diet and wants to eliminate from his food certain ingredients, this aspect can be also specified.

Using these filters, our mobile application sorts the recipes based on the following criteria:

- *The percentage of available ingredients must be greater than a specified value* (the default value is 80 %, but it can be changed). The results will be displayed in descending order of the percentage of ingredients available for a recipe.
- When the results are similar, the *cooking time and the number of calories* can be used as a delimiter to rank the retrieved recipes.
- When the number of results is greater than 20, an additional sorting criterion is used: the user can change the weights for *the source of the recipes*, thus rearranging the results. The default order if Martha Stewart, Jamie Oliver, Traditional Food and Other.

Following suggestions from the testers of our application, we added the "Missing Ingredients" option to each recipe in the list of returned recipes. This way, we highlight the fact that the recipe required ingredients which were not marked as available by the user. These are usually just a few (1 to 4), but sometimes it is highly relevant, especially if they are important ingredients. We considered important ingredients those that appear in the name of the recipe. When it is missing from the available ingredients, a "Missing Ingredients" tag will be displayed in red.

---

[110] https://www.marthastewart.com/
[111] https://www.jamieoliver.com/

*Preparing Food*

Once the user has selected a recipe, the application displays the steps needed to prepare the food in a textual format, with a movie associated with it. If the user chooses to play the movie, the AREasyCooking application allows him to control the movie with his voice or eyes (see Figure 40). The introduction of voice and eye controllers was dictated by the need to control an instruction movie while preparing a recipe by a person who has his hands occupied.

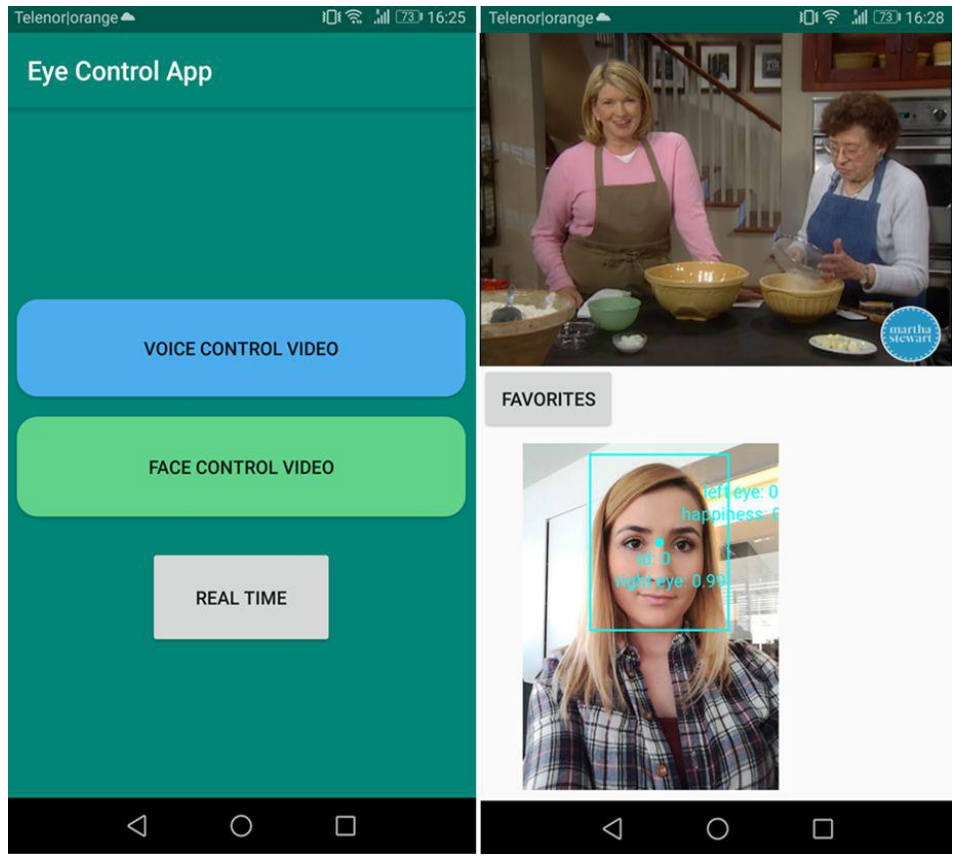

**Figure 40:** Options to play the movie with voice or with face control (left) and Controlling the movie with eyes control (right)

*Control Movie with Voice Control*

In the last years, big companies like Google and Android launched voice control applications to help people with limited mobility navigate their phones (Carman, 2018), (Hesse, 2018), (Vertelney et al., 1990) or even to control applications with the help of Siri, Google Assistant, or Cortana (Hartt, 2018) and (Matthews, 2018). On another side, researchers investigated controlling critical mobile device operations in real time (Omyonga and Shibwabo, 2015) and even the entire Android operating system (Zhong et al., 2014).

In the case of our application, we use the Google Assistant SDK[112]. When installing the application, the volunteer can train the voice recognition module by pronouncing several required keywords, but this step is not mandatory. However, we noticed that pre-training yields better results. We have implemented the recognition of the following commands:

- "*Play*" - starts to play a movie;
- "*Stop*" - stops for the moment the playing;
- "*Right*" - skips a number of seconds from the movie (between 5 and 10 seconds, a number set by the user, default to 10);
- "*Left*" - goes back a number of seconds.

*Control Movie with Eye Control*

Similar to voice control, the eyes control applications are constantly spreading lately, especially for people with disabilities. In (Suraj et al., 2011), the authors use a camera to detect the eye movements and then transmit control signals over a wireless channel. In (Singh et al., 2014), authors use motion based eye gesture for automatic controlling of video frames for various applications in their computer. One interesting aspect related to eye control is presented in (Hameed and Ahmed, 2018), where the authors discuss the positive and negative aspects resulting from the use of eye tracking in security systems. Windows 10 lets us use eye-tracking technology to control the mouse pointer, type using an on-screen keyboard, and communicate with other people using text-to-speech[113].

For the AREasyCooking application, we recognize a set of eye movements using the Mobile Vision API[114] from Google and map them to the four commands used to control the movie. Thus, both eyes open is equivalent to "*Play*", so the movie will be played. On the contrary, both eyes closed, or a situation when both eyes are not detected by the camera, for more than a number of seconds (number set by the user) is equivalent to "*Stop*" and stops playing. If the user closed the right eye is equivalent to "*Right*" and skips a number of seconds from the movie. Similarly, if the left eye is closed equals to "*Left*" and goes back a number of seconds.

---

[112] https://developers.google.com/assistant/sdk/
[113] https://support.microsoft.com/en-ca/help/4043921/windows-10-get-started-eye-control
[114] https://developers.google.com/vision/

**Conclusions**

This section presents the AREasyCooking application, designed to help users (1) *quickly select a recipe fitted for the available ingredients* and (2) *guide them in preparing it*. We used augmented reality in order to recognize ingredients available in the fridge and implemented support for a bar code reader in order to facilitate the introduction of packed and labeled ingredients. Once ingredients are identified, the application selects recipes from our database, using additional filters for allergenic or dietary restrictions. After selecting a recipe, AREasyCooking displays the basic steps for the preparation of the dish, along with the video instructions. Using new technologies, the user can control this movie through voice commands or with the help of the eyes, in the context of having the hands busy preparing the food.

## III.1.4 Smart Museums

**Introduction**

Nowadays, the evolution of technology is crucial to humans for several reasons, most of them will require time and training. Augmented Reality (AR) occupies a top place in the list of the most used technologies as much in the gaming industry as well as for applications that are based on the educational software concept applied in the cultural heritage. Smart Museums project aims to attract people of all ages to visit museums more often. The proposed application uses the capabilities of AR technologies to transform a visit to a museum in an attractive experience full of memorable memories. Basically, we can have access to textual or audio information about an artwork or its creator, or we can have access to image galleries with works by the same author.

The development of Smart Museums application started in 2017 at Faculty of Computer Science of Iasi, as a group project, meant to encourage teamwork between students and also make them discover and use new technologies (Porfireanu et al., 2019). Now we have reached the point where our project has a practical dimension and can be easily integrated in the experience of visiting a local museum.

*How*? Simply view the exhibit you are interested in through your phone or your tablet screen and instantly multiple pieces of information will pop up on the screen. The application user will be able to choose between a text description, audio files, browsing through an image gallery of related exhibits, thus finding out even more details of the author's work. In a society where technology has been gaining ground in more and more areas, we believe our application is a great way to ensure a friendly and smart experience, regardless of the visitor's age or the contact he has had with similar applications until then.

**System Architecture**

The architecture of the system is based on the Client-Server model (see Figure 41).

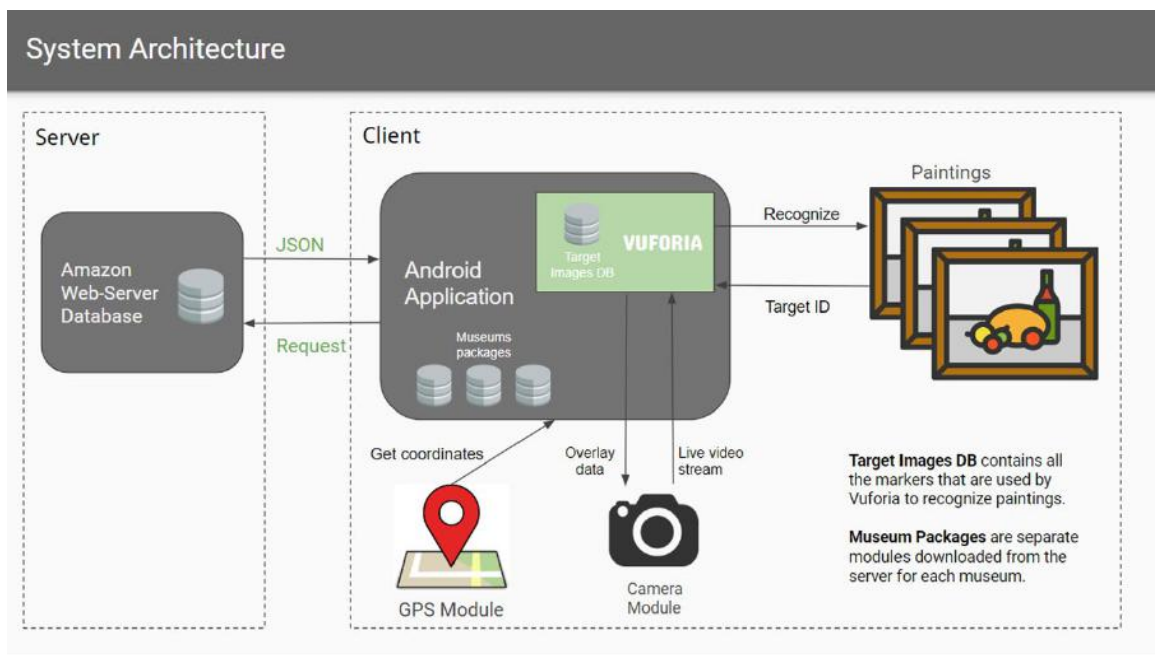

**Figure 41**: Smart Museums - System Architecture (Porfireanu et al., 2019)

In the process of creating the application, we used Vuforia, Unity and a client-server protocol to achieve three of the main functionalities.

*Client-Server Model*

How we can see in Figure 41, the **Server** component (written in C#) provides information to the Client according to the following types of requests: *login* (only the admin of the museum and the user of the application), *get-museum* (allow to the user of application to get only one museum at a time and to download it on his device), *insert-museum* (the administrator can insert new museums in the application), *delete-museum* (also at the administrator level, allow deletion of museums from database) and *modify the database with exhibits* (allow administrator to change details for one exhibit).

The **Client** component was written also in C# and it was integrated into the Android application, in order to communicate with the server and to get information. User interaction is through the graphical interface. Also, at Client side we have a Java application, for the administrator of museum who can manage the information about the museum (*insert*, *update*, *delete*). The client is a compatible Android application on all mobile devices that have a camera and Android support. When we start the application, the first screen of the Android application offers the main options: *Visit museum* or *Download museums*.

In order to use the option *Visit Museum*, the user must:

- *Download the package for the museum* on the mobile device. The package can be downloaded in advance at home or in other places or can be downloaded in the museum. Of course, we recommend doing this in advance, because in some cases it is possible to last a bit depending on the speed of the Internet or the availability of the data transfer.
- *Have the location enabled* so that the application allows it to log in and use this option.
- *Be within the museum* to allow the application to use the information about this museum.

Once it enters the application, the camera can recognize the targets using an API from Vuforia. Targets are artworks, sculpture paintings or exhibits from the museum. When a target has been recognized, the application shows a number of buttons for doing various activities.

At the top of the application, we have the 3 buttons to help the user to do different things (see Figure 42 left):

- *Audio* if the user wants to hear the text from above option instead of reading it;
- *Text* if the user wishes to read details about the opera or about the artist;
- *Gallery* if the user wants to see pictures of similar exhibits with the recognized one.

At the center of the screen, the user sees the name of the exhibit that was scanned with the phone (see Figure 42 in the middle). In the bottom of the application are two buttons. The left button opens a Quiz. This questions the user about the exhibit and tests his knowledge. The button on the right opens a panel of other works of art similar to the one scanned by the user (see Figure 42 on the right).

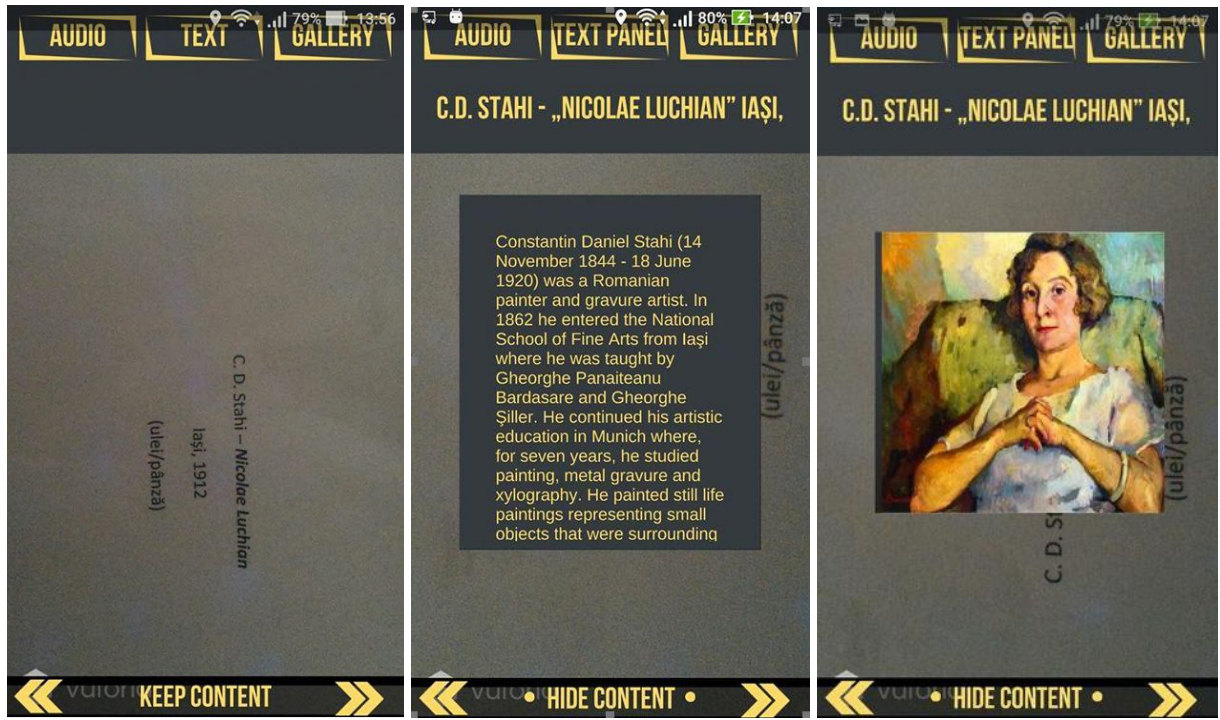

**Figure 42**: The screen based on AR component (left), Text details about the artist Daniel Stahi (middle), Using the Gallery option (right) (Porfireanu et al., 2019)

**Conclusions**

This section presents our work in Smart Museums project, which has the aim to attract more people (children and adults) in museums from Iasi. Augmented Reality can help the visitor to better understand an artwork or to obtain easier information about creator or to see similar artworks. Current and future work is focused on creation of games to verify the attention of visitors or to involve them in collaborative puzzling activities.

## III.1.5 Conclusions

Augmented reality has begun to be more and more in a variety of fields. This comes with new facilities that allow us to see another facet of the world we live in, the facet augmented with suggestive images, animations, movies, sounds, etc. They help us to make learning activities more interactive, more dynamic, more attractive for new generations of children, who respond positively to using such applications. It is interesting that the teachers also see in these tools help in what they do and appreciate the effort made to ease their work.

Starting with the first application created in 2010 (Arusoaie et al., 2010), the activity with the students has led to the creation of applications for fields such as eLearning, games, interior design, gastronomy, intelligent museums, etc. for their bachelor or master thesis. Many of these led to the publication of articles, currently having 11 articles published with students, which use augmented reality[115].

## III.1.6 Future Work

Future work will continue the activities we started in the eLearning area, and we want to create applications for other subjects as well. We also want to apply to raise funds that will allow us to equip the classes with tablets and to make products that we will make available to teachers and students. In the gastronomy area, we intend to continue the project started, to test it on a larger group of users and to make it available for Android phones and tablets to be used by the people who cook at home. In the SmartMuseums area we want to finalize the project to make it available to visitors and thus attract more children but also adults to come to visit museums.

Besides the existing directions, we also consider two new fields: the medical field (and here we are in discussions with the UMF[116] professors and colleagues in the ImagoMol cluster) and the field of botanical gardens (here we are in discussions with those at the Botanical Garden in Iasi).

---

[115] https://profs.info.uaic.ro/~adiftene/publications.html
[116] "Grigore T. Popa" University of Medicine and Pharmacy - Iasi

# III.2 Virtual Reality

**Introduction**

Virtual reality is a cutting-edge technology that is becoming more and more popular. Through virtual reality we can create a reality that does not exist, it is separated from the physical one. The virtual reality can be seen as an imaginative reality, the virtual term assuming the absence of visible, tangible limits, everything is related to our imagination.

In technical terms, virtual reality describes a three-dimensional environment generated by a computer, an environment in which we can transpose with the help of applications. Once transposed into this virtual world we can interact with objects, with virtual actors, but also with other people transposed as we do in this virtual world to perform certain tasks together.

Virtual reality can be applied in various fields, such as: architecture (modeling, simulation and visualization), sports and medicine (experiments and simulations), simulators (pilots, astronauts, drivers), art, entertainment (games and movies) and not lastly education (Vlada and Popovici, 2004), (Cruz-Neira and Fernández, 2018) and (Salah et al., 2019). We can reach new discoveries in areas that have a direct impact on our daily lives (Bălan et al., 2013). This technology is increasingly common and we can expect more and more innovations in this sphere in the coming period (Hagl and Duane, 2018).

In recent years, smartphones, tablets, and other mobile devices have been equipped with sensors that allow us to simulate the virtual world. The most common sensors are the magnetometer, accelerometer, and gyroscope. They are very important for the generation of virtual reality because they allow us to simulate the movement in space that can be transferred into the virtual world.

**Areas of Applicability**

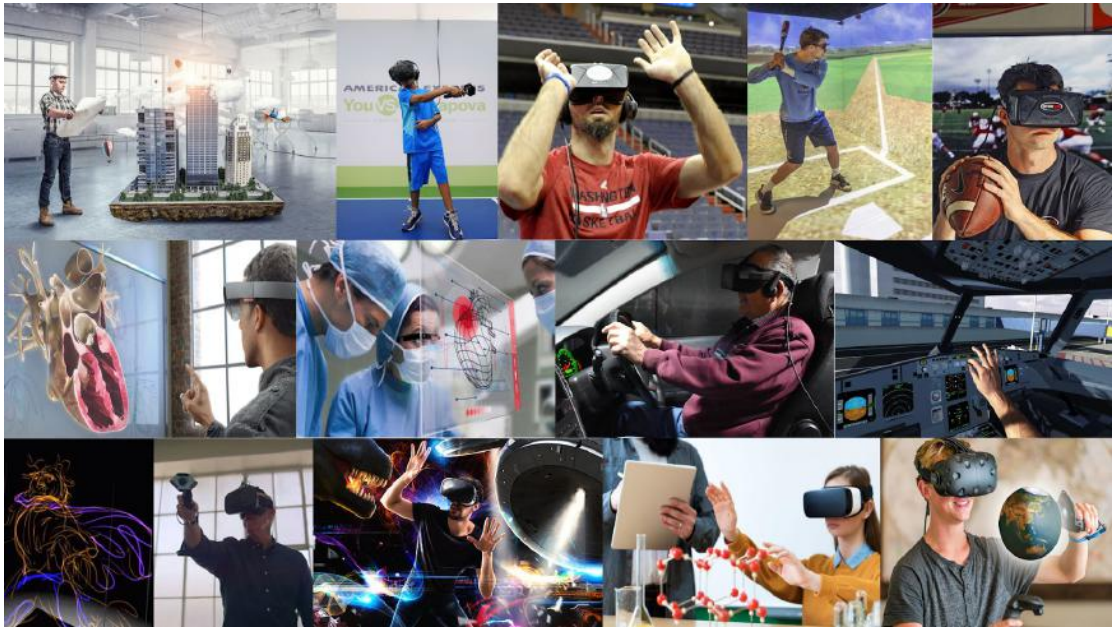

**Figure 43**: The most important domains in which VR applications were created

- **Architecture** - modern applications that use virtual reality allow rapid modeling and visualization in future 3D images of future constructions. It saves time, and 3-D models can be easily sent for analysis and approval to customers or colleagues anywhere in the world.
- **Sport** - you can participate in tennis, basketball, American football, soccer, hockey, golf, squash games at home or in a specially arranged room, where you can interact with your friends from a distance. Use special tools equipped with sensors or special sports articles can be used, which make the experience as close to reality as possible.
- **Medicine** - learning, experimenting, operations become more intuitive and can be performed anywhere (in the classroom, in the laboratory, but also at home or elsewhere) with the help of virtual reality. Zoom operations allow them to see details that would not be easy to see with the naked eye, and the internet allows experienced doctors to passively or even actively participate in remote operations. One aspect that is not to be neglected is the costs that decrease very much, because 3-D models can be analyzed and studied by students, teachers, without expiring and without being damaged by the passage of time.

- **Virtual Simulators** - help those who want to learn to drive a car, plane, bike or motorcycle, creating an experience very close to reality. On the one hand, the necessary costs are reduced, but on the other, their safety and protection are ensured.
- **Art** - virtual reality helps us to visit museums and art galleries from anywhere in the world or allows us to create our own works of art in a virtual world dedicated to the artist's imagination. The user can make these trips and virtual visits with friends or other users interested in the beautiful.
- **Virtual Games** - transpose the user into a fascinating world described in the smallest details, which make the interaction very realistic and attractive. Games that use virtual reality require significant hardware resources, even seven times more compared to classic computer games, according to a study by NVIDIA. They show that if PC games require a resolution of 1920 X 1080 at 30 FPS, those using virtual reality require a resolution of 3024 X 1680 at 90 FPS.
- **Movies** - are starting to use virtual reality more and more, and now we are witnessing an explosion of animated films and not just using this technology. The great novelty of the classic films comes from the fact that they allow a unique experience of 360º creating a favorable context for a new industry that will be growing in the next period.
- **Education** - with the help of virtual reality is more intuitive and more attractive for students and children. Compared to classical methods, technology comes with new elements such as animation, movies, 3D models, nature sounds, etc. The purpose of these technologies is not to replace the classical methods used in education, but to help them and make them more attractive to students and teachers. One reaches the situation in which new concepts are learned without being aware of it, and in subjects such as chemistry, physics, geography, one can experience both individually and in a team.

**Similar Applications**

Solar System Explorer (Guşă et al., 2019) is an application that uses motion sensors of the mobile phone, to create a virtual environment, which offers the possibility of knowing the Solar System. The application is an educational one, but also fun through the user's interaction with it. It can be used in particular to help students improve their knowledge of the Solar System and learn about each planet individually. Obviously, the application can be used by users of any age, who want to improve their knowledge about our solar system.

Also, users can evaluate their knowledge gained through the tests provided and can opt for more degrees of difficulty. The application offers the possibility to browse the list of planets and select the desired one, search through a search engine or voice commands, or directly access by clicking on the screen. For each object in the Solar System, information, statistics, video documentaries taken from YouTube, 3D animations, as well as current Twitter posts are available.

*Star Chart*

The Star Chart[117] app is created for Android and allows you to view the stars and planets on your mobile device like viewing inside a planetarium. Provides an animated and illustrated the visual experience. The application allows you to enlarge the images to observe the details.

The application needs the location of the device, and users have to enter the latitude and longitude, or they can provide them with GPS access. Thus, when the device moves, the space map also changes, giving the impression of a real exploration of the sky.

*Solar System Explorer 3D*

The Solar System Explorer 3D[118] application is also created for Android and offers the possibility to explore the Solar System. There is a list of planets and depending on the object chosen on the screen, its image is displayed along with the information and options for it. Also, the Solar System map can be navigated by dragging or rotating the image with your fingers on the screen.

The app also offers the Flight option to explore the space, with a left-right, up-down button, and a button to rotate the image.

*Sky Portal*

The SkyPortal[119] application is made for both Android and iOS. It is one of the best star simulators. We can look at the sky above us or we can move the device to explore space. There is also an option to search for objects and get all the information about them. The application allows you to choose different dates and times than the current one to see what changes over a certain period.

---

[117] https://play.google.com/store/apps/details?id=com.escapis tgames.starchart&hl=ro
[118] https://www.amazon.com/Burlock-IT-Pty-Ltd-Explorer/d p/B00PE9ISMY
[119] https://play.google.com/store/apps/details?id=com.celestro n.skyportal&hl=en_US

The application also contains a Night mode to protect the eyes. Another feature offered by this application is the use of the compass to position us where we want on the map.

*Sites in VR*

Sites in VR[120] is built for both Android and iOS. It uses virtual reality and allows the user to transpose into new places, different cities and see unique landscapes. The application is not particularly focused on space exploration, but there is this option, besides the possibility of seeing museums, palaces, castles, or other tourist attractions in different cities of the world.

The desired location can be chosen from several countries, or cities made available by application, but also from several tourist attractions that can be found in some areas of the world. Once the desired landscape has been selected, the user will be able to explore the surroundings by moving the device to the desired area or with Google Cardboard-type virtual reality glasses.

**Solar System Explorer**

Solar System Explorer combines elements of the Android platform with frameworks and APIs for various services, such as Virtual Reality, YouTube, Twitter, making it an easy-to-use application (Guşă et al., 2019). To make a model of the solar system, Autodesk Maya was used, which has 3D object modeling software. The app contains two parts of virtual reality: one that was made using mobile sensors, and the other was made with Google VR[121]. After launching the application, the sensors begin to receive movement. When changes occur, they are transmitted and the image is drawn to the new coordinates. The part of virtual reality created with Google VR is faster in detecting movement and allows the use of cardboard glasses. The app also offers a way to connect to the latest solar system news and events by posting messages on Twitter. Thus, the user is connected to an important source of information, which is already centralized and filtered. The most interactive part of the application is the test provided, which allows the evaluation of the acquired knowledge. There are three levels of difficulty: light, medium and heavy. The score obtained helps to track the progress of the study.

---

[120] https://play.google.com/store/apps/details?id=air.com.erca ngigi.sitesin3d&hl=en_US
[121] https://vr.google.com/

The application could be used as an evaluation system for astronomy classes. The user can search for a planet using the first menu button or selectable item from the list provided. Also, the second menu button is the voice control function that can be used to search for planets. The user has to say the name of the planet and will receive more details about it. After a planet is selected, the menu has changed. It will provide object-specific features. Here are some documentaries that were downloaded using a YouTube API. The application offers 3D animations for a clear view of the planets and the space around them. Also, real and important information is retrieved from the NASA website, and the links link directly to a secure webpage for detailed articles.

*3D modeling of the solar system*

Autodesk Maya software is software for 3D object modeling, animation, simulation, rendering, and composition. It offers complete functionality and creativity focused on a versatile production platform. Maya brings professional and efficient tools for creating characters and effects, as well as tools for increasing productivity in modeling, texturing and creating shaders in a fluid, clear and customizable environment. Maya is mainly used in media and entertainment.

In building the 3D model we used the Polygon Sphere option for drawing the sun and planets from the Solar System. Thus, we obtained one sphere for each object, the one for the sun positioned in the center of the plane with $(x, y, z) = (0, 0, 0)$, and the others around the first one using the translations on $(x, y, z)$. We used scaling so that the ratio between the dimensions of the planets is the same as in reality.

To get a model as close to the real Solar System, for planets we used textures from a site that offers them for free, but also other effects and coloring games created from Maya. Also, with the Paint Effects option, we have completed the model by adding stars and other effects that give the impression of a real image from space. Following the combination of all the techniques, we obtained a scene in Maya, which represents the 3D model of the Solar System, as can be seen in Figure 44.

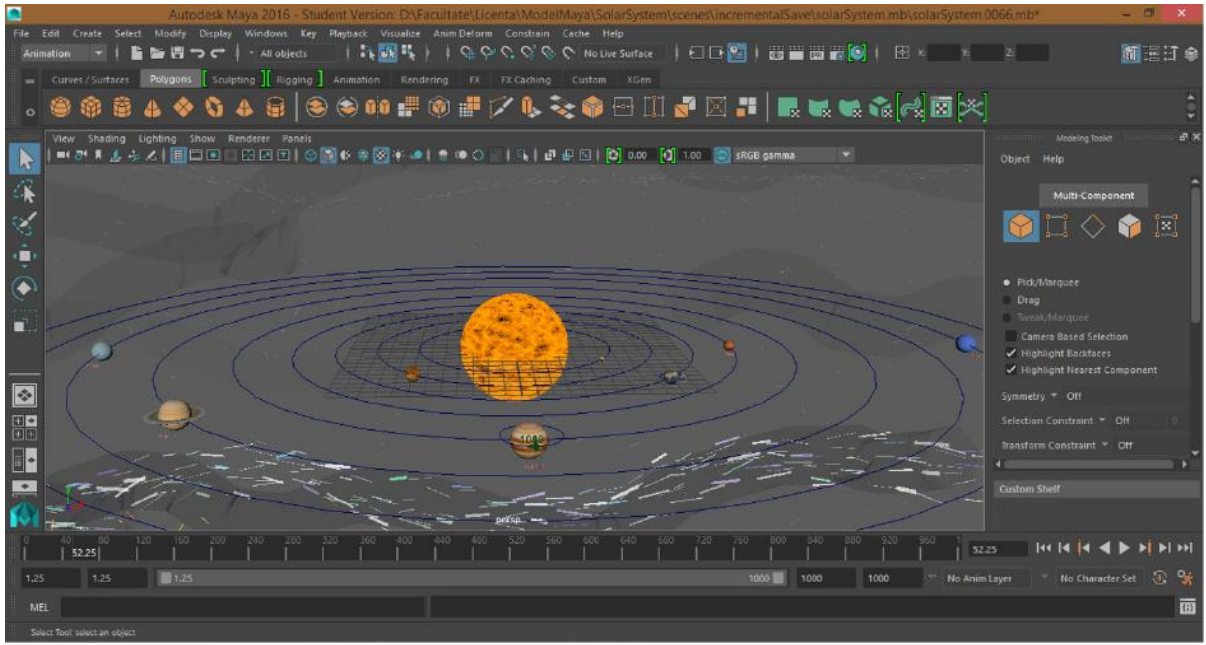

**Figure 44**: The final scene obtained in Maya, which represents the 3D model of the Solar System (Guşă et al., 2019)

Further, to use the model obtained in the Android application we used the Render View option. Thus, we managed to export images with the model viewed from any angle, at the desired resolution selected from Maya. An example of rendering can be seen in Figure 45.

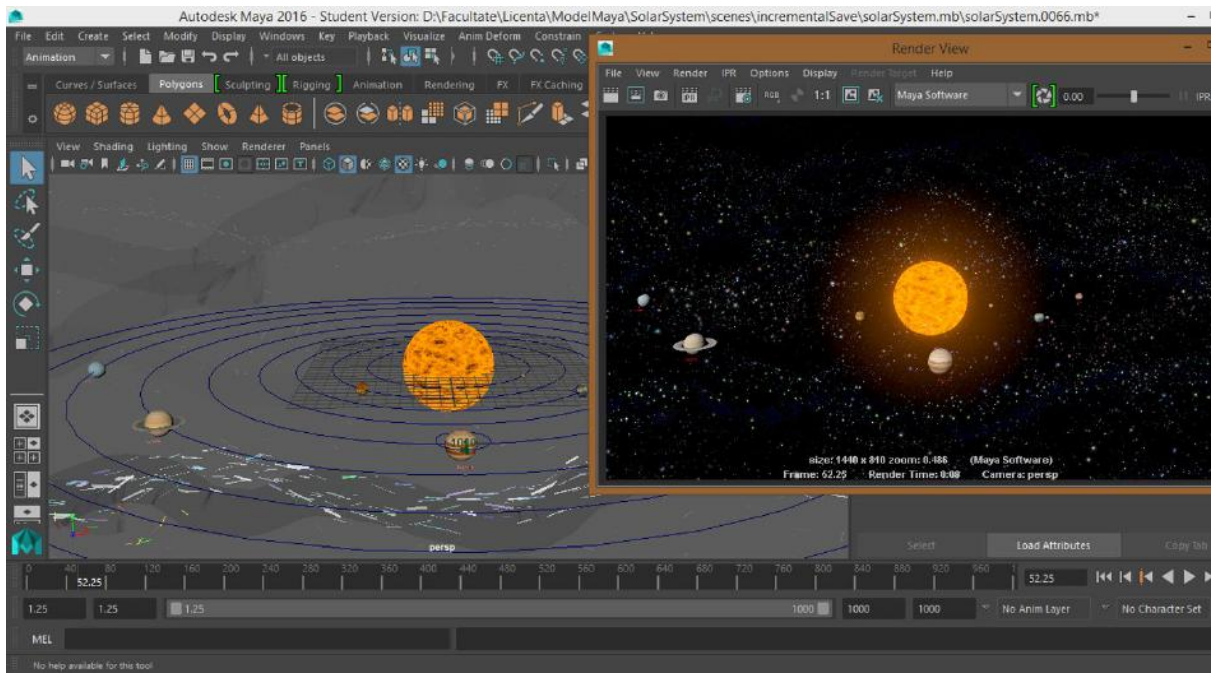

**Figure 45**: Example of rendering in Maya (Guşă et al., 2019)

*Application Architecture*

**The Main Module**

When the application starts, the main activity is launched, which contains the menu buttons, as well as the two parts of virtual reality. The first is created using sensors and image redesign when changes are transmitted from them, and the second is created using the Google VR SDK, also offering the ability to use Cardboard glasses. The application can be used in Full-Screen mode, and the orientation set is Landscape, except for the activity that provides posts on Twitter, where Portrait orientation is allowed. In Figure 46, you can see how the application looks after it is launched, with all the details mentioned above.

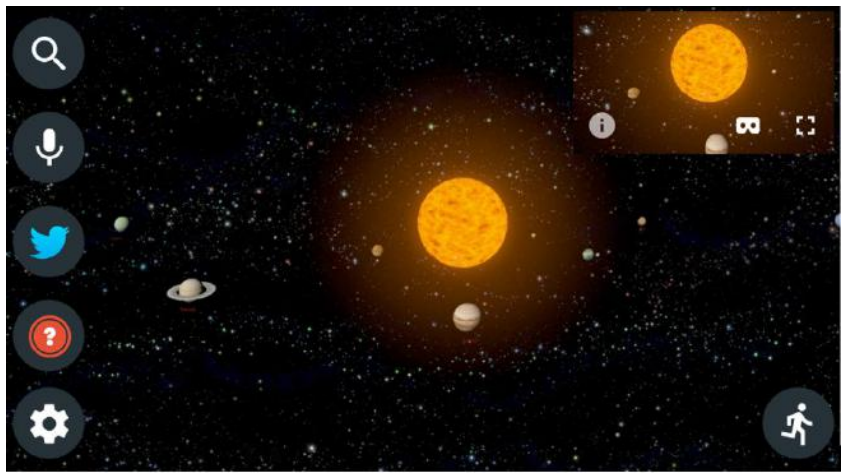

**Figure 46**: Solar System Explorer homepage (Guşă et al., 2019)

*VR using Device Sensors*

Also, once the application is started, the sensors of the phone are started to receive the movement. A system of axes is used, in which a point has 3 coordinates: *x*, *y*, *z*. Thus, at each movement detected on the Oy axis signals will be sent to redesign the background image to the new coordinates, giving the impression of its movement.

The main image on the screen will be the one used by the sensors, being the image of the Solar System drawn with the Maya program. Users will be able to zoom in and out on the image (zoom in - zoom out), for example, Figure 47, but they will also be able to access each planet with a simple touch of it on the screen.

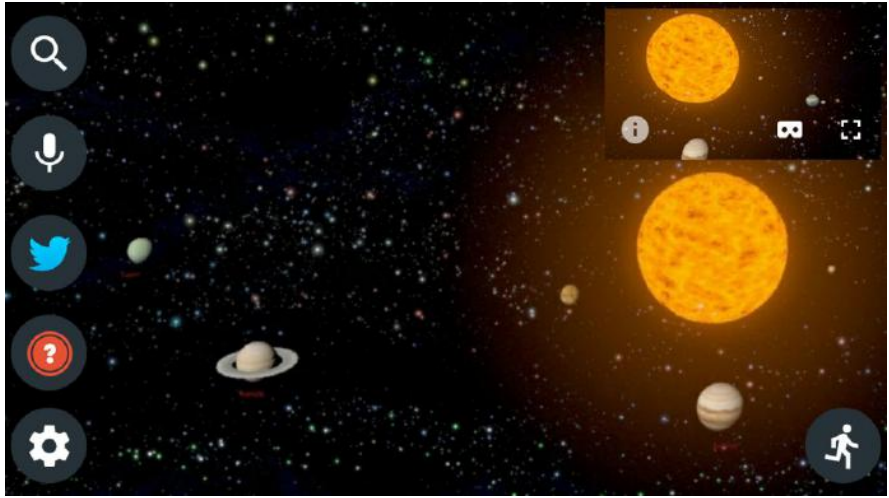

**Figure 47**: Zoom in (Guşă et al., 2019)

After clicking on a planet, another activity will start, and another menu with the image specific to each object will appear on the screen. As can be seen in Figure 48, the Sun was selected and the menu and main image changed. Thus, when the user is interested in finding out more details about a particular planet he can at any time click on it and document himself. The menu specific to the objects of the Solar System will be described in the following section.

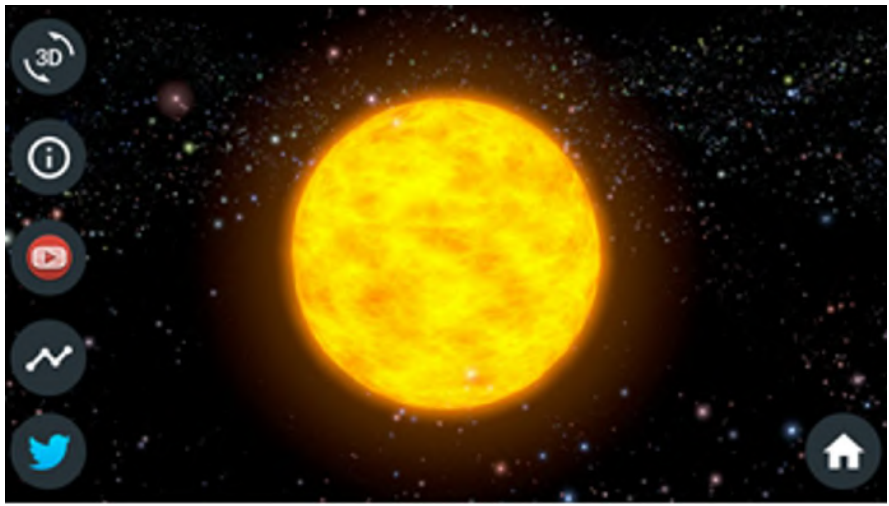

**Figure 48**: Launch of Sun-specific activity after being selected from the main image (Guşă et al., 2019)

*Google VR*

Returning to the main screen, in the upper-right corner, a box in which the same image of the Solar System is used and which also moves when the coordinates of the device on which the application runs are changed. In this part of the screen is presented a variant of virtual reality offered by those from Google VR. In Figure 49 on the left, we can see how this virtual reality is viewed in Full - Screen mode.

Thus, the user has two options to explore the Solar System. Although the second one, built using Google VR tools, is faster in receiving motion and moving the image to new coordinates, the first version built using device sensors and redesigning the image to signal a change of position, offers other possibilities such as accessing a planet or enlarging and shrinking the image.

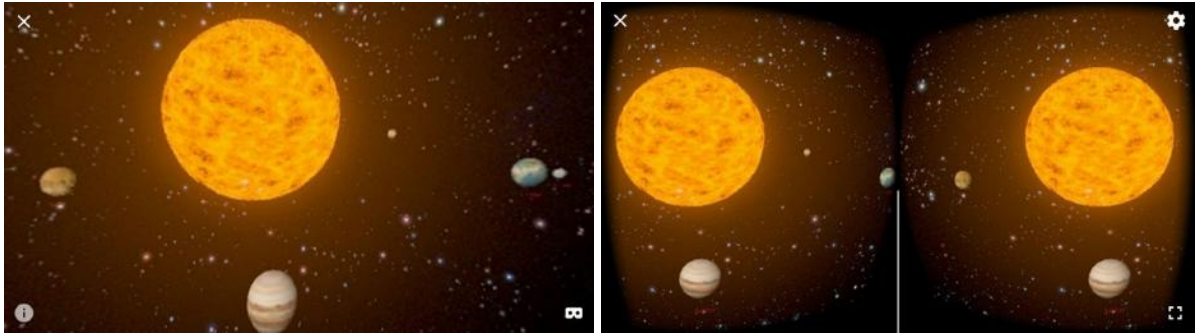

**Figure 49**: Virtual reality using Google VR (left), Using Google Cardboard Glasses for Virtual Reality (right) (Guşă et al., 2019)

The second important aspect to the virtual reality built with Google VR tools is that it offers the possibility to explore this fictional world with the help of Cardboard glasses. Therefore, the user can easily choose how he wants to view the Solar System. Figure 49 on the right shows the viewing option using Cardboard glasses.

**Other Modules**

- *Search the list of objects* - Exploring the menu on the left, the first button is the *Search* function of the planets. A box is provided (at the top of the screen) where you can enter text, representing the name of the planet you are looking for. Thus, a planet can be selected by scrolling through the list presented, or it can be searched faster by entering part of the name of the planet we are looking for in the Search bar.
- *Voice Commands* - The second button in the main menu of the Solar System Explorer application is the ability to search for an object in the Solar System using voice commands. Thus, at the push of a button, the user will be able to see a box that confirms that the microphone of the device is going to record the heard commands. When it is desired to obtain information about a particular planet, its name must be spoken in English and the activity-specific to it will be opened. This search method gives the user an easier way to find what they want. When the name of a planet is spoken, the device's microphone may not receive the command due to the distance to it, or because of the surrounding noise. In this case, a message will be displayed asking the user to repeat the command. Therefore, it must make sure that it is spoken clearly, close to the microphone, and there are no other noises that could be received.

- *Twitter Posts* - The third menu button gives the user the possibility to interact with the social network Twitter. Through the Solar System Explorer application, the user is given access to tweets related to the Solar System planets, posted in real-time, on accounts such as those of NASA, without the need to create a Twitter account or connect to the personal one. At the same time, the posts are already filtered and centralized, so you can access tweets about each planet individually, or using the *All* option to browse all existing posts related to the Solar System.

- *Quiz* - The fourth button on the left side of the menu is the most interactive part of the Solar System Explorer application. The functionality of the button is given by the redirection of the user to a test, which aims to help him verify his knowledge about the Solar System, acquired after using the application. Thus, 10 questions will be made available, meant to emphasize the most important aspects that anyone should know about the planets. The tests offer 3 levels of difficulty: *Easy*, *Medium* and *Hard* so that the same set of questions can be used to evaluate the knowledge according to the desired degree of difficulty. At the same time, depending on the level of difficulty chosen, the question-answer model differs. For the *Easy* level, the user is provided with a variety of single-choice answers, and he must tick only one option. The *Medium* difficulty level is a means of assessing average difficulty because the user must find the answer alone, without offering him a list of possibilities. The *Hard* level is the most difficult level, using the same model question - they answer as at the Medium level, but also, a timer was added.

**Conclusions**

The virtual reality has experienced a real explosion in the last years, being more and more used, both in the games industry, but also in new fields such as architecture, medicine, education, sports, entertainment, simulators, etc. The users of the applications that use the virtual reality are delighted by the ease of use of these new technologies and the high degree of attractiveness, the experiences lived being unforgettable by combining the animations, the sounds and the images very detailed.

Solar System Explorer is an application that uses virtual reality with which the user can study the Solar System. Two virtual reality methods are available that make the interaction with the application much more interesting, due to the possibility of exploring the space by receiving the movement of the mobile device. At the same time, accessing the planets and deepening their knowledge about them is done in an easy and enjoyable way, with the possibility of simply clicking on them, using the Search function, or using the voice command. The application also provides a method of connecting to the latest news and events related to the Solar System, by presenting some posts on a popular social network, Twitter. Thus, the user does not have to log in to an account or search for specific tweets, as the information is already centralized and filtered. There is the possibility of obtaining specific posts for a particular planet or all those related to the Solar System. For the second variant is presented a statistics of the tweets, and thus we see how often things were posted about one planet compared to the others.

The most interactive part of the application is the test provided, which allows the evaluation of the knowledge gained during its use. There are three levels of difficulty: easy, medium and hard which determine how the questions will be asked and the answers were given. For Easy, single - choice answer variants are used, for Medium, EditTexts are used, in which the answer must be written, and for Hard, the same model is used as for Medium, the difficulty being added by the presence of the stopwatch. The score obtained helps to observe the progress made after the study.

In the future, the application could be developed for other platforms such as iOS. Also, stars, constellations, satellites, or other elements of the Solar System could be added for a more elaborate knowledge base. The application could also be used as an evaluation system for astronomy classes, possibly adding the possibility of creating an account for both students and teachers, and the latter having access to student results and being able to ask questions.

## III.3 Amazon Alexa

### III.3.1 Context

The Amazon Echo will always be known for bringing intelligent functionality into the world of speakers. And it was all due to Alexa, Amazon's smart voice, who brought the Amazon Echo speaker to life (Davis, 2017). Obviously, it was a trick at the beginning, but Amazon was very vocal about improving Alexa, and now it's a vital and important part of the notion of a smart home. Initially, Alexa was only available through the devices that the company made, but is now ubiquitous, as many speaker manufacturers were allowed to use Alexa - the largest recent user is Sonos One[122].

*What can we do with Alexa*? We can ask Alexa *what time it is*, *what the weather is like* (we will have to enter in advance where we live), we can ask for *a brief presentation of the news*, ask her to *tell us a joke*, *set a stopwatch*, *order a taxi*, *order a product on Amazon*. There is a whole list of things we can do without creating a special ability for Alexa (Skill) (which are essentially Echo components of our phone's applications) (Chacksfield, 2018). We can make the Echo *sing* by calling on Alexa's abilities. In the area of music, we have where to select skills: Amazon Music, Spotify, Sonos system, radio stations, etc. You can also *read a book* by calling on the contents of the Kindle.

Amazon also allows developers to create services that can be integrated with Alexa. Alexa Skill Kit[123] (ASK) allows developers to create services that make Alexa smarter. A *skill* (or ability) has two parts: *the interface* and the *skill service*. In order to have a functional ability, the two parts must be implemented. Currently, there are over 60,000 skills among Amazon services available from large companies Netflix, Uber, etc., but also from small companies and even programming enthusiasts[124]. Among the areas present here is education and it can help with the homework[125].

---

[122] http://www.techradar.com/reviews/sonos-one-review
[123] https://developer.amazon.com/docs/ask-overviews/build-skills-with-the-alexa-skills-kit.html

## III.3.2 Educational Applications

A series of educational applications are available, implemented as skills for Alexa[126]. The *Question of the Day application*[127] proposes a new question every day, extracted from different domains, such as science, art or entertainment. The major difference with our system is the fact that the collection of questions and hints for answers is a closed collection, not dynamically adapted to the knowledge level of the user.

Another educational game is *AmazingWord Master Game*[128], an application introducing a chain game. Starting from a random word, the user has to name a word that starts with the last letter of the initial word, and so on. The difference between this game and our application is that here the user competes with the application; while in Bob each user competes with himself or his friends.

The *Tricky Genie*[129] is a different kind of application, emphasizing the comprehension of English. Thus, the game forces the user to make a choice between three possible solutions for a challenge presented as a story. The major drawback of this game is that it has a limited number of predefined stories, and the learning does not adapt to the level of English that the user needs to improve.

**Learn Geography and History of Music with Amazon Alexa**

Next, we will see an ability that can be used in learning the geography and history of music (Filimon et al., 2019a) with the help of Bob application. The proposed ability can create questions by using well-known external sources, such as DBpedia and Wikidata. An interesting feature is related to the possibility to adapt the questions to the students level using CAT

---

[124] https://techcrunch.com/2019/01/02/the-number-of-alexa-skills-in-the-u-s-more-than-doubled-in-2018/
[125] http://www.perkinselearning.org/technology/blog/6-ways-amazon-alexa-can-help-homework
[126] https://www.ford.com/technology/sync/
[127] https://www.alexaskillstore.com/Question-of-the-Day/B01N6QUAXX
[128] https://www.alexaskillstore.com/Amazing-Word-Master-Game/B017OBSCOS
[129] https://www.alexaskillstore.com/Tricky-Genie/B01IKR3OAC

(Computer Adaptive Testing) (Weiss and Kingsbury, 1984). The service can be called with both an Amazon Echo device and a smartphone.

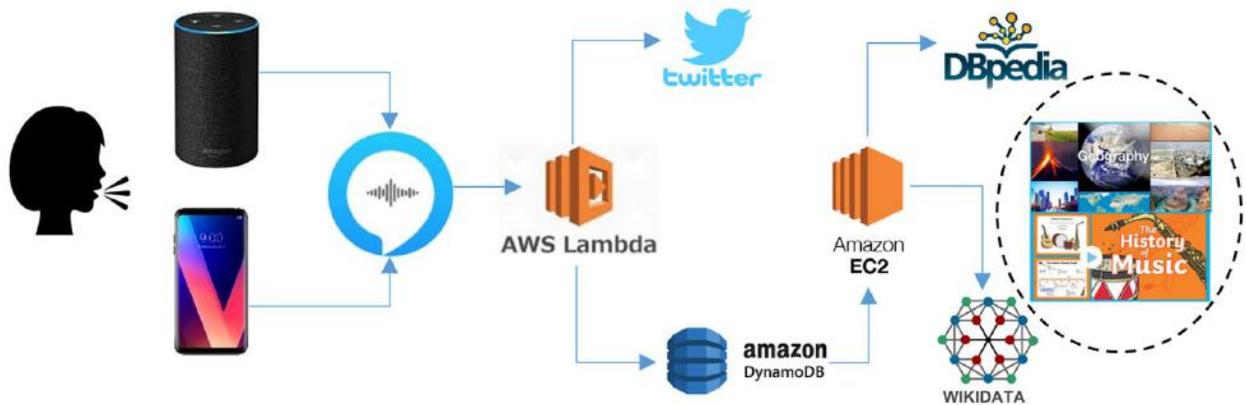

**Figure 50**: Architecture of the Bob application (Filimon et al., 2019a)

The architecture of the proposed system can be seen Figure 50. The main actions available through the voice interface are:

- *answers general culture questions* in geography or music history;
- *starts general culture tests*, which make an assessment of the participants' level of knowledge;
- *searches the position from the general ranking* and the number of points accumulated;
- *posts on Twitter* the obtained result.

The first step in designing the application was to obtain the user's spoken input through a voice service. Subsequently, templates were established for the interaction between the user and the system, specifying the *intents*, *slots* and *utterances*. The AWS (Amazon Web Services) Lambda service was then used to develop the main functionalities and interaction mode with Alexa. The questions and answers were extracted from DBpedia[130] and Wikidata[131] and then saved to a DynamoDB[132] database on Amazon's servers.

*Intents Schema*

---

[130] https://wiki.dbpedia.org/
[131] https://www.wikidata.org/
[132] https://aws.amazon.com/dynamodb/

The Intent Scheme is a JSON object that contains a list of objects with information about each intention that the skill will be able to handle. Objects containing information about an intent will store its name along with an optional list of objects indicating the slots that can appear within the intent, along with their type. The names for intents can be offered by ASK, defined by the developer, or presented as slot types. Our Bob application has several intents:

- *AMAZON.CancelIntent* is defined in ASK, and under this name are collected a series of replies that the user can communicate when they want to stop their tests;
- *BobBeginTest* is defined by the developer, and it includes replies that the user can tell when they want to start a new general culture test;
- *BobAnswer* is defined by the developer, and it contains input phrases that the user can provide when he wants to answer a general culture question. Because there are many possible answers, a number of slots are used.
    - *AnswerL*, which has the AMAZON.Landform type, defined by the user;
    - *AnswerC*, which has the type AMAZON.Country, defined in ASK.

*Slot Types*

A slot represents a variable name, which is assigned by the user during the speech. The type of a slot is defined by listing all its possible values. These are responses that contains possible answers to questions, such as:

- *AMAZON.GB_CITY* contains the names of all cities that compose correct answers to questions related to *cities*, in the form which is most commonly used in Great Britain;
- *AMAZON.Landform* contains the names of landforms which can appear as correct answers to questions related to *lakes*. In a further stage of the application, other types of landforms will be added (*mountains*, *hills*, *valleys* etc.);
- *AMAZON.Country* contains the names of *countries* that can be correct answers to questions about countries.

*Utterances*

Utterances are added to the Sample Utterances section of the interaction model. Through these utterances, we specify each possible input that comes from the user and precedes them by the intent it relates to.

*Skill Service*

In order to develop the application, the AWS Lambda service was chosen because it provides support for Alexa skills and facilitates calls to other AWS services through a lambda function (https://aws.amazon.com/lambda/). The programming language is NodeJS and has been chosen since it is recommended for small server-side applications where non-blocking operations are very important. Creating the Link between AWS Lambda and Voice Service: In order to create a lambda function, the developer needs to access the management section in the AWS account. In the section dedicated to create lambda functions, the first step involves setting a blueprint, i.e. the running environment, which for Bob is Node.js 6.10, along with a template for the function to be developed. The second step is to configure the trigger function, in our case, the voice interaction through Alexa Skill Kit. The configuration function section provides information about permissions, memory limits, timeout, etc. After creating the function in the voice service configuration section, the function ID must be provided. This step assures sending the information from the user, after processing, to the lambda function that will decide on the answer it will provide. In order to have a functional skill, the code needs to be deployed, in our case through the AWS Command Line Interface (CLI).

*Usability Testing*

In order to evaluate the application, we invited seven participants (three female and four male) and organize different playing sessions (Filimon et al., 2019b). We have recorded their experience with the game while they were performing usual tasks. During the tests, we took into account their social interaction activity, having 4 participants with high activity, 2 with moderate and 1 with reduced.

The conducted usability test consisted of an *introduction*, *five tasks* and a *short interview*. We ask the participants to think out loud and express their thoughts and sentiments during the test. From our observation, the users found very quickly the needed actions and were able to successfully connect to application. On the other hand, the participants had some trouble while they pronounced complicated or compound words. The main issues were that the speed of speaking and speaker volume could not be configured during a test. Another issue is related to the flexibility to choose the domain of the questions during a test. Also, the participants with a high level of knowledge asked for the possibility to choose, at the beginning of a test, the level of difficulty, in order to avoid the simple questions in the first part of the test, while the system is guessing the user's level.

Recordings analysis clearly showed that:

- *Communication and interaction between participants and the application* was very attractive, exceptions a few points where it was a little bit confusing;
- Difficulties had been encountered when the *questions were not targeting country of birth* of the participants;
- There was a continuous interaction between the participants in the experiment, so that in addition to learning the geography, the *participants improved their communication skills and English speaking skills*.

*Error analysis*

Most problems occurred related to understanding and evaluating answers to questions that Bob is addressing, which affects the test side. These components of the system are most error-prone because the user is in the position to say complicated words that may not be correctly understood by Bob, especially if the user in a non-native speaker. Unwanted behavior occurs because the user response is not clear enough and fails to be associated with an existing intent, or the associated intent is not expected. These problems are based on the fact that the voice service component, when a user speaks a word-by-word, will compare them with the items in the replica list that the system accepts and identifies the one that seems closest to the one uttered by the user. Problems arise when the user did not pronounce correctly certain words or spoke too quickly, or if his interaction with Alexa was interfered with by other sounds or an unexpected event of the user (for example, coughing or sneezing).

If the user interaction with Bob fails to be assigned to any intent, it will be ignored. Naturally, the user will repeatedly say the most likely answer, which may give the impression that the application has been blocked. If the interaction was assigned to another intent, which is unlikely, then the application will give an unexpected response. For example, if during a test the correct answer sounds very similar to "*Post on Twitter*", the application will tell the user that the score has been posted on Twitter. One of the consequences of this behavior is the interruption of the test. At this time, the problems with processing the input from the user cannot be solved by the Bob application because this component is provided by Alexa. In the future, in order to improve the understanding of the user's pronunciation, we intend to make a learning module in which Bob learns minimal information about certain geographic elements, and his task is to repeat the name of that element.

### III.3.3 Iasi City Explorer

**Introduction**

There are skills developed to exploit the tourist potential of big cities which provide recommendations for restaurants, cafes, car rentals, information about weather. Also, these skills suggest activities suitable for a certain time of the day, including main attractions and places where you can pleasantly spend your time. Skills for Alexa, similar to Iasi City Explorer, are:

- *Bucharest Guide*[133]: offers information about 33 points of interest in Bucharest.
- *Rome tour*[134]: A guide to tourist attractions in the capital of Italy. It contains the most popular locations in Rome, such as monuments or museums. Skill sends location addresses to companion application on mobile or TV.
- *NYC Guide*[135]: Provides information to tourists in New York such as tourist attractions or information about transportation.
- *MyParis Guide*[136]: provides detailed information about the top 5 things to do in Paris.

---

[133] https://www.amazon.com/Catalin-Batrinu-Bucharest-Guide/dp/B074WBNM7R
[134] https://www.amazon.com/Chris-Cinelli-Rome-tour/dp/B07475HGSS
[135] https://www.amazon.com/AlexSantisteban-New-York-City-Guide/dp/B076QGS2PS
[136] https://www.amazon.com/Adrien-Chan-MyParis-Guide/dp/B01MT2T4O6

The main purpose of Iasi City Explorer is to enhance the tourist experience in Iasi and also to help the newcomers to explore the city and find location easier, without the need to make intermediate search. Iasi City Explorer is available on the Amazon Echo device, but also on web browsers, Android and iOS platform, via Amazon Alexa app, where the user can interact with the application through voice commands. The main reason for picking Alexa as a development platform is the increasing popularity of the Artificial Intelligence, implicitly of the digital assistants and voice first applications.

**Proposed Solution**

At the development level, the application follows a cloud-based architecture, orchestrating services provided by Amazon Web Services[137] with Google Maps Platform and Yahoo Weather. The interaction between services can be seen in Figure 51.

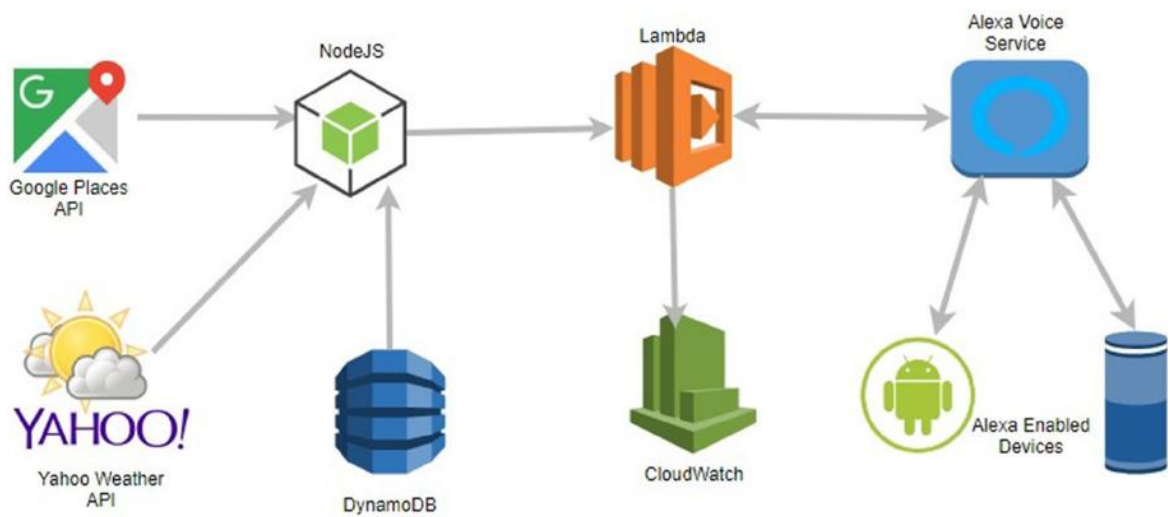

**Figure 51**: Iasi City Explorer - System architecture (Miluţ et al., 2019a)

The user interacts with Alexa-integrated devices such as Amazon Echo, mobile applications, etc. The query is intercepted by Alexa Voice Services, and is forwarded to the Alexa Skills Kit that launches the Lambda function[138]. Depending on the request received, Lambda takes data from Yahoo Weather API[139], Google Places API[140] or DynamoDB[141]. DynamoDB data is also obtained

---

[137] https://docs.aws.amazon.com/IAM
[138] https://searchaws.techtarget.com/definition/AWS-Lambda-Amazon-Web-Services-Lambda
[139] https://developer.yahoo.com/weather/
[140] https://developers.google.com/places
[141] https://docs.aws.amazon.com/amazondynamodb/latest/developerguide

through the Google Places API. To monitor the Lambda function, CloudWatch metrics are recorded.

To access DynamoDB and CloudWatch, the Lambda function gets some permission through Identity and Access Management (IAM). The response processed using the Lambda function is passed back to the user through the Alexa Voice Service and the device through which the interrogation was made. In the following, we will present the technical details of this process.

*Intent Scheme*

The customized intentions in City Explorer are as follows (Miluț et al., 2019b):

- *AboutIntent*: to get a brief description of the city. Launched by the expressions "*about*" and "*tell me about this place*".
- *AttractionIntent*: to receive a tourist attraction recommendation and learn about it. Triggered by the words: "*recommend an attraction*", "*give me an activity*", "*what to visit*", "*tell me about a place to visit*".
- *FoodIntent*: Restaurant recommendations based on a preference or generic recommendation, if preference is not specified. Triggered by expressions containing the "dish" slot, for a specific recommendation: "*dish*", "*where can I get some dish*", "*I would like some dish*", and for a general recommendation "*food*", "*I want to eat*".
- *ActivityIntent*: Suggests the user a place to perform a certain activity (e.g. *I want to swim*). Examples of phrases: "*recommend location*", "*suggest location*", "*I want location*", "*I want location*", "*give me a location*".
- *CarIntent*: Suggest companies that can rent cars. Activated by the expressions "*rent a car*", "*I would like a car*".
- *RecommendIntent*: location recommendations based on time of day. Triggered by: "*what should I do*", "*where should I go*", "*give me an idea*", "*give me something to do*".
- *GoOutIntent*: Time and weather. Turned out by: "*go out*", "*go out-side*", "*how is the weather*", "*weather in Iasi*", "*weather*", "*what time is it*".

*Utterances*

An utterance is a word or expression attributed to an intention to trigger it. The list of utterances helps the Alexa Skill Interface process the words spoken by the users in intentions.

*Slots*

Slots used in the application are built-in, which means they are offered by the Alexa Skills Kit. City Explorer uses 2 slots, namely *AMAZON.Foods*, and location - like *AMAZON.LocalBusinessType*. These two types of slots include dishes (e.g. *chocolate, cake, scrambled egg, Campbell's low sodium chicken broth*) and a list of location types (*medical clinics, food store, auto rental store, dry cleaning*).

*Extracting Data*

To populate the database, a separate section of the application was created. This section includes a Node JS server that contains a DynamoDB client linked to the PLACE LIST table and a list of location types made by Google Places queries through Place Search. In order to get more information on the locations obtained from the query, they are provided as a Place Details function. The results are filtered and serialized in JSON format. Finally, serialized results are inserted into the table through the DynamoDB client.

*Usability Testing*

In order to analyze the real life applicability of application and gather user feedback, we performed a series of usability tests. They targeted potential users, belonging from two age categories: the elders with visual deficiency and youngsters (Calancea et al., 2019). The latter category participants were blindfolded to better impersonate people that have no means of using visual aids.

*Methodology* - The performed test was composed of a presentation regarding the problem context, several tasks to be conducted during the experiment, a satisfaction questionnaire and a brainstorming session to gather ideas for other useful functionalities. Each participant interacted with the assistant in a different environment, without communicating with other users prior to the experiment. These measures were taken with the purpose of obtaining mostly unbiased results. Everyone had to perform on the provided smartphone application 5-6 tasks, i.e. voice interactions with the application (from the set "*tell me about this place*", "*recommend an attraction*", "*Where can I get some dish*", "*recommend a location*", "*I want to rent a car*", "*give me something to do*" and "*weather in Iasi*"), which need to take approximately 3-4 minutes per session.

*Elders with visual deficiency*

*Participants*: We collaborated with a group composed of 4 people who have little experience with technology. Their age was between 50 and 56 and all of them experienced various eyesight deficiencies. Regarding the usage of smartphones, 100% percent of the participants have used a smartphone before, while only 50% interacted with a smart assistant (ex. Siri, Google Assistant).

*Results*: As a first impression, the participants enjoyed the simplicity of interaction with the application, mainly because the authentication method is fast and reliable, there are no intricate interface elements like small buttons or text input fields and the vocal interaction substitutes entirely the need for glasses or fairly good eyesight. After the proposed test scenario was executed successfully, each of them was asked to rate their general experience with a grade from 1 to 10, where 1 stands for confusing/ frustrating experience, and 10 for clear/pleasant experience. The average rating for the user experience was 7,75, since three of our participants were not fluent English speakers, which often resulted in misunderstanding of the responses to some part of the vocal enquiries. This questionnaire was followed by a feedback and brainstorming session meant to highlight the current issues and the most desired future improvements. We concluded that an increase in the collection of available languages was a development priority in the near future, alongside with the possibility to repeat answers in case of need.

*Young people without visual disabilities*

*Participants*: In order to assess the opinion of another age segment, we repeated the experiment with 6 student peers, aged between 18 and 23, which can be categorized as experimented technology users. All of them use their smartphones on a daily basis and require the help of Iasi Smart City application for easy tasks at hand. Since we needed to simulate that the proposed use case is performed by people with visual deficiencies, this segment of participants used blindfolds or similar methods to take full advantage of the applications features.

*Results*: Analyzing the gathered preliminary observations, the participants did not seem to experience difficulties while performing the requested tasks. Even though the smartphone usage experience was opposed to the usual interaction, they found enjoyable that the level of access to information remained similar to the one they were accustomed to. After completing the test scenario, they were asked to rate their general experience with a grade from 1 to 10, similar to the previous set of participants. The average rating for the user experience was 8, since young people are more inclined to embrace technology. When asked for their feedback and desired future features, most of them opted for adding options for social media access, as well as the possibility to send and read text messages. These improvements would surely increase the popularity of the application among this segment of the population.

*Remarks*

When comparing the two sets of participants, we noticed a certain reticence among the elder group in regard to the usage of technology in completing their daily tasks. Nonetheless, the youngsters seemed to be far more receptive to the idea of allowing a smart application to help them in any situation they might find themselves into. To make the evaluation process more transparent, we chose to use SUS (Drew et al., 2018), in which the responses were multiplied by 2.5, thus obtaining a scale from 0 to 100 from the original 0 to 40 scores. These are considered to be percentile ranks (Rogosa, 1999). Therefore, we concluded that Iasi City Explorer application can support and help people with visual deficiencies in fulfilling their activities when they visit Iasi city. Also, the application would have a great impact among all ages, regardless of the severity of their condition.

### III.3.4 Smart Home

**Introduction**

The development potential of using Amazon Echo in IoT (Internet of Things) domain is huge and varied, so there are applications in most areas, for example: *medical*, *social* and *logistics*. Applications fall into three broad categories, namely *smart*, *industry-friendly cities*, and those used by regular users: *healthcare* and *intelligent homes*. Intelligent systems that help citizens are implemented in the big cities, so the level of comfort increases. Some examples of such applications: *intelligent tax payment systems*, *intelligent parking systems* (which can find and manage free parking spaces in a car park), *ticketing systems in the means of transport*. The medical field also uses intensive technologies for: *tracking medical equipment*, *securing salons in hospitals*, *tracking people with disabilities* or diagnosing them at a distance. Below are some applications for improving daily living, used by ordinary users.

- **Ring Alarm:** Ring is a company selling various security solutions for the interior and exterior of homes. They sell a wide range of sensors such as motion sensors, flood sensors, contact sensors and a wide variety of cameras, all to ensure residents' safety. In the Amazon skill store, there is also the Ring app[142], which by adding it to the user's account, can control the above mentioned smart devices.
- **Smart Life** is another popular app in the Amazon store, often used by people who want to interconnect smart home appliances. The application involves buying smart devices from the manufacturer's website, such as plugs and switches. They connect to the wireless internet. In addition to the products on the site, other industry companies such as Koolertron and Oittm produce many devices that mate with Smart Life[143], including: intelligent cameras and light bulbs. The skill developed for this application has support for most devices, which are very easy to control by voice interaction.
- **Smart TV Remove**[144] works with the user's phone, turning it into a smart TV remote control. To use the skill, you must download the application with the same name in the Android store. The application has some limitations, including: it only works with Android devices, and the most important thing: the device must have an infrared sensor, and unfortunately the new devices are no longer equipped with such a sensor.

**Proposed Architecture**

Figure 52 presents an overview of the application architecture.

---

[142] https://shop.ring.com/collections/security-system
[143] https://www.amazon.com/Tuya-Inc-Smart-Life/dp/B01N1ZVI7M
[144] https://play.google.com/store/apps/details?id=com.adi.remote.phone&hl=ro

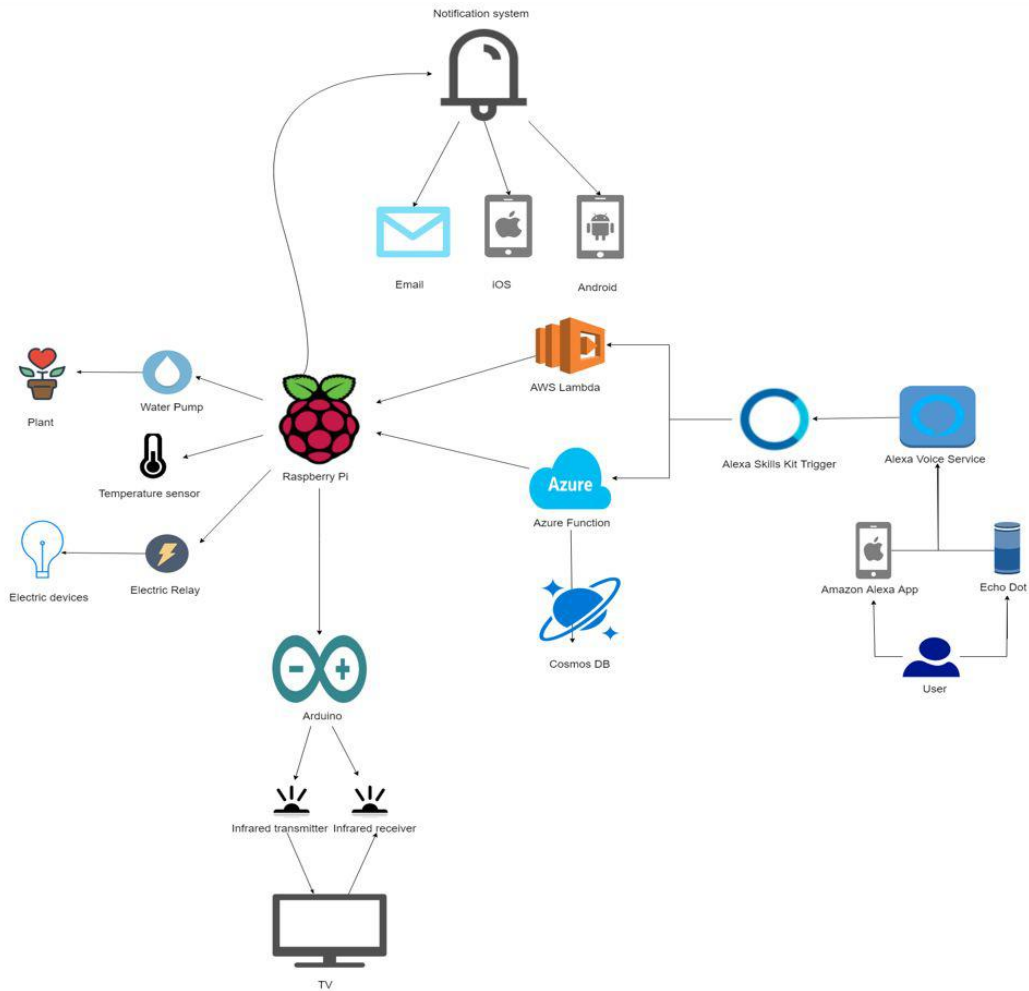

**Figure 52**: Smart Home System Architecture (Matei and Iftene, 2019)

*Voice Service*

As mentioned earlier, this part deals with the processing of natural language and its transformation into commands that the business part of the application understands. The application's name serves as a way to invoke it within the application, so the user starts the skill with one of the following invocations "*open smart home*", "*ask smart home*".

*Intents Definition*

The main actions of the application include the way to *stop the application* (*AMAZON.Stopintent*) that *stops an action on the fly* (*AMAZON.Cancelintent*) and the one to *ask for help* (*AMAZON.Helpintent*). Besides, we needed the following to build this application: *TurnOnintent* and *TurnOffintent*, used in the TV but also in the case of the Bulb, CustomCommand, used to execute the various start/stop commands for the TV, Get Information's (to request information on ambient temperature and humidity) and the last, WaterFlower, is the one to be used to operate the water pump.

*Slots Definition*

In order to ease our work and not duplicate the code, as well as other intents, we have used the feature provided by the portal called slot. By doing so, we have defined a number of devices that the actions receive as a part of it, so we managed to gain time and structure the code better. The slots contain a list of devices, in the case of *Device* and commands in the case of *Command*. In order to reuse *TurnOnintent* and *TurnOffintent*, we chose to send them a *Device* that lists the following: TV and light, so we took full advantage of it and did not bring unnecessary complexity. The next slot contains a list of commands the user can give to the TV, in addition to those on/off, these are: volume up, volume down, channel up, channel down, mute and unmute. Within these slots, a dictionary for synonyms can also be defined to cover a wider range of phrases.

*Utterances Definition*

It should be noted that they are as diversified and accept as many variations as possible from the basic invocation, for example: "*Turn on the TV*", "*turn on the television*" the user does not have to memorize the phrases, and these comes naturally. It is noteworthy that depending on the programming language version, how these phrases should be said and how they need to be related to the application name can easily vary, so a parallel of this kind is debated in the following sections.

*Raspberry Pi*

With Raspberry Pi, the issue of requests and answers has been solved in several ways: First, there is the option of connecting it to a network, either Wi-Fi or cable, via Ethernet. Secondly, the variety of languages allows to easy program the device. Also, the communication via Raspberry Pi was made easy through the VNC Viewer application by entering its address and the account created on the machine. Figure 53 represents a non-finite state of the system. As you can see below the water pump or the electric relay was not attached yet.

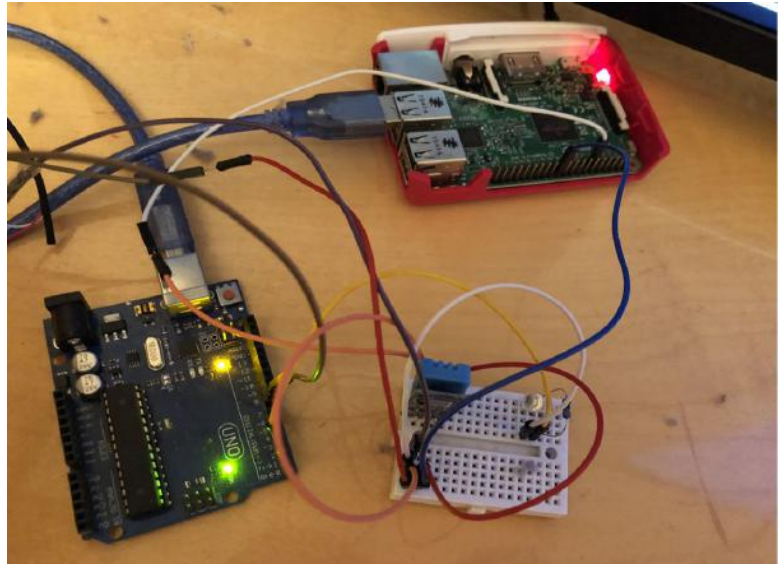

**Figure 53**: Raspberry Pi Architecture (Matei and Iftene, 2019)

## III.3.5 Conclusions

Alexa was chosen as the application development platform because voice first applications are becoming increasingly popular, and one of the reasons is ease of use. An application that can be used by voice interaction (Țucă and Iftene, 2017) can be easy to use by children, by visually impaired people or with motor disabilities or by people who are reluctant to interact with a touchscreen (the elderly for example).

Another advantage of using Alexa as a development platform is portability. Besides the popular Echo devices, Alexa is integrated with many IoT devices, but also with classic smartphones, making it accessible at all times and in any place

## III.3.6 Future Work

Future work will focus on a few main directions:

1. *eLearning* - We want to expand Bob's knowledge database to related fields such as history, biology, literature, as well as other fields such as art, sports, cinema, music, etc.

In the future, Bob can be a way to create entertainment when we wait for traffic or when we cook.

2. *City Explorer* - We aim to further promote the application and thus to increase the popularity level of the city. Also, we want to come with other information from Iasi about parks, universities, botanical gardens, sports fields, etc.
3. *IoT* - The development potential for the application presented is only the number of devices and sensors with which it can integrate. As a result of massive development in this segment and the emergence of new electrical devices and sensors, the application has a huge potential, given the type of architecture chosen for it. Some examples might be: connecting to several devices that support infrared transmission: air conditioning, projectors, integration with other household appliances: washing machines or clothes, smart vacuum cleaners or the central thermostat, or integration with some sensors that can be added to the house: humidity, carbon monoxide, gas or motion.

## III.4 Conclusions

New technologies are very attractive to those who create applications, but also to those who use applications. These open up new horizons and make us think differently about future applications. The challenges will be given by the speed of processing (Alexandru et al., 2019), but also by how we will think the architecture of these applications (Alboaie et al., 2019), (Baboi et al., 2019).

In addition to augmented reality, virtual reality and voice-based applications, students are also interested in applications for devices they carry with them daily (smart watches, smart bracelets, smart glasses), devices that track their gestures (video cameras or leap-motion, etc.

## III.5 Future Work

Future work aims first and foremost to address new areas that await applications such as those presented in this section, both for education, entertainment and for areas such as botanical garden and medical field. We have started several projects with the professors from the UMF and with the teachers from the UAIC and their evolution will be seen in the upcoming years.

We also plan to combine these areas so that we can use, for example, voice or gesture control in virtual reality or augmented reality. In this way, we intend to further facilitate the user interaction with the applications we create, but we intend to come up with new ways to do 3D modeling (in the field of architecture), or modeling of applications (in the field of software engineering).

## III.6 References

# IV. Final Conclusions and Future Work

## IV.1 Final Conclusions

The habilitation thesis presents the main directions of research of the author after 2009 (the year when the author defended his doctoral thesis). With over 150 papers written during this period, main directions include:

1. *Exploiting data from social networks* (Twitter, Facebook, Flickr, etc.) - creating resources for text and image processing (classification, retrieval, credibility, diversification, etc.);
2. *Creating applications with new technologies*: augmented reality (eLearning, games, smart museums, gastronomy, etc.), virtual reality (eLearning and games), speech processing with Amazon Alexa (eLearning, entertainment, IoT, etc.).

The work was validated with good results in evaluation campaigns like CLEF (Question Answering, Image CLEF, LifeCLEF, etc.), SemEval (Sentiment and Emotion in text, Anorexia, etc.). Because students have often been involved in these assessments for a year, or at most a year and a half, often the results have not been as expected. In the future we will need to find a way to continue the work started at some point, without having to resume certain steps from scratch.

## IV.2 Future Work

The activity from the following period will continue the current activity that is related to social networks and the use of new technologies.

**Social Networks**

Regarding social networks, we aim to try to continue to **build resources** that are useful in the areas where we started working lately (such as the medical field for example), but at the same time we want to **identify in real time if a news item is true or false**. The challenge of determining in real time whether a news item just posted on the network is true or false comes from the fact that to classify it we cannot use the classic features we use in classifying posts:

number of likes, retweets, comments, feelings from comments, etc. In such situations we must use other approaches related to the exploitation of existing semantic information in resources that provide knowledge about the world such as Wikipedia, DBpedia, Wikidata, YAGO, etc. For example, for the statement "*Pope has given birth to a child*", we can find out from YAGO that Popa has <hasGender> equal to "*male*" and from this we can deduce that the statement is *false*, because a man cannot give birth to a child. What do we do, however, if information such as "*Pope endorses Trump*"[145] appears? In this case, it is necessary to look for information on the sites of major newspapers, on the Vatican page or even on Donald Trump's Twitter page, because such news has high chances of appearing on these channels. If the information does not appear we can assume that the information is *false*.

The problem is quite complex, especially when you want to get information quickly, even in real time. Existing resources are quite large (for example, YAGO files are of the terabyte order), and their organization and exploitation will in itself be a rather complicated problem.

A new direction in this area, discussed with colleagues from UMF, is related to the **exploiting of information from Twitter related to medical domain**. They were interested in posting on a Google map of the posts on Twitter that mention diseases or medical terms. In this way, you can see the diseases that are discussed more, the areas where these discussions take place, their frequency, etc. Such an application would be useful when we want to travel somewhere to know what to expect and possibly to take measures to counter the effects of flu, for example. In Figure 54, we can see the distribution of tweets in English from period "27.05.2019 - 03.06.2019" which contain the word "cancer".

---

[145] https://medium.com/newco/how-to-detect-fake-news-in-real-time-9fdae0197bfd

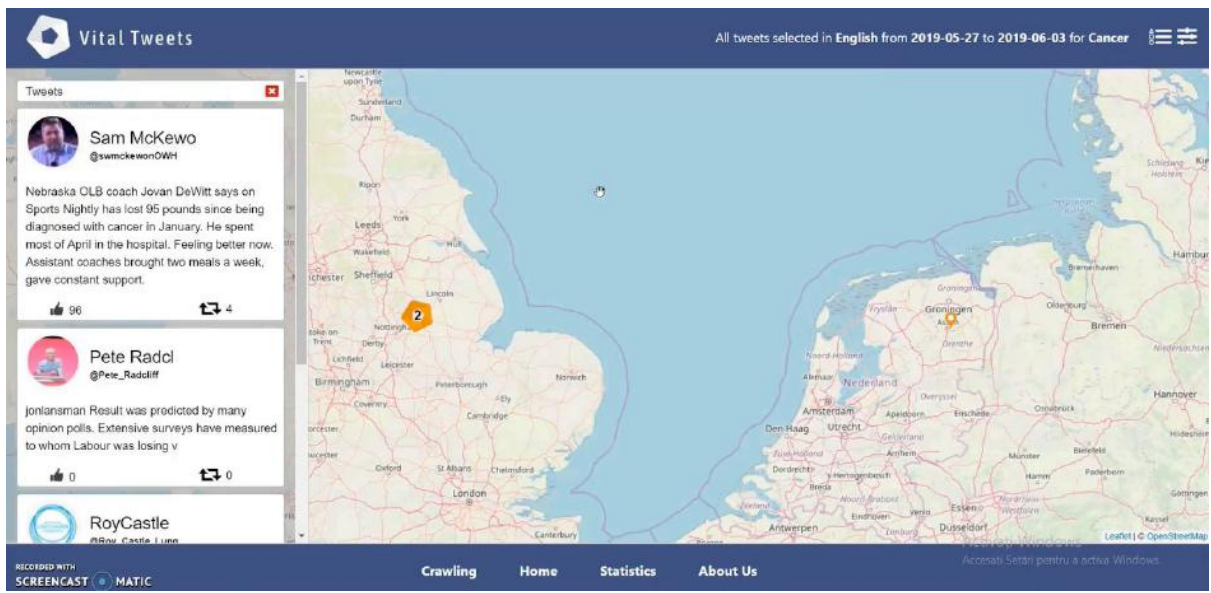

**Figure 54**: Details about two tweets which contain word "cancer"[146]

The identified problems are related to the situations when words in the medical field are used with other meanings than their basic meanings. For example, the "ciuma roşie" ("red plague"), which was widely used but with a different meaning during the online protests against the Romanian government, may lead us to the wrong conclusion that we have a plague epidemic in Romania. For this reason, filtering the relevant data and removing unnecessary information becomes a very important problem, which will require more attention from us.

**New Technologies**

The use of new technologies in the creation of applications is a direction that every year attracts new students, who want to be constantly updated with the latest discoveries and technologies. Although for this it is often necessary to invest money to buy their devices and time to learn how to use them, students find ways to get involved in such projects. The big advantage is that most of the IT companies in Iasi are involved in such projects, on the one hand, they are interested in increasing such skills to our students, on the other hand, they already have research laboratories within companies, where they can involve students during internships. Now, we are in preparation with two projects (1) **Digital Botanical Garden** and (2) **Using new technologies and artificial intelligence in medical domain**.

---

[146] At this project were involved students from group A2 at the programming engineering labs in 2019

**Digital Botanical Garden** project intents to attract more people to visit it, but also to help them with useful information during the visit. Thus, we want to include a module of augmented reality which will be able to recognize fauna and flora, based on markers in the case of flowers, which are in the central part of the park, which have identification plates with their name, or based on image recognition for others or for birds and insects. Another module, will try to recognize the bird after trill, using sound processing, especially during the evening, when they are not visible. Our intuition is that by combining visual processing with sound processing, we will improve the quality of the bird recognition component. The problems will be related to finding the optimal balance between the number of resources we will use (the more quality the better) and the speed with which we want to achieve the results (the more resources, the longer it will take to process them). Other components will be responsible for guiding the visitor to the points of interest in the garden (using the GPS of the smartphone) and notifying the users of the news and events that take place in the botanical garden.

**Medical Domain**

The active involvement in the ImagoMol cluster has led to the beginning of collaborations with professors from UMF, who wish to benefit from the latest research in computer science, new technologies and artificial intelligence. Thus, they want to create eLearning materials with augmented reality, that allow interaction and control with voice and gestures.

Another direction of collaboration with our colleagues from ImagoMol cluster is related to REVERT (taRgeted thErapy for adVanced colorEctal canceR paTients) project, where we are involved in work package 3. In WP3, we will create a platform that allows the stratification of patients at risk according to their individual response to a treatment with already approved drugs. This will be the basis for developing a decision-support system that identifies relevant signatures within the molecular and clinical data of the patient. A mobile app will be designed to make predictive models available to the final users to collect data and send alerts and messages in real time. The challenges from this project will be related to (1) *identification of new ways to collect useful patient data* using sensors from the mobile phone, smart bracelets and why not with the help of devices that we will design with colleagues from UTI (Technical University Iasi), (2) *use* *this data in algorithms that use artificial intelligence to make predictions* related to which is the best treatment for a particular patient and related to the evolution of the patient's health.

## IV.3 Research Activity

**Published Results**

Until now, the research activity has led to publication of over 200 papers in conferences and journals[147] (h-index is 11 on Google Scholar[148], 10 on AMiner[149], 5 on Scopus[150] and 4 on Web of Science[151]). Some of them are indexed in international databases:

- 85 papers are indexed in DBLP database;
- 67 papers are indexed in SCOPUS database;
- 41 papers are indexed in ISI Web of Knowledge database;
- 11 papers appeared in International Journals;
- 14 papers are Published by IEEE Computer Society;
- 15 papers are published in LNCS by Springer.

Many of the reported results from this thesis were accepted and presented at important conferences such as: ISD (International Conference on Information System Development) rang A (4 papers), KES (International Conference on Knowledge-Based and Intelligent Information & Engineering Systems) rang B (7 papers), Cicling (International Conference on Computational Linguistics and Intelligent Text Processing) rang B (1 paper and 2 posters) and SemEval Workshop at ACL (International Workshop on Semantic Evaluation at Association for Computational Linguistics) workshop at A* conference (5 papers). 23 other papers were published at conferences of rang C or in LNCS volumes: SYNASC (International Symposium on Symbolic and Numeric Algorithms for Scientific Computing), CLEF (Cross-Language Evaluation Forum), ICCCI (International Conference on Computational Collective Intelligence), ICCP (IEEE International Conference on Intelligent Computer Communication and Processing) and LREC (International Language Resources and Evaluation). 14 papers were published in the following journals: Romanian Journal of Human-Computer Interaction, Computer Science Journal of Moldova, International Journal of Computers, Communications & Control and Research in Computing Science Journal.

---

[147] http://profs.info.uaic.ro/~adiftene/publications.html
[148] http://scholar.google.ro/citations?user=p2ScknsAAAAJ&hl=en
[149] https://aminer.org/profile/adrian-iftene/53f44585dabfaeecd69a9430
[150] https://www.scopus.com/authid/detail.uri?origin=resultslist&authorId=23397232600&zone=
[151] http://apps.webofknowledge.com/UA_GeneralSearch_input.do?product=UA&search_mode=GeneralSearch&SID=R2yh3jUmsSiJsHWzg6Q&preferencesSaved=

**Personal Contributions in Published Results**

The personal contributions of the author of this thesis of habilitation in the scientific papers published by him come from the following:

- initiator of the research activities and the publication activities of the scientific articles;
- basic contributor to state-of-the-art section. These were most often a continuation of the studies done by him in the PrivateSky project[152], what were published on iTransfer platform[153];
- active participant in the design of the architecture of the systems presented in these papers. These have often been discussed with students in the laboratory hours of programming engineering (undergraduate) or advanced techniques of programming engineering (at the master's degree). During the degree and master's degree theses of the students, these architectures evolved and were supplemented with elements that improved the quality of the solutions from certain points of view (quality, speed, usability, etc.);
- proposals for evaluating solutions and comparing them with existing solutions or using metrics known in the specialized literature (on the social network side);
- testing applications that use new technologies and conducting usability tests with students, who have not been involved in the development of software products (to eliminate their subjectivity) or with students from primary, secondary and high school classes;
- analysis of problems and errors that appeared in the proposed solutions and the proposal of ways to solve them.

The former students actively contributed to the implementation of the architecture presented in the works, came up with ideas for improving these solutions and participated actively in the testing of the applications. The other co-authors of the papers participated in discussions with the students, came up with ideas and actively contributed to the writing of scientific papers.

---

[152] https://profs.info.uaic.ro/~ads/PrivateSky/
[153] http://itransfer.space/author/adriftauth/

**Research Projects and Mobilities**

As a member of the NLP group[154] from Faculty of Computer Science, or as a member of Euronest cluster[155] and ImagoMol cluster[156], the author of this thesis was involved in writing many proposals of projects for almost all types of calls from the last 15 years.

Until now, the author was **member** in 25 projects (enetCollect, CyberParks, PrivateSky, Promote scientific research in the field of forensics in judicial activity, SIMPAS, STAGES, Investing in human resources – quality and efficiency in agriculture and services (IRU-CEAS), The training of professors in .Net technology, MetaNet4U, LiSS, ALEAR, SIR-RESDEC, eDtlr, eManage, GRAI, InterOB, LT4eL, RolTech, ROTEL, AMASS, Methods inspired from nature in graph coloring problems), **project manager** in MUCKE project and in CompetIT&C project, **partner responsible** in EVALSYS, **scientific advisor** in STAGES project.

Participant in Erasmus+ program with mobilities at the following universities: (1) Dostoevsky Omsk State University, Russian Federation, September 2019, (2) Shizuoka University, Hamamatsu campus, Japan, April 2018, (3) Bar-Ilan University, Tel Aviv, Israel, June 2016 and in eMerge Erasmus Mundus at (4) "Alecu Russo" University, Bălți, Republic of Moldova, April and June 2015.

**Participation and Involvement in Events**

Participant at over 100 events, the author was involved active in organizing over 25 events, being 3 times the chair of the conferences (RoCHI-2016 (International Conference on Human-Computer Interaction), EcoMedia-2016 and MFOI-2019 (Conference on Mathematical Foundations of Informatics)), 2 times co-Chair (MFOI-2016 and MFOI-2018), 2 times Associated Chair (RoCHI-2018 and RoCHI-2019), 1 time Tutorial Chair (ECIR-2015 (European Conference on Information Retrieval)) and over 20 times in the organizational committee (ConsILR (International Conference on Linguistic Resources and Tools for Processing the Natural Language), BringItOn, Eurolan, RoCHI and MFOI).

---

[154] http://nlptools.info.uaic.ro/index.jsp
[155] http://clustereuronest.ro/
[156] https://www.imago-mol.ro/

At some of the events the author had invited lectures:

- *Identification of fake news on the Internet,* The 3nd Conference on Legal perspectives on the Internet, "Law Evolution Through Technology", Law Faculty, "Alexandru Ioan Cuza" University, October 26, 2019, Iaşi, Romania
- *Current threats to cyber security*, The 2nd Conference on Legal perspectives on the Internet, Law Faculty, "Alexandru Ioan Cuza" University, October 27, 2018, Iaşi, Romania
- *New trends in Computer Science. What will be new*?, Dies Academici, "Alexandru Ioan Cuza" University October 26, 2017, Iasi, Romania.
- *Using Text Processing in a Multimedia Environment*, The 9th Conference on Speech Technology and Human-Computer Dialogue (SpeD 2017), July 6-9, Bucharest, Romania.
- *Using augmented reality in eLearning*, Teach for future Conference, May 12, 2016, Iasi City Hall, Iasi, Romania
- *If you want your talk be fluent, think lazy*!, The 6th International Conference on Speech Technology and Human-Computer Dialogue (Sped2011), May 18-21, 2011, Brasov, Romania (with Dan Cristea).

He has also been invited to review committees at many conferences. We mention here conferences like the ACL (Association for Computational Linguistics), EMNLP-IJCNLP (Empirical Methods in Natural Language Processing and International Joint Conference on Natural Language Processing), COLING (International Conference on Computational Linguistics), IJNCLP (International Joint Conference on Natural Language Processing), ECIR (European Conference on Information Retrieval), SemEval (International Workshop on Semantic Evaluation), KEPT (Knowledge Engineering: Principles and Techniques Conference), IMRSMCA (International Workshop on Machine Reading for Social Media Content Analytics), RANLP (Recent advances in Natural Language Processing), FSDM (International Conference on Fuzzy Systems and Data Mining), BCI (Balkan Conference on Informatics), INISTA (International Conference on INnovations in Intelligent SysTems and Applications), CSCS (International Conference on Control Systems and Computer Science), MITI (International Conference on Mathematics, Informatics and Information Technologies), RoCHI (International Conference on Human-Computer Interaction), ConsILR (International Conference on Linguistic Resources and Tools for Processing the Natural Language), MFOI (Conference on Mathematical Foundations of Informatics) and BringItOn.

# List of Figures

# List of Acronyms

| | |
|---|---|
| AI | Artificial Intelligence |
| API | Application Programming Interface |
| AR | Augmented Reality |
| ASK | Alexa Skill Kit |
| AWS | Amazon Web Services |
| CAT | Computer Adaptive Testing |
| CLEF | Cross-Language Evaluation Forum |
| CLI | Command Line Interface |
| ECRC | European Computer-Industry Research Center |
| EXIF | Exchangeable Image File Format |
| GDPR | General Data Protection Regulation |
| GPS | Global Positioning System |
| HMD | Head Mounted Displays |
| HMS | Helmet-Mounted Sights |
| HUDs | Head-Up Displays |
| IAM | Identity and Access Management |
| IoT | Internet Of Things |
| JSON | JavaScript Object Notation |
| K-NN | K-Nearest Neighbor |
| LIRE | Lucene Image REtrieval |
| MARTA | Mobile Augmented Reality Technical Assistance |
| MREAL | Mixed Reality System |
| MUCKE | Multimedia and User Credibility Knowledge Extraction |
| NLP | Natural Language Processing |
| RAM | Random-Access Memory |
| REVERT | taRgeted thErapy for adVanced colorEctal canceR paTients |
| SASM | Semantic Analysis in Social Media |
| SIFT | Scale-Invariant Feature Transform |
| SOLO | Structure of Observed Learning Outcome |
| UAIC | "Alexandru Ioan Cuza" University of Iasi |
| UMF | "Grigore T. Popa" University of Medicine and Pharmacy of Iasi |
| UTI | Gheorghe Asachi Technical University of Iasi |
| VR | Virtual Reality |
| YAGO | Yet Another Great Ontology |